# Exploration and Coverage with Swarms of Settling Agents

Rappel, Ori
Faculty of Aerospace
Engineering
Technion, Haifa 32000,
Israel

Ben-Asher, Joseph
Faculty of Aerospace
Engineering
Technion, Haifa
32000, Israel

Bruckstein, Alfred
Computer Science
Department
Technion, Haifa
32000, Israel

*We consider several algorithms for exploring and filling an unknown, connected region, by simple, airborne agents. The agents are assumed to be identical, autonomous, anonymous and to have a finite amount of memory. The region is modeled as a connected sub-set of a regular grid composed of square cells. The algorithms described herein are suited for Micro Air Vehicles (MAV) since these air vehicles enable unobstructed views of the ground below and can move freely in space at various heights. The agents explore the region by applying various action-rules based on locally acquired information Some of them may settle in unoccupied cells as the exploration progresses. Settled agents become "virtual pheromones" for the exploration and coverage process, beacons that subsequently aid the remaining, and still exploring, mobile agents. We introduce a backward propagating information diffusion process as a way to implement a deterministic indicator of process termination and guide the mobile agents.*

*For the proposed algorithms, complete covering of the graph in finite time is guaranteed when the size of the region is fixed. Bounds on the coverage times are also derived. Extensive simulation results exhibit good agreement with the theoretical predictions.*

# 1 Introduction

In many real-life scenarios, one is interested in deploying agents over an unknown region in order to explore it and possibly create an environment capable of detecting disturbances, localized activities and various events of interest. Often such scenarios are characterized by no a-priori knowledge on the region to be explored and covered, movement on the ground may be hampered by rubble and debris, agents lacking a global reference frame and limited communication capabilities with other agents and/or with the deploying station. For example, operating in a post-earthquake, collapsed building, rescue workers would obviously benefit from having a map of the free space inside the wreckage and a sensor network that can detect survivors. In general, unknown indoor environments present the major challenges of difficult ground movement, lack of a global reference frame (due to the impossibility of receiving the GPS signal) and severe degradation of RF communication due to multi-path and absorption effects. Sensing ranges, in most real-life scenarios, are also severely limited due to irregularities in the terrain. These irregularities might also hamper and slow down the movement of ground-based exploration and rescue robots.

We present here the concept of exploring an unknown region by a swarm of flying drones (agents) that eventually uniformly cover the region. The agents – identical, autonomous, and anonymous - enter the region one after another from entry points and explore the region in a distributed manner according to some predefined action rules based on locally sensed information. While flying drones (MAV's) have many advantageous in search and rescue scenarios, their main drawback is a high power consumption during flight combined with limited, on-board energy (i.e., battery). This is especially critical in the aforementioned scenarios in which agent size is severely limited and consequently so are the battery weight and agent capabilities. The solution is to design exploration and coverage methods that put an emphasis on minimizing the flight time. In addition, the payload capacity of small MAV's is limited and the deployment of dispensable beacons is not relevant. Therefore, as a meta-rule, some of the agents settle and become beacons that can guide the exploration by the remaining mobile agents. The resulting sensing and covering infrastructure, effectively a directed acyclic graph of the region (or a directed forest if more than one entry point exists) composed of settled agents, may be used to generate implicitly a map and an ad-hoc sensor network to aid and guide subsequent operations.

*Exploration* is commonly defined in robotics research as the process of visiting of every part of a region by at least one agent at least once. In the context of this article, *uniform filling* (dispersal) refers to uniformly distributing agents in an unknown (but bounded) environment hence a uniformly filled region is obviously an explored region. *Deployment* is the process through which the agents enter the region, and the combination of deployment with the filling process must result in uniform coverage of the region of interest with agents Note, some researchers refer by "uniform filling/ dispersal" to the combined deployment and uniform distribution of agents.

The exploration strategies described in this paper are based on two basic meta-concepts. The first concept is the use of agents in the Dual Role of (mobile) Explorers and of Physical Beacons or Pheromones that guide exploring agents while also creating the uniform coverage. This concept is especially relevant for small, payload-limited flying agents as explained above. The idea draws upon the concept of stigmergy in biology which is defined by Kennedy et al [1] as *"communication by altering the state of the environment in a way that will affect the behaviors of others for whom the environment is a stimulus"*. Our implementation of stigmergy utilizes the settled agents as the medium to change the state of the environment, in the form of beacons. The moving and exploring agents use these changes in the environment to guide their future actions. This process obviates the need for other markers and/or the need to rely on the environment to store information. An agent entering the region starts as a mobile agent and hence may move, once at every time step, to one of its neighboring cells in the region. The agents can also decide to become stationary and settle (i.e., become a beacon) if the proper conditions are met. The change from a mobile agent to a stationary one is considered irreversible. Once settled, an agent serves as a "beacon" to other, mobile agents thereby guiding their movement. The conditions for a change of the state and the actions taken in each state differentiate between the algorithms described herein. Once an agent settles its energy consumption decreases significantly. Thus, the Dual Role meta-concept helps decrease the energy consumption of the agents, generate the coverage, and guide the exploring agents. Another way to look on the covered surface is as a "shared memory" serving the mobile agents. And just like we want a shared memory to be reliable and scalable we require the same from the coverage formed by the settled agents.

As described above, the users deploying the swarm need an indication when the exploration and coverage of the region terminates (i.e., all the region has been explored and is uniformly covered). In addition, directing drones to empty regions can mitigate the effect of limited available energy. The second meta-concept, termed Backward Propagating Closure, realizes the requirement for a deterministic indication of termination and will be shown to effectively funnel drones to empty regions. Settled agents in fully explored, and therefore completely covered areas signal that further exploration is unnecessary by displaying *Closed* to already explored sections of the region. This signal, starting with settled agents closest to the boundaries of the region propagates "backwards", like a wave front, towards to the deployment point/s. Thus, when the settled agent at the entry point/s displays Closed signal it indicates to the user that all the cells of the region are already filled hence the exploration is completed.

The algorithms developed and analyzed in this paper are divided into two families - the Dual Layer Coverage, DLC for short, algorithms and the Single Layer Coverage (SLC) algorithms. The DLC algorithms result in two layers of uniform coverage of the region – one of settled agents and one of mobile agents hovering above - whereas the SLC algorithms result in a uniform coverage of the region by single layer of settled agents

In the first algorithm presented the agents do a random walk over the region and continue entering the region until each cell is occupied by two agents – one mobile and one settled. A long cover time is the main drawback of such a random process and therefore the subsequent algorithms utilize "gradient-based" exploration strategies. In these processes, mobile agents are guided to empty areas by moving up the "gradient" implicitly generated by the signaling settled agents. These algorithms assume the agents can read the gradient displayed by the settled agents and keep in memory a value corresponding to the value displayed by the beacon below. Using these capabilities, the cover time is significantly reduced. Termination of the exploration and coverage process in the DLC algorithms is indicated by waiting for the next agent deployment a sufficiently long time. The SLC family of algorithms utilize (in addition to the concepts described above) the Backward Propagating Closure in order to both generate a deterministic indication of process termination and funnel the agents to empty regions. Consequently, these algorithms also show a reduction in the number of agents used in the coverage process and in the combined energy consumption compared to the DLC algorithms. However, these improvements come at the cost of additional sensing and signaling capabilities by the agents. Extensive experimental results show good correlation with theoretical predictions of performance.

In the following section, a review of related work is presented. Next, definitions and assumptions pertinent to all the algorithms discussed in this paper are presented. In sections 4 and 5, the DLC and SLC algorithms are described in detail and analyzed. Experimental results are discussed in section 6.

The main contributions of this paper are as follows: Seven algorithms for the exploration of unknown indoor regions by uniformly covering the region with agents are presented. A modified definition of the term termination time is introduced in which termination is achieved only when an external observer can positively detect termination. A novel concept to propagate information about coverage completion is introduced. It is used to give the external observer the needed indication and guide the agents during the coverage process. The mutual interaction between the entry process of the agents into the region and their movement in the region is analyzed. Theoretical and experimental results show the termination time is linear with region size.

## 2  Related Work

This work further develops themes and ideas that originally appeared in [2] and [3]. Preliminary results on the coverage of unknown indoor regions by a swarm of MAV's using DFS and the efficiency of a backward propagating information diffusion process was presented in [2]. In [3] the authors present and rigorously prove results about "DLLG", a dual-layer coverage algorithm, and present "SLUG", a backward propagating single-layer algorithm. The authors of [3] further compare the performance of these algorithms in terms of termination time and energy use. The concepts at the base of these algorithms are further developed in this work and additional algorithms are presented. We compare these empirically and analytically and present mathematical proofs and conjectures about their performance.

Numerous papers have been published on coverage and/or exploration of unknown regions using swarms of agents. The strategies used to spread the swarm of agents in a region can be divided into three broad categories - using virtual/artificial physics [4], separation into Voronoi partitions based on a utility function [5] and those using graph-theoretic concepts such as Depth First Search [6]. The various possibilities for required outcomes, combined with the large number of possible assumptions about the capabilities of the agents (memory, sensing, swarm size, etc.'), along with the various activation (i.e., timing) and deployment schemes and the assumed topologies of the unknown environment resulted in rich and diverse research topics and thousands of papers and reports. Comparison between the various strategies is henceforth quite problematic. A review of the common assumptions and required outcomes is given by Das in [7, p. 403]. In the next several paragraphs a brief review of the prior research most relevant to our work is presented. A more extensive overview of the literature can be found in Flocchini et al [8].

Filling and more specifically uniform filling (or dispersal) has been extensively studied and continues to be the focus of many researchers. In the field of mobile swarm robotics this problem is studied as a means to achieve a global behavior such as exploration, coverage, or a specific formation pattern. This vast body of work is broadly divided between those using centralized control (i.e., driven by a central entity that receives reports from agents and consequently commands their motions) control or decentralized control (i.e., based on designing independent behavioral rules for the agents resulting in the desired global outcome).

This work follows and extends previous research by Wagner et al. [9]. In their work, the authors assume a group of mobile agents tasked with exploring, but not uniformly covering, all the vertices of a graph-like environment guided by virtual pheromones [10]. The problem of uniform filling was formulated by Hsiang et al. [11] for the case of ground robots. In this paper the authors describe several distributed algorithms based on the 'follow-the-leader" concept and Depth First Search or Breadth First Search strategies to guide the filling process. The agents enter the region via a door and the objective of the algorithms is to minimize the filling time defined as the time until all the cells are filled with an agent (called the "make span"). The paper proves an upper bound for the filling time based on three noteworthy assumptions. The first is that the number of agents in a cell is limited to one. The second is that the robots are synchronous and the third that the agents have a finite memory. All of these methods assume an initially unknown but bounded and connected region modeled as a subset of a regular grid.

Theoretical bounds on the minimal capabilities (sensing range, memory) of the agent required to achieve uniform filling are derived by Barrameda et al. [12] [13]. The authors show agents must have some persistent and finite memory in order to cover a non-simply connected region. In [12] the authors assume the robots are asynchronous and consider the filling of simply connected regions (i.e., without holes). The authors then present lower bounds on the visibility range and amount of memory needed to reach uniform coverage. The results are based on the 'meta-rule' that a robot never backtracks. In [13] the authors present bounds on the memory and visibility range of robots needed to fill non-simply connected regions. They use the follow-the-leader paradigm and in order to deal with holes they use robots to temporarily block the multiple possible ways of advance. A similar concept is presented in [14].

Some researchers assume that the initial location of the agents is inside the region based on some arbitrary spatial distribution [15], [16]. A more common assumption it that the initial location of the agents is outside the region, as is the case in this research. If this is the case, a deployment process is commonly defined [17] [18], [19], [20]. If this is the case a deployment process should be defined (i.e., the agents should enter the region of interest according to some well defined timing strategy).

Additional ways to divide the body of work is according to the way the region is modeled. The region to be filled may be a part of the two-dimensional plane (i.e., continuous) [17], [4], [21] or modeled as graph (usually as grid graph) [15]. Moreover, the region may restricted to be simply-connected which means none of the obstacles is completely surrounded by empty cells [12], [15] however a more realistic and common assumption is for the region to be non-simply connected [21], [22], [23], [24]. In this research we assume the region of interest can be modeled as a non-simply connected grid graph since the presence of pillars and/or piles of rubble cannot be ruled out.

In the vast majority of the research the concept of recurring action-cycles is used to describe the behavior of an agent. Hence an additional way to organize the body of work is according to the scheduler (fully synchronous, semi-synchronous, or asynchronous) used to define the relationship between the action cycles of the different agents. Défago et al. [25] present a review of the various models including the time scheme used in this research.

While most research assume the agents are ground robots, the advantages of airborne robots in exploration (i.e., increased field of view) have been recognized by some researchers in the last decade. The idea of using airborne drones as a beacon was first mentioned by Stirling et al. [17], [18] and later by Aznar et al. [26], [27], with the objective of exploring (not uniformly filling) a region. In [17], [18],[19], [26], [27] each drone has a unique Id with the exploration strategies based on either depth first search [17], [18] or on a potential field generated by virtual pheromones in [26], [27]. In Rappel et al. [2] the authors present a DFS-based uniform coverage strategy by airborne agents however the issue of collision avoidance, critical when using airborne agents, is problematic since the number of mobile agents in each cell is not limited.

Energy is the prime constraint on the performance of swarms composed of flying agents. Consequently, most papers that assume flying agents also discuss the energy consumption and use it as one of the performance metrics. Stirling et al. [17] proposed an energy model that takes into account the energy consumption of a drone before its entry into the region, during movement and when operating as a beacon. Based on this model the total energy consumption is computed and sensitivity to different parameters analyzed. Aznar et al. [26] also use the total energy consumption with recurrence equations used to predict the total energy.

A settled agent can signal other agents using various mediums – RF, acoustic, and visual amongst others. Possibly the best solution for the scenarios described above is visual as it is both localized and secure from interference by neighboring agents. First suggested by Peleg [28] as a means of signaling with lights without explicit communication, the idea has been since extended for the lights to be used only as a memory [29]. Das et al. [30] investigate additional capabilities imparted to agents with lights. Prior to this work the idea was used by Poudel et al. [31] that present a uniform scattering algorithm that is linear with time for synchronous robots with lights, a compass and a visibility range of 2. It is important to note that comparison of the performance metrics (e.g., cover time) of the various approaches, especially the issue of energy consumption, is problematic due to the differing assumptions about the agent's capabilities, the modeling of the region and the time scheme.

## 3  Preliminaries

In this section definitions and assumption that are pertinent to all the algorithms are described.

### 3.1  Problem Formulation

In the *coverage problem*, a swarm of drones seeks to enter and uniformly cover an a-priori unknown region **R** composed of interconnected cells in a decentralized manner. In the terminal configuration every cell is occupied by two drones in the *dual layer coverage* problem or by one drone in the *single layer coverage* problem

The unknown region, **R**, to be covered is a connected region that is modeled as an orthogonal grid graph and access to the region is via well-defined *entry point/s*. It is an arbitrary, possibly non-simply connected grid composed of equally sized cells forming the empty region to be explored surrounded by walls and incorporating obstacles. Each cell of **R** can contain a single mobile agent and/or a single settled agent. The agents enter the region in a prescribed order from an "agent source" or a deployment point that is adjacent to the entry point and that has a sufficient number of drones to

The region **R** can also be therefore represented by graph a $\mathbb{G}(V, E)$ with the vertices, *V*, representing the empty cells and the edges, *E*, representing the connections between adjacent cells. In the rest of this paper, the terms cell and vertex are interchangeable. Unless stated otherwise, $n = |V|$ i.e., the cardinality of the graph is equal to the number of empty cells in the region and $m = |E|$. An example, non-simply connected region with one entry point, eight rooms and a double corridor is depicted in Figure 1.

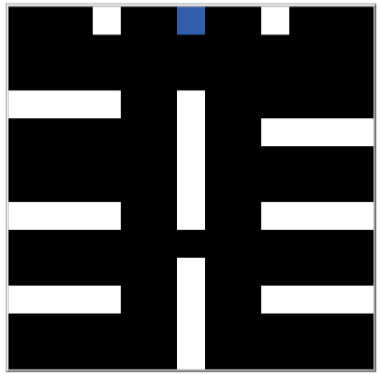

Figure 1  Typical Region. The empty cells are marked in black, the obstacles in white and the entry point in blue

## 3.1 Timing model

Time, $t$, is discretized such that $t = 0, 1, 2, 3, \ldots, k, \ldots$ and each step is further subdivided into M (a very large integer) equal sub time-steps, $dt$, such that $dt = 1/M$. Each agent wakes up once in every time step at a randomly selected sub time-step as given by Definition **1** and Eq. (1). This is the most probable time model in real–life, decentralized multi-agent systems in which each agent has its own internal timer that controls when it wakes up. For sufficiently large values of M the scheduler is asynchronous and the probability that two agents will wake up in the same instance is negligible. The time between successive action cycles of a specific agent is bounded by (0,2) and the number of activations of agent $a_j$ between two successive action cycles of agent $a_i$ is in the range [0:2]. Such time models are called fair 2-bounded asynchronous schedulers in the literature [25].

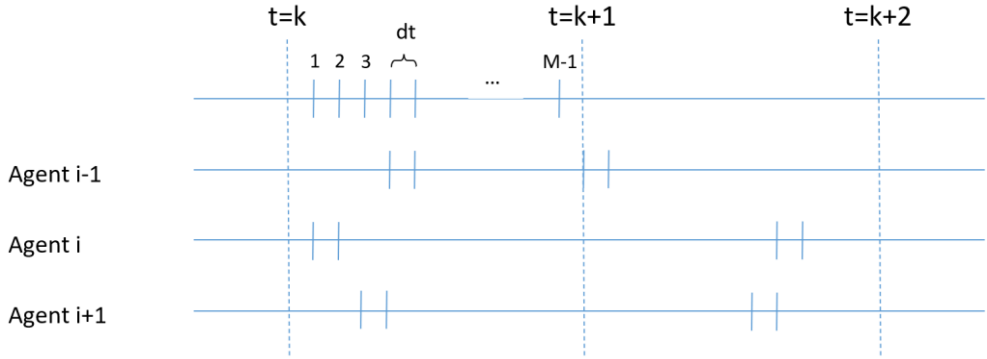

Figure 2   An illustration of the asynchronous time model. The top row shows the subdivision of each time step into sub time steps of size $dt$. Each agent wakes up exactly once per time step $t$ at one of these sub-steps

**Definition 1**: The wake-up time of agent $a_i$ at time step $k$ is denoted by $t_{i,k}$ and defined as:

$$t_{i,k} = t_k + \frac{U(1, M-1)}{M} \qquad (1)$$

With $U(1, M-1)$ denoting the uniform distribution over the integers $\{0, 2, \ldots M-1\}$.

## 3.2 The Agents

The agents are assumed to be *Miniature Air Vehicles* (MAV's) capable of motion in the horizontal plane (i.e., parallel to the to the surface) and in the vertical plane (i.e., perpendicular to the surface). In the horizontal plane the agents move in one of four principal directions (i.e., "Manhattan Walk") whereas in the vertical plane the agents can only move straight down (i.e., parallel to the gravity vector). Hence, at every time step, every agent is either in flight and searching for a place to land, or has already landed somewhere in **R**. The former type of agent is called *mobile* whereas the latter is called *settled*. Agents always start as mobile agents and may settle at some time step. Settled agents are assumed to remain in place forever, providing useful information to the mobile agents. Each agent can sense the presence of agents in any adjacent cell at Manhattan distance 1 from it (i.e., the sensing range of agents is 1). Agents can sense whether a neighboring agent is mobile or settled. While mobile agents can sense both mobile and settled agents, the settled agents can only sense neighboring settled agents. Figure 3 depicts the sensing volume and possible directions of movement.

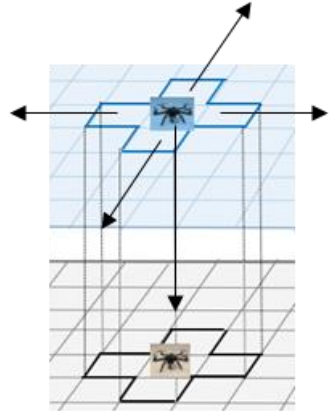

Figure 3  Sensing and moving region of an agent. The ground, occupied by a settled agent is marked in grey while the plane in which the flying agents move is marked in blue. The mobile agent, a drone, is shown both in the air (dark blue) and settled on the ground (brown). The arrows denote the possible directions of movement

**Definition 2**: The set of all entities – agents and/or walls - sensed by agent $a_i$ at a cell (or node) $u \in V$ at time $t$ is denoted by $\xi_u(i, t)$. This set is also called "the neighbors of $a_i$ at time $t$ at cell $u$". For brevity, we omit the reference to both a specific time and agent whenever possible. The set may include up to 8 agents since each cell may be occupied by both a settled agent and a mobile agent. As $\mathbb{G}(V, E)$ is a grid graph and the agents perform a Manhattan-type walks, the positions of the neighbors of agent $a_i$ are enumerated systematically according to Figure 4. A mobile agent in cell $u$ is enumerated as "5" while a settled agent in cell $u$ is enumerated as "0". Furthermore, when discussing the agent directly underneath a mobile agent in cell $u$ reference is made to position "0".

The set $\xi_u(i,t)$ is hence the 10-tuple $\xi_u(i,t) = \{\xi_0, \xi_1, \ldots, \xi_9\}$ where $\xi_j$ has a value from the set $\{-1, 0, S_j\}$ corresponding to {wall, empty, state of agent j}. The set $\xi_u^G$ refers to the sub-set of sensed entities on the ground, $\xi_u^G = \{\xi_j : j = 0:4\}$ while the set $\xi_u^A$ refers to the sub-set of mobile neighboring entities in the air, $\xi_u^A = \{\xi_j : j = 5:9\}$. Again, for brevity, we omit the reference to a specific cell, agent or time step when it is obvious from the context.

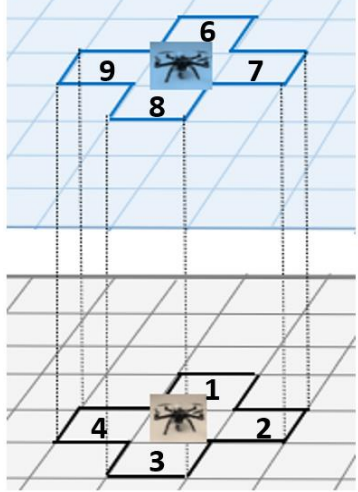

Figure 4  Neighboring region enumeration

The agents are identical, anonymous (i.e., have no identifiers) and autonomous (i.e., have no central controller). The agents may also utilize a small, limited memory–depending on the algorithm. At every wake-up, the agent does a Look, Compute and Action (LCA) action cycle [32] in which it senses its surroundings, makes a decision and implements the action. All actions are assumed to be instantaneous (i.e., atomic). The agent may do several types of actions. The first is "move" from one cell to another; the second is "settle in place", the third is "settle in a neighboring cell" which is composed of a moving to the relevant cell and settling in place and the fourth is "signal" in which a settled agent becomes a beacon and projects (e.g. using lights) some information.

The energy consumption of an agent is the sum of the energy consumed while mobile and the energy consumed while settled.

$$E_i = E_m^i + E_s^i \tag{2}$$

Assuming constant power consumption in the mobile and settled states Eq. (2) can be re-written as:

$$E_i = P_m t_m^i + P_s t_s^i \tag{3}$$

Assuming $P_m \gg P_s$ we get that

$$E_i \cong t_m^i \tag{4}$$

**Definition 3**: The vertex location of the agent $a_i$ at time $t$ is denoted by $v(a_i, t)$

### 3.3 Agent Entry Model

When dealing with the covering problem, there is the issue of the initial location of the agents. In real life applications such as Search and Rescue it is obvious the agents are initially outside the region. Clearly and fully defining the interaction between the deployment process and the coverage algorithm is critical when trying to develop tight bounds on the coverage time. Several deployment strategies are discussed in the literature ([18], [19], [20]). Using the terminology coined in [18], the first and simplest strategy is Linear-Temporal Incremental Deployment (LTID) in which agents attempt to enter the region at a constant rate. The second is Single Incremental Deployment (SID) in which an agent is deployed at a constant rate but only if the previously deployed agent has become a beacon. Hence, this strategy requires either propagation of messages across the network of beacons with the additional delays and complexity such a communication protocol entails or long-range wireless communication with the accompanying problems of multi path attenuation in the enclosed region. Experimental results from [18] indicate this strategy is advantageous when DFS is used since the entering agent will travel the shortest path to the exploration frontier. When greedy exploration is used the advantages of this strategy are doubtful. The third strategy Adaptive Group Size (AGS) deploys agents into the region depending on the density of mobile agents. The mobile agents are assumed to measure the density of mobile around in their sensing neighborhood. The main advantage is localized reductions in the number of mobile agents but at a cost of additional sensing and complex logic. All three deployment strategies described above were developed in conjunction with a DFS exploration strategy. In [19] the authors present a hybrid entry model called FGID combining SID and AGS that allows the number of mobile agents to be greater than one and does away with the dependency on the density of mobile agents. In [20] an agent at the exploration frontier signals that additional agents are to enter the region depending on the number of empty cells it detects. While this model is adaptive to the explored environment it requires communication between the agent and the entry point/s (which is not specified in the paper). In the three papers mentioned above as well as many others the blocking of an entry point by another agent is not discussed. It is dealt with by either limiting the entry rate or (implicitly) neglecting the phenomena.

In this research the agents are initially outside the region and enter it one-at-a-time. We use the constant rate entry model in which the *time interval between successive entry-attempts*, denoted as $\Delta T$, is constant – see Figure 5. When $\Delta T > 1$, the time window in which agents may enter the region is limited to the first time step and in the other $\Delta T - 1$ time steps only the agents inside the region are active. In other words, no agent will enter the region during the interval $\Delta T$ if entry does not occur during the first time step. Whether an agent actually enters or not depends on the occupancy of the entry point. Since movement of the mobile agents out of the entry point/s depends on the filling strategy, the actual entry rate depends on the progression of the filing process and vis-versa. The model described is similar to the LTID strategy described in [18] however unlike LTID this strategy takes into account the occupancy of the entry point/s.

The effect of $\Delta T$ on the covering process and specifically the termination time is investigated both theoretically and experimentally.

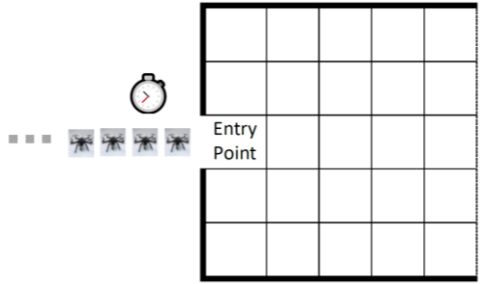

Figure 5  The Constant Rate Agent Entry model. Agents, located outside the region, may enter every $\Delta T$ time steps

### 3.4   Performance Metrics

We are interested in the performance of the swarm as a whole as it solves the coverage problem and therefore define the following three metrics – termination time, total energy used, and number of agents used. The definition of the termination time is kept intentionally vague since the two families of algorithms – DLC and SLC – have different termination criteria and these will be defined in their respective sections. As described above we assume that the power consumption of a mobile agent in flight is constant and that once an agent settles its power consumption is zero. Hence the energy consumption is linear with respect to the time an agent is mobile, and the total energy used is the sum of the flight times of all the agents. Both the total energy used and number of agents used are measures of the effort to cover the region while the termination time measures its result. In addition to these three aggregative metrics, we are also interested in the maximal energy consumption of any of the agents since it relates to the energy limit of the agents.

**Definition 4:**  The *termination time*, $T_C(\mathbf{R})$, of a coverage process done by a multi-agent system with respect to a region $R$ is the first time step that the process of covering $R$ with agents terminates.

**Definition 5:** The set of agents participating in the coverage process at time step $t = k$ is defined by $A(t) = \{a_1, a_2, \dots, a_{N(t)}\}$ and the cardinality of $A(t)$, is denoted by $N(t) = |A(t)|$. Note, when referring to the number of agents without explicit mention of the time the meaning is the number of agents at $T_C$, $N = N(T_C)$.

**Definition 6:**  The *total energy consumption* used by a multi-agent system with respect to a graph environment $\mathbf{R}$ is equal to $\sum_i E_i$ – the sum of all individual agents' energy consumptions. Total energy consumption is also denoted $E_{Total}(\mathbf{R})$.

**Definition 7:** The *maximum specific energy consumption* by any agent at the termination of the process, is denoted by $max(E_i)$, and defined by $max(E_i) = max(E_i: i = 1:N)$

## 4     Dual Layer Coverage (DLC) Algorithms

Uniform coverage in the Dual Layer Coverage family of algorithms is achieved when every cell in the region contains exactly two agents – one settled and one mobile hovering above it. All the algorithms are based on the basic concept of the dual use of agents – as explorers and as beacons – but different, local information-based action, rules are used. In the following sub-sections, the algorithms are described in detail and their performance is analyzed

**Definition 8**: The *termination time*, $T_C(\mathbf{R})$, of a coverage process done by a multi-agent system using a Dual Layer Coverage algorithm with respect to a region $\boldsymbol{R}$ is the first time step every cell in $\boldsymbol{R}$ is occupied by two agents – one settled and one mobile.

### 4.1    Dual Layer Coverage by Random Walk (DLRW)

Single agent and multi-agent random walk on graphs and grids has been investigated in numerous prior works. Notably in Alon et al. [33] the authors analyze the length of time until every cell in a graph is visited (at least once) by one of several agents doing a random walk starting from a common vertex. This length of time, defined in [33], as the cover time does not require any specific distribution of the agents when it is reached. A more relevant characteristic of a random walk is the hitting time defined as the expected time it will take an agent to move from vertex $u$ to vertex $v$, $u \neq v$; $u, v \in \boldsymbol{V}$. Patel et al. [34] derived expressions for the hitting time of random walks on graphs when the number of agents in a cell is not limited. It is, however, these constraints on the number of agents in a cell and the final agent distribution that are critical to uniform filling process.

The acronym DLRW stands for Dual Layer, Random Walk and thus describes both the required coverage of the region coverage and the essence of the local action rules used in the algorithm to achieve it. The action rules of this simple algorithm state that when there is an empty cell in the sensed region of a mobile agent it will settle in it thereby expanding the region covered. Preference is given to settling directly underneath as may happen when the first agent enters an empty region or if a settled agent fails underneath a mobile one. If settled agents already occupy all the neighboring cells, then a mobile agent will check if there are cells without a mobile agent in them. If there are such cells, then it will arbitrarily select one and move to it. If all the neighboring cells are filled by two agents (a settled agent and a moving agent), then the agent will stay in its current cell. Once an agent settles (in other words, lands) it will stop moving and become a beacon. As a beacon, its very presence indicates there is an agent in the cell and nothing more. In accordance with Definition 2 the state of agent $a_i$, denoted by $S_i$, is the set {mobile, settled}. The state machine of an agent is described in Figure 6, the flowchart of the algorithm in Figure 7 and the local action rules for agent $a_i$ at time $t$ are described by the pseudocode in Figure 8.

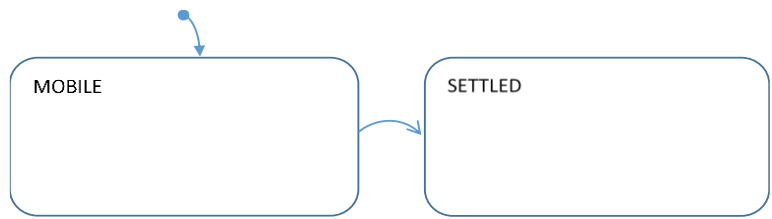

Figure 6  Dual Layer, Random Walk algorithm – agent state machine

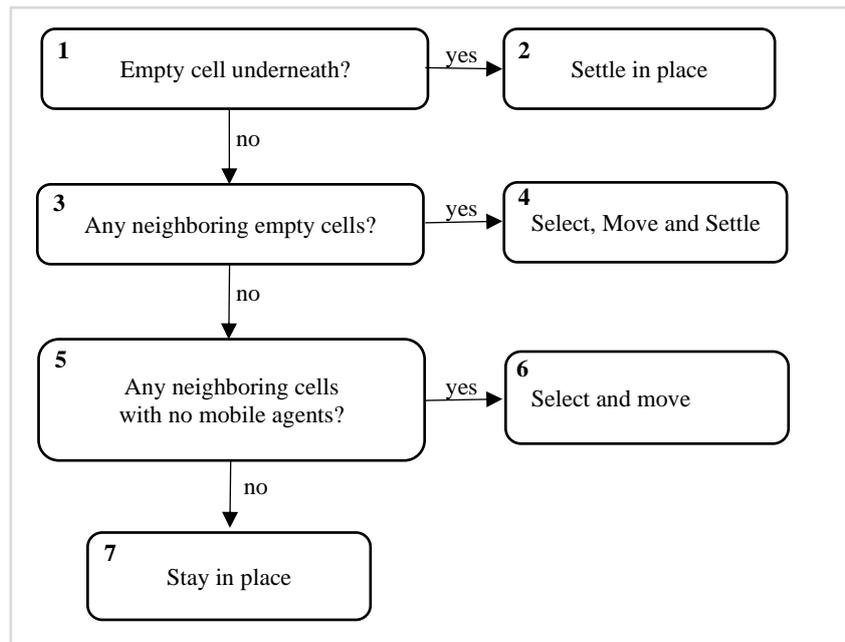

Figure 7  Flowchart of the Dual Layer, Random Walk algorithm – mobile agent action rules
Each block is numbered and referred to by this number in the pseudocode below

**Initialization**: for all agents $a \in A$ located at $t=0$ at the agent source set $S_i =$"mobile"

```
If S_i = "mobile"
    If ∃(ξ^G) = 0                              /*if there are empty, neighboring cells
        If ξ_0^G = 0                           /*if the cell directly underneath is empty [1]
            u ← u;   S_i ← "settled"           /*settle there [2]
        Else                                   /*if not [3]
            ξ̃^G ← {ξ_j^G = 0, j = 1:4}         /* define set ξ̃^G as all empty neighboring
                                               /*cells [4]
            v ← integer(random(|ξ̃^G|))         /*arbitrarily select a cell from ξ̃^G
            u ← v;   S_i ← "settled"           /*settle at selected empty cell & update state
        End If
    Else                                       /*if there are no empty, neighboring cells
        ξ̃^A = {ξ_j^A = 0, j = 1:4}             /*define set ξ̃^A as all neighboring cells with
                                               /*no mobile agents
        If |ξ̃^A| > 0                           /*if there are neighboring cells with no
                                               /*mobile agents [5]
            v ← integer(random(|ξ̃^A|))         /*arbitrarily select cell from ξ̃^A [6]
            u ← v;   S_i ← "mobile"            /*move to the selected cell
        Else                                   /* if all surrounding cells are filled by 2
                                               /*agents
            u ← u;   S_i ← "mobile"            /*stay in place [7]
        End If
    End If
End If
```

Figure 8   Dual Layer, Random Walk pseudocode

*4.1.1   The Absorbing Markov Chain representation*

The use of Markov Chains in analyzing random walks by a single agent is common (e.g. [35], [34]) . The state of the Markov chain in this case is the location of the agent. This is a possible since if the agent doing the random walks is in a certain cell, then all the other cells are necessarily unoccupied. However, this definition is irrelevant in the case of multiple agents doing a random walk and instead we define the state of the process as a vector composed of the state of each cell. When dual layer coverage is used, the state of a cell is defined by the number of agents in it.

We model the algorithm described above as an absorbing Markov chain in order to prove its eventual termination. In addition, the absorbing Markov chain representation allows calculation of the expected cover time however that is practically impossible to do so except for the simplest regions.

We denote the state of cell $i$ at time step $t$ by $S_i^t$ and the state of the coverage process at time step $t$, $S^t$, as the tuple $\{S_1^t, S_2^t, \cdots, S_n^t\}$. The state of a vertex has a value from the set $\{0, 1, 2\}$ corresponding to {empty; filled by one settled agent; filled by one settled and one mobile agent}. As described above, in the DLRW algorithm, the possible actions of each agent (and their corresponding probabilities) depend solely on its current position and neighborhood. Therefore, by definition, the possible transitions from the current state of the coverage process depend only on the current state of the agents. Since this is true for all cells and at any time step $t$, changes in the state of the coverage depend only on the current state of the process. Hence the coverage process defined above has the Markov property as defined in [36] and is thus a Markov process.

In [37], a state of a Markov chain is defined as absorbing if it is impossible to leave it. Following from above definition, a Markov chain is defined in [37] as absorbing if (a) it has at least one absorbing state, and (b) if from every state it is possible to reach an absorbing state (not necessarily in one step). Using above definition of $S^t$, the states at the beginning and at the end of the coverage process are $S^0 = \{0, 0, \cdots, 0\}$ and $S^{T_c} = \{2, 2, \cdots, 2\}$ respectively. Moreover, based on the above definition of absorbing, $S^{T_c}$ is an absorbing state since it can be reached from any state, nor can any agent leave it. Once reached, no additional agents can enter the region. Having defined the starting and absorbing states for the algorithm we can now prove that the DLRW terminates with probability 1.

**Proposition 1:** The DLRW algorithm terminates with probability 1.

Proof: See [37], Theorem 11.3

Note, that while the number of possible transient states is combinatorically upper bounded by $3^{|V|} - 1$, the actual number of transient states and the possible transitions between them may be much smaller and depends on the region's topology and the time scheme used.

Using the formulation in [36] and [37], for the case of one absorbing state and $T$ transient states we define the transition matrix, P, as:

$$P = \begin{pmatrix} Q & R \\ 0 & I_a \end{pmatrix} \quad (5)$$

Such that $Q$ is a $T \times T$ matrix describing the probability of transitions between transient states, $R$ is a $T \times 1$ vector of the transitions from the transient states to the absorbing state and $I_a$ is a $1 \times 1$ identity matrix. With these definitions, $N$, the fundamental matrix for P, is a $T \times T$ matrix defined as:

$$N = (I - Q)^{-1} \quad (6)$$

Following from Eq. (3), the expected number of steps from each transient state to the absorbing state, denoted by $\tau$, is given by Eq. (7) where $c$ is a unit column vector. Note that since $S^{T_c}$ is the absorbing state, the expected number of steps from $S^0$ until the absorbing state is reached is $\tau(1)$. In other words:

$$\tau = Nc \quad (7)$$

and

$$T_C = \tau(1) \tag{8}$$

To clarify the above result, we shall consider a simple example of a 2 by 2 region and with the entry point being in the cell marked as 1 – see Figure 9. The asynchronous time scheme described above is used and the agents enter the region once every time step. The states are defined in Figure 10 and the state transition diagram is shown in Figure 11. Possible transitions are marked by the directed edges along with their probabilities. The probabilities depend on the time scheme as it determines the possible orders of movement between the agents.

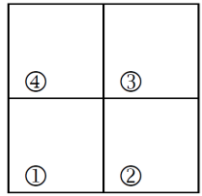

Figure 9  Absorbing Markov chain example – the region to be explored is a 2 × 2 graph with the entry point at ①

| State | $S_1$ | $S_2$ | $S_3$ | $S_4$ |
|---|---|---|---|---|
| a | 0 | 0 | 0 | 0 |
| b | 1 | 0 | 0 | 0 |
| c | 2 | 0 | 0 | 0 |
| d | 2 | 1 | 0 | 0 |
| e | 2 | 0 | 0 | 1 |
| f | 2 | 1 | 0 | 1 |
| g | 2 | 2 | 0 | 1 |
| h | 2 | 1 | 0 | 2 |
| i | 2 | 2 | 1 | 1 |
| j | 2 | 1 | 1 | 2 |
| k | 2 | 1 | 2 | 2 |
| l | 2 | 2 | 2 | 1 |
| m | 2 | 2 | 1 | 2 |
| n | 2 | 2 | 2 | 2 |

Figure 10 Absorbing Markov chain example – definition of the possible states.

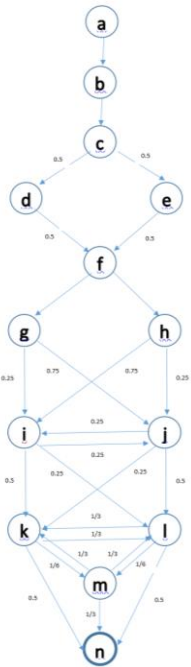

Figure 11 Absorbing Markov chain example - State transition diagram

The transition matrix corresponding to the above example is:

$$P = \begin{pmatrix} 0 & 1/2 & 1/2 & 0 & 0 & 0 & 0 & 0 & 0 & 0 & 0 & 0 & 0 & 0 \\ 0 & 0 & 0 & 1 & 0 & 0 & 0 & 0 & 0 & 0 & 0 & 0 & 0 & 0 \\ 0 & 0 & 0 & 1 & 0 & 0 & 0 & 0 & 0 & 0 & 0 & 0 & 0 & 0 \\ 0 & 0 & 0 & 0 & 1/2 & 1/2 & 0 & 0 & 0 & 0 & 0 & 0 & 0 & 0 \\ 0 & 0 & 0 & 0 & 0 & 0 & 1/4 & 3/4 & 0 & 0 & 0 & 0 & 0 & 0 \\ 0 & 0 & 0 & 0 & 0 & 0 & 3/4 & 1/4 & 0 & 0 & 0 & 0 & 0 & 0 \\ 0 & 0 & 0 & 0 & 0 & 0 & 0 & 0 & 1/4 & 1/2 & 1/4 & 0 & 0 & 0 \\ 0 & 0 & 0 & 0 & 0 & 0 & 0 & 0 & 1/2 & 1/4 & 0 & 1/4 & 0 & 0 \\ 0 & 0 & 0 & 0 & 0 & 0 & 0 & 0 & 0 & 1/4 & 1/2 & 1/4 & 0 & 0 \\ 0 & 0 & 0 & 0 & 0 & 0 & 0 & 0 & 1/4 & 0 & 1/4 & 1/2 & 0 & 0 \\ 0 & 0 & 0 & 0 & 0 & 0 & 0 & 0 & 0 & 0 & 0 & 1/3 & 1/6 & 1/2 \\ 0 & 0 & 0 & 0 & 0 & 0 & 0 & 0 & 0 & 0 & 1/3 & 0 & 1/6 & 1/2 \\ 0 & 0 & 0 & 0 & 0 & 0 & 0 & 0 & 0 & 0 & 1/3 & 1/3 & 0 & 1/3 \\ 0 & 0 & 0 & 0 & 0 & 0 & 0 & 0 & 0 & 0 & 0 & 0 & 0 & 1 \end{pmatrix}$$

The analytical solution of the above example yields the expected cover time of 10.333 with a standard deviation of 2.54 versus an expected time of 10.5 with a standard deviation of 2.7 by in using the simulation. Note that there are 13 relevant transient states out of the possible 80 combinatorial states. The main takeaway arising from the above example is that the absorbing Markov chain analogy makes it possible to accurately compute the expected cover time however the process is arduous and requires exact knowledge of the region and might be impossible to practically implement due to the complexity of bookkeeping all possible, probabilistic transitions.

*4.1.2 Algorithm Analysis*

To bound the expected coverage time, we look at the two processes that together determine the coverage time – the filling process and the deployment process.

**Definition 9**:
  (a) The hitting time from a vertex $u \in V$ to the entry point, *e*, is denoted by $h(u, e)$. The *maximum hitting time* from any vertex to the entry point is denoted by $h_{max}(e)$.
  (b) The (absolute) maximum hitting time over all ordered pairs of nodes is denoted by $h_{max}$ such that $h_{max} \geq h_{max}(e)$.
  (c) The minimum hitting time from any vertex to the entry point is denoted by $h_{min}(e)$.

**Proposition 2:** The *coverage time* of the region **R** with a single entry point denoted as *e* satisfies

$$2n \times \Delta T \leq coverage\ time$$

and

$$coverage\ time \leq 2n \times h_{max}(e) \times \Delta T$$

*Proof*: For the lower bound, we analyze the coverage process as if it's composed of two process occurring in parallel and therefore define the absolute lower bound as $max(T_{entryProcess}, T_{coveage\ Process})$. The rationale - interaction between the two processes will always increase the lower bound of each of the processes. When agents do not enter the region because of the deployment process, (e.g., large $\Delta T$) the filling process will lengthen and when the entry point is occupied due to the random walks performed by the agents the deployment process will lengthen. Hence, this definition of the lower bound will always be smaller than the lower bound that takes into account the interaction between the processes. The shortest time for the deployment process is the required number of agents, $2 \times n$, multiplied by $\Delta T$. Similarly, the shortest time for the coverage is $2n \times h_{min}(e)$ however since $h_{min}(e)=1$ in any region the shortest time for the coverage is $2n$. Therefore, the lower bound for the expected coverage time is $2n \times \Delta T$.

For the upper bound, we first analyze the problem for $\Delta T = 1$ and then expand the result for $\Delta T > 1$. As the number of agents in the region increases so does the probability that the movement of an agent will be blocked due to the presence of other, mobile agents (and the limit of one mobile agent in a cell). Hence, as the number of agents increases so does the expected time to reach a cell on the graph (and settle in it if it is empty). Consequently, the expected time to reach the last empty node in the region is the longest. Therefore, our analysis focuses on the situation in which the region is fully covered by settled agents and there are $n - 1$ mobile agents. The single cell, *u*, occupied only by a settled agent may be located anywhere in the region. That cell is called – the *empty cell*. In this situation, only the mobile agents neighboring the empty cell may move and do so only to the empty cell. Obviously, during the random walk process the position of the empty cell changes. Imagine a token that is located at the empty cell. When an arbitrary, neighboring agent, *a*, of the empty cell moves to the empty cell, it exchanges its position with the token's. Hence, at the end of the action cycle of agent *a*, the agent is in the previously empty cell and the token (that is the empty cell) in the previous position of agent *a*. This process continues until the token reaches the entry point. Clearly, the token is performing a random walk on the graph. Furthermore, analysis of the multi agent random walk is therefore equivalent to analysis of the single token random walk. Since the graph is undirected the expected time until the token reaches the entry point is equivalent to the expected time for the last agent to enter the region to reach the last empty cell - $h(u,e) = h(e,u)$. Using Definition 9, $h(u,e)$ is upper bounded by $h_{max}(e)$. Thus, the upper bound on the expected coverage time when $\Delta T = 1$ is given by $2n \times h_{max}(e)$.

The main drawback of the above upper bound is difficulty in calculating $h_{max}(e)$ even when the region and the entry point are known. An alternative, looser upper bound uses the fact that $h_{max} \geq h_{max}(e)$ and the known values of $h_{max}$ for certain graphs. Specifically, for the 2 dimensional grid and linear graphs the upper bounds are $\Theta(n^2 \log n)$ and $O(n^3)$ respectively ([33], [38]).

When $\Delta T > 1$, agent entry into the region, as defined above, is possible only during the first time step of $\Delta T$. Hence, it is possible that when the token reaches the entry point no agent will enter the region. In such a case, the random walk done by the token will continue until the time step when both the token reaches the entry point again and it is the first time step of $\Delta T$. In order to extend the upper bound to these cases we will use the fact that these are two independent random processes. The entry model can be thought of a disc with a single hole, adjacent to the entry point and rotating at $1/\Delta T$ turns per time step. The last agent will enter only when the hole in the disc and the empty cell are aligned. Since the two processes are independent of each other the joint expectation of the two processes is given by the multiplication of the expectations of each process [37] and we get that

$$coverage\ time \leq 2n \times h_{max}(e) \times \Delta T \blacksquare$$

## 4.2 Dual Layer, Limited Gradient (DLLG)

In the Dual Layer, Random Walk algorithm the settled agents only signal their presence and thus the mobile agents can only perform a random walk in their search for a cell to settle in. In the DLLG algorithm (the second of the DLC family of algorithms), the mobile agents advance up a gradient of *Step Counts* (SC for short) projected by the settled agents. Once a mobile reaches the empty parts of the region, it settles. Mobile agents move up the gradient when the difference between the step counts of the current and destination cells is one which is equivalent to a gradient steepness of one. Specifically, at every time step, every mobile agent $a_i$ with step count $sc_i$ attempts to move to a neighboring location with step count $s_{ci} + 1$ (and no mobile agents). Hence the name of this algorithm – Dual Layer, Limited Gradient (DLLG). The step count of a settled agent is the number of steps (not time steps) taken by the agent from the entry point until the time it settled and became a beacon. Moreover, the step count of every mobile agent necessarily equals the step count of the settled agent (the "beacon") it flies over, thus mobile agents are always "climbing" the gradient defined by the beacons in single-step increments. The result is a directed movement (unlike the random movement in DLR) of the mobile agents to the empty, still unexplored parts of the region. Mobile agents become beacons over time. As a first priority, whenever a mobile agent sees a neighboring empty cell, it attempts to land in that cell, changing its state to "settled" and thereafter projecting its step count to any mobile agent that sees it.

DLLG is a distributed algorithm that is implemented via straightforward local action rules, which all of the agents in **R** execute at every time step. As a first priority, and as in DLRW, a mobile agent will settle down if there are neighboring empty cells and expand the filled region. In the case an agent enters the region and there are no settled agents directly underneath it, the step count it will project is "1". If there are no empty cells to settle in, the agent will attempt to advance (i.e., move up the gradient) to a cell with a step count greater by one than its current step count. When a mobile agent cannot advance up the gradient, due to whatever reason, it stays in its current location. When there are several relevant cells to advance to, an agent will arbitrarily select one. The result is a directed acyclic graph (DAG) created by the settled agents and in which the edges are the possible directions a mobile agent may move. Mobile agents, once all the empty cells are occupied, continue advancing up the gradient of step counts stopping only when blocked by a wall or by another mobile agent. The result is a filling of the upper layer from the boundaries of the region towards the entry point with a subsequent, monotonous reduction of the area yet uncovered. These more complicated action-rules (compared to DLRW action rules) require agents with the following additional capabilities.

- Ability to project the step count to the mobile agent moving above.
- A finite memory of $\log n$ bits to store the step count.

Hence, in accordance with Definition 2, the state, $S_i$, of agent $a_i$ is determined by the combination of its physical state (i.e., mobile or settled) and step count.

Implementation of DLLG requires agents that can store as well as signal their step count. Since the signaling is by the beacons to the flying agents, multiple lights facing upwards controlled by color, intensity, etc., seems the most logical solution [7, p. 252]. In the case of a quadcopter, the lights at the ends of the rotor booms can be used for this.

**Definition 10**: The step count of agent $a_i$ at time step $t = k$ is the number of steps the agent moved from its Entry Point to its location at time $t = k$.

**Definition 11**: The state of an agent $a_i$ using the DLLG algorithm is defined by the 2-tuple $S_i = (s_1, s_2)$ where $s_1$ has a value from the set {mobile, settled} and $s_2$ is the step count.

The local action rules of the Dual Layer, Limited Gradient algorithm implemented by each agent at every time step are defined by the flowchart in Figure 12 and the pseudocode in Figure 13. The blocks in the flowchart are enumerated for easy reference in the pseudocode.

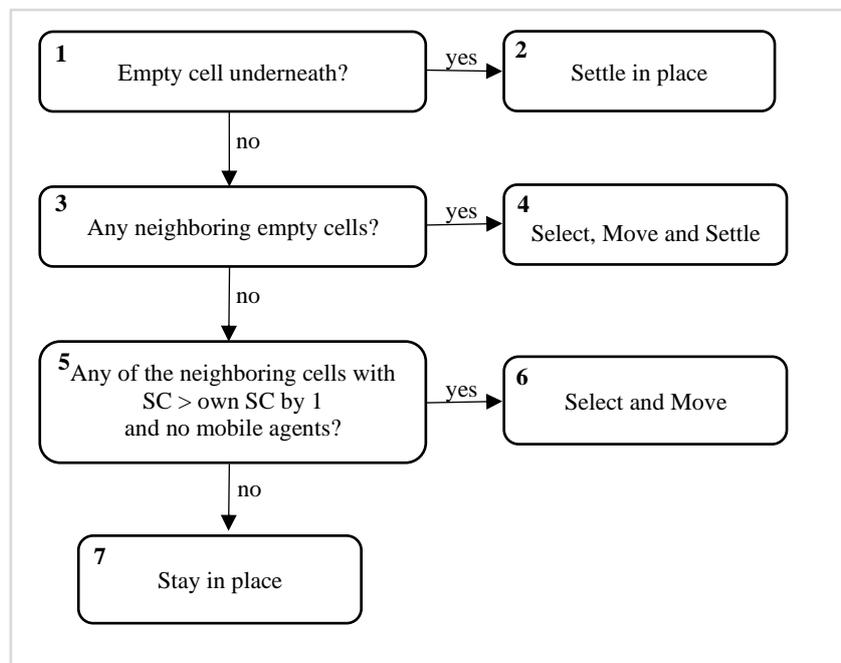

Figure 12 Flowchart of the Dual Layer, Limited Gradient algorithm – mobile agent action rules. Each block is numbered and referred to by this number in the pseudocode below

**Initialization**: for all agents $a \in A$ located at $t=0$ at the agent source set $S_i = (s_1, s_2) = (\text{"mobile"}, 0)$

```
If s₁ = "mobile"
    If ∃(ξᴳ) = 0                                          /* if there are empty, neighboring cells
        If ξ₁ᴳ = 0                                        /* settle if the cell directly underneath is
                                                          /* empty [1]
            u ← u;  s₁ ← "settled";  s₂ ← 1               /* set step count as 1 [2]
        Else                                              /* if not [3]
            ξ̃ᴳ = {ξⱼᴳ = 0: j = 1:4}                        /* define set ξ̃ᴳ as all empty neighboring
                                                          /* cells [4]
            v ← integer(random(|ξ̃ᴳ|))                     /* select a cell from ξ̃ᴳ
            u ← v;  s₁ ← "settled";  s₂ ← s₂ + 1          /* settle at selected empty cell, update state,
                                                          /* increase step count by 1
        End If
    Else                                                  /* if there are no empty, neighboring cells
        possiblePos ← ∅                                   /* define possiblePos - the set of possible
                                                          /* locations
        destSC ← s₂ + 1                                   /* define destSC is the step count of the
                                                          /* destination
        coveredCells ← ξᴳ with ξᴳ.s₁ = "settled"
        possiblePos ← relevantCells with ξᴳ.s₂ = destSC and ξᴬ = 0
                                                          /* the subset of relevantCells with SC > own
                                                          /* SC by 1 and no mobile agents
        If |possiblePos| > 0                              /* if possiblePos is not an empty set,[5]
            v ← random(possiblePos)                       /* arbitrarily select a cell from set [6]
            u ← v;  s₁ ← "mobile";  s₂ ← lowestSC         /* move to new location, update step count
        Else                                              /* if possiblePos is an empty set
            u ← u;  s₁ ← "mobile";  s₂ ← s₂              /* stay in place [7]
        End If
    End If
Elseif s₁ = "settled"                                     /* if settled
    Project (s₂)                                          /* continuously project the step count
End If
```

Figure 13    Dual Layer, Limited Gradient algorithm pseudocode

*4.2.1 Example*

An example of the coverage of a non-simply connected region with 18 cells using the DLLG algorithm is depicted in Figure 14 in the form of snapshots at different times. The snapshots offer a top-down perspective of the process and in order to clarify the process different colors are used as follows - empty cells are black, obstacles are white, cells filled by a settled agent are yellow with the step count in red and mobile agents are green "X"-s. The snapshots show the gradual filling of the region by settled agents - Figure 14(a) – (d) and the filling of the upper layer by mobile agents from the region's boundary towards the entry point in Figure 14(e). The final state is shown in Figure 14(f).

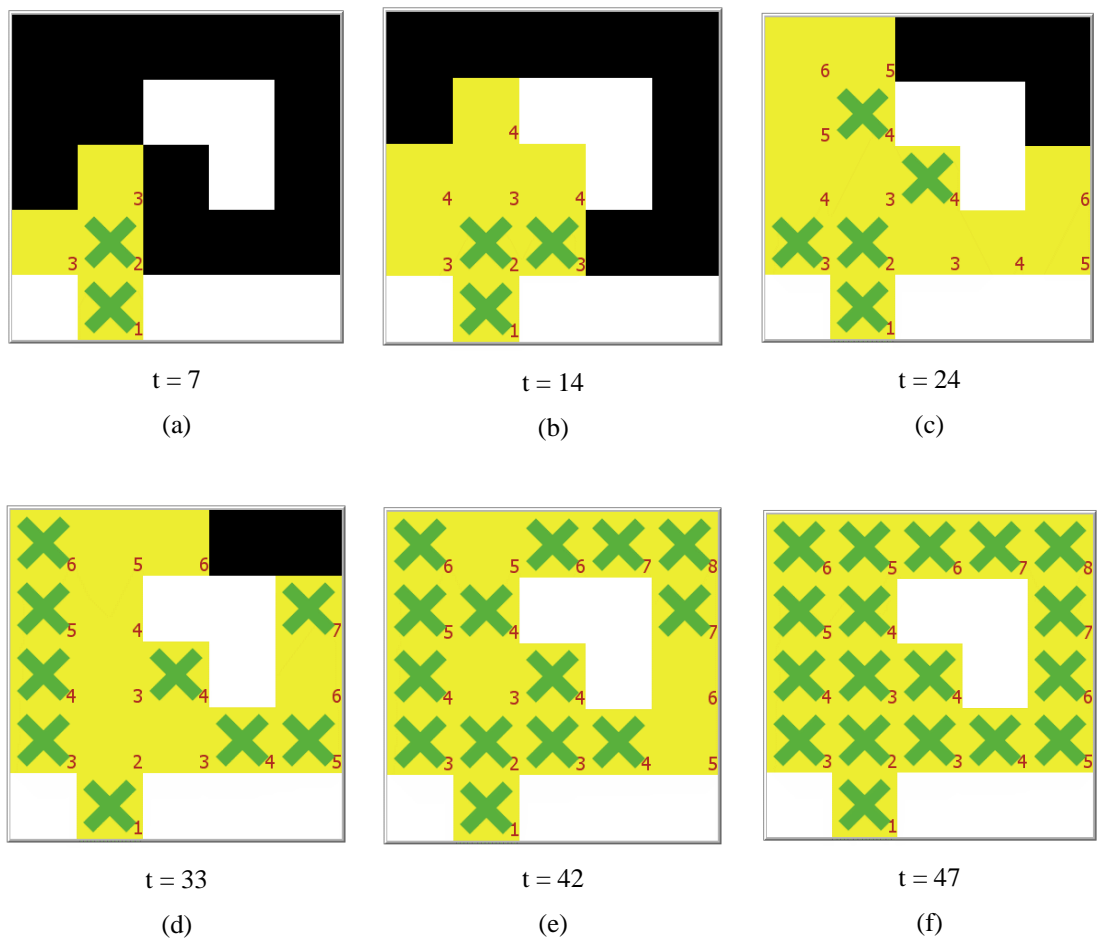

Figure 14 Snapshots of the coverage process on a region with 18 cells using the Dual Layer, Limited Gradient algorithm at different time steps as noted beneath each snapshot. Steps are ordered left to right, top to bottom. Color code: empty cells are black, obstacles are white, cells filled by a settled agent are yellow with the step count in red and mobile agents are green "X"-s.

*4.2.2 Algorithm Analysis & Performance*

Analysis of the coverage process needs to take into consideration the coupling between the entry model and the local action rules. The agents enter the region according to the entry scheme described above once every $\Delta T$ time steps. The termination time for the case of $\Delta T \geq 2$, derived in [3] is given by:

$$T_C(\mathbb{G}) = (2n-1)\Delta T \tag{9}$$

**Proposition 3**: For a linear graph $\mathbb{G} = \mathbb{G}(V, E)$, with $n$ cells the termination time of the coverage process using the Dual Layer, Limited Gradient algorithm and $\Delta T = 1$ is upper bounded by:

$$T_C(\mathbb{G}) \leq 4n - 5 \tag{10}$$

As described above, the agents start outside the region and the outcome of the DLLG algorithm are two layers of agents in the region. In order to derive the upper bound on the termination time we will use an adversarial scheduler that defines the relative wake up order of any two neighboring agents in a time step. This is useful in a linear graph in which movement of agents closer to the entry point is only affected by agents further away. Then, based on the adversarial scheduler and the agent entry model the upper bound on the termination time is derived.

**Definition 12**: The distance of agent $a_i$ from the entry point is given by $\delta(\text{EP}, j)$.

**Definition 13**: The scheduler that will result in the largest termination time on a linear graph is defined as the wake-up order of the agents, in time step $t = k$, that begins with agents at the entry point and ends with the agents at the expansion frontier. We call this the *adversarial scheduler*. When zooming in on two adjacent, mobile agents, this means that at time step $t = k$ the agent closer to the entry point will wake up first.

$$t_{i,k} < t_{j,k} \quad \forall i,j \text{ such that } \delta(EP,j) \geq \delta(EP,i) + 1 \tag{11}$$

*Proof*: We start by analyzing the movement of two neighboring mobile agents, denoted by $i$ and $j$, located over settled agents in the graph – see Figure 15. Agent $i$ will move to the right only after agent $j$ moves since only one mobile agent can occupy a cell at any given time. The relative wake-up defined by Definition 12: The distance of agent $a_i$ from the entry point is given by $\delta(\text{EP}, j)$.

**Definition 13** means agent $i$ will always move one time step after agent $j$. Note that in all other wake up schedules, agent $i$ might wake up after agent $j$ and thus both agents will move in the same time step hence the adversarial scheduler defined above will result in the worst termination time regardless of the distribution of mobile agents.

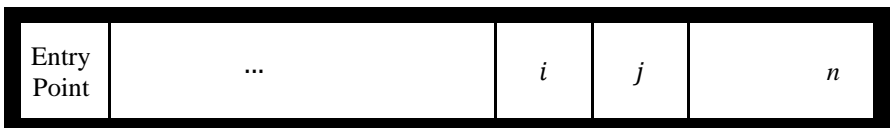

Figure 15 Typical linear graph with *n* cells and the entry point in the left.

We can now analyze the coverage process of the linear graph $\mathbb{G}$. At $t = 0$, the first agent will enter the region at the entry point. At $t = 1$ the 1st agent will settle according to the DLLG action rules, and the 2nd agent will enter the region. At $t = 2$, the mobile agent at the entry point will move to the right and an additional mobile agent will enter. At $t = 3$, due to the adversarial scheduler only the right-most mobile agent will move (and the agent entry point will stay in place). Only at $t = 4$ will the agent at the entry point move after which another mobile agent will enter. This sequence continues until $2n$ enter the region and is depicted in Figure 16 that is divided into 3 sections. The left most with the time step, the center section shows the positions of the mobile and settled agents and the right section shows the total numbers of mobile, settled and all agents in three columns. Every row in the center section is sub-divided into two and shows the locations of all the agents.

| t | EP | | | | | | #mobile | #settled | total |
|---|---|---|---|---|---|---|---|---|---|
| 0 | M | | | | | | 1 | 0 | 1 |
| | | | | | | | | | |
| 1 | M | | | | | | 1 | 1 | 2 |
| | S | | | | | | | | |
| 2 | M | | | | | | 2 | 1 | 3 |
| | S | S | | | | | | | |
| 3 | M | M | | | | | 2 | 2 | 4 |
| | S | S | | | | | | | |
| 4 | M | | | | | | 2 | 2 | 4 |
| | S | S | S | | | | | | |
| 5 | M | M | | | | | 2 | 3 | 5 |
| | S | S | S | | | | | | |
| 6 | M | | M | | | | 2 | 3 | 5 |
| | S | S | S | | | | | | |
| 7 | M | M | | | | | 2 | 4 | 6 |
| | S | S | S | S | | | | | |
| 8 | M | | M | | | | 2 | 4 | 6 |
| | S | S | S | S | | | | | |
| 9 | M | M | | M | | | 3 | 4 | 7 |
| | S | S | S | S | | | | | |
| 10 | M | | M | | | | 2 | 5 | 7 |
| | S | S | S | S | S | | | | |

Figure 16    Dual Layer, Limited Gradient - coverage process of a linear region with $\Delta T = 1$

It is easily discernable that with the adversarial scheduler an additional agent enters the region once every two time steps. Moreover, the number of agents at the odd time step is given by:

$$N(t) = 0.5t + 2.5 \tag{12}$$

From which we get However, in DLUG ~~a mobil~~

$$t = 2N - 5 \tag{13}$$

And after rearranging and setting $N = 2n$ we get:

$$T_C^{adversarial}(\mathbb{G}) = 4n - 5 \tag{14}$$

And consequently

$$T_C(\mathbb{G}) \leq 4n - 5 \quad \blacksquare \tag{15}$$

The lower bound on the termination time is the shortest time until $2n$ agents enter the region. When $\Delta T = 1$ it is simply $2n - 1$.

Experimental results verify the two upper bounds. Moreover, the experimental results show the termination time in connected regions (not necessarily linear regions) when $\Delta T = 1$ is upper bounded by $2 \times (2n - 1)$. Based on these results we write the following conjecture.

**Conjecture 1**: The termination time of the Dual Layer, Limited Gradient algorithm over any connected region **R** and for $\Delta T \geq 1$ is given by:

$$T_C(\mathbb{G}) \leq 2 \times (2n - 1) \tag{16}$$

## 4.3 Dual Layer, Unlimited Gradient (DLUG)

In the Dual Layer, Unlimited Gradient (DLUG) algorithm settling agents signal their step count and the mobile agents attempt to move up the gradient as fast as possible. This means that at every time step, a mobile agent will move up the gradient as long as it is increasing and regardless its steepness. Specifically, at each time step the mobile agent will select as its destination the neighboring cell with the step count that is closest, but larger, than its current step count that is also not occupied by a mobile agent. This difference from the Dual Layer, Limited Gradient algorithm described above is beyond technical. Conceptually, in the DLUG algorithm, preference is given to moving towards the empty parts of the region as fast as possible at the cost of a less ordered gradient. The rationale being to minimize the instances a mobile agent is staying in place since the energy consumption is the same regardless of whether it's moving or staying in place.

The purpose of the step count is to guide the mobile agents. However, defining the step count as the "number of cells an agent has travelled since it entered" (see Definition 10) may create local maxima. These local maxima will slow down or even stop the movement up the gradient since agents do not move down the gradient. Hence, the definition of the step count is modified to "at every time step the step count of a mobile agent equals the step count of the settled agent (the "beacon") it flies over". Consequently, whenever a mobile agent settles it will signal the step count of the last beacon it flew over incremented by one.

**Definition 14**: (a) The step count of a mobile agent $a_i$ at time step $t = k$ is the step count of the settled agent directly underneath - $sc_i = \xi_0 \cdot s_2$.
(b) The step count of a settled agent $a_i$ at time step $t = k$ is the step count of the last settled agent it flew over incremented by 1.

According to Definition 14 an agent senses the step count directly underneath it at the beginning of every action-cycle rather than storing it in memory and incrementing by one at every time step the agent moves. When the agent settles it sets the projected light pattern accordingly. This alternative implementation negates the need for a memory and means the agent is oblivious. The modified definition of the step count reduces to the original definition when the gradient steepness is limited to one since the step count of every settled agent is equal to its distance from the entry point.

Besides this fundamental difference in the local action rule the Dual Layer, Unlimited Gradient algorithm is similar to the DLLG algorithm. The state of an agent is defined using Definition 11 and as a first priority mobile agents settle in empty neighboring cells. Furthermore, the two algorithms can be implemented by the same agents. The result of the DLUG algorithm is a directed acyclic graph (DAG) with its root at the entry point and its vertices denoted by their step count. However, in DLUG a mobile agent $a_i$ in cell $v$ may move to any neighboring cells with a higher step count and consequently the outgoing degree of cell $v$ will always be equal or greater than the degree of the same cell when DLLG is used. The DAG resulting from the DLUG algorithm will always have at least as many directed links as the DAG resulting from the DLLG algorithm.

At every time step all the agents in **R** execute the local action rules defined by the Dual Layer, Unlimited Gradient algorithm and described in the flowchart in Figure 17 and the pseudocode in Figure 18.

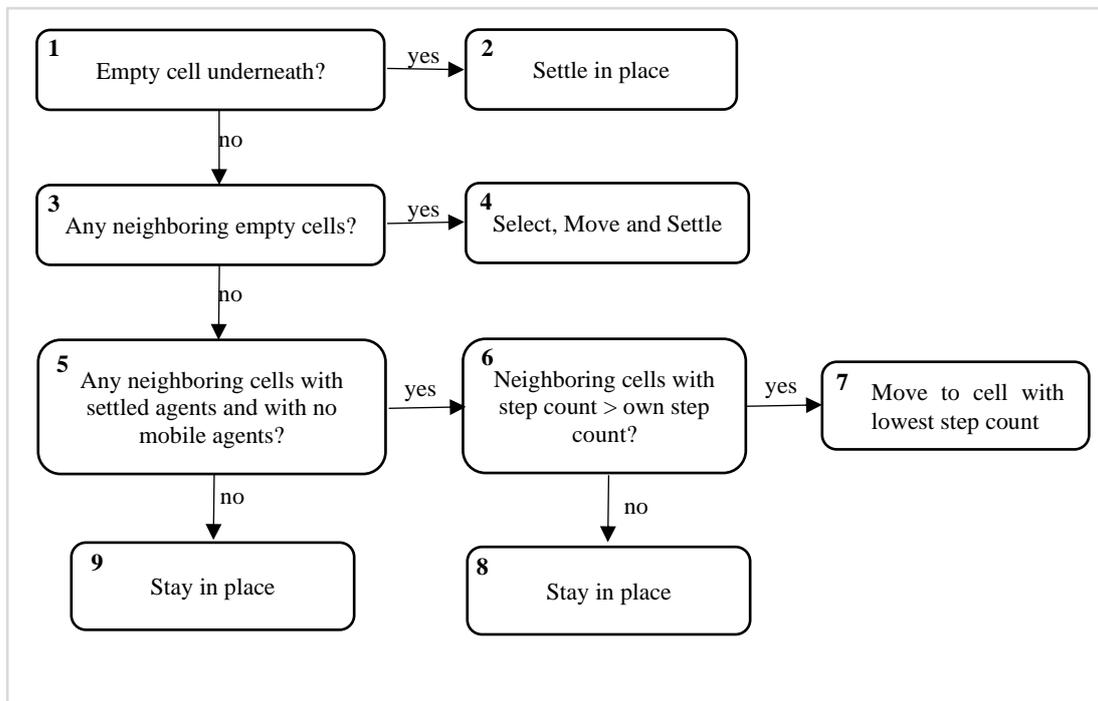

Figure 17 Flowchart of the Dual Layer, Unlimited Gradient algorithm – mobile agent action rules. Each block is numbered and referred to by this number in the pseudocode below

**Initialization**: for all agents $a \in A$ located at $t=0$ at the agent source set $S_i = (s_1, s_2) = (\text{"mobile"}, 0)$

---

If $s_1 = $ "mobile"
  If $\exists(\xi^G) = 0$ then        /* if there are empty, neighboring cells
    If $\xi_1^G = 0$ then        /* settle if the cell directly underneath is
                                                   /* empty [1]
      $u \leftarrow u$; $s_1 \leftarrow$ "settled"; $s_2 \leftarrow 1$     /* set step count as 1 [2]
    Else         /* if not [3]
      $\tilde{\xi}^G = \{\xi_j^G = 0: j = 1:4\}$     /* subset of $\xi^G$ containing the empty cells
      $v \leftarrow random(\tilde{\xi}^G)$     /* select a cell from $\tilde{\xi}^G$
      $u \leftarrow v$;   $s_1 \leftarrow$ "settled";   $s_2 \leftarrow s_2 + 1$   /* settle at selected empty cell, update state,
                                                                  /* increase step count by 1 [4]
    End If
  Else         /* if there are no empty, neighboring cells
    $possiblePos \leftarrow \emptyset$     /* define possiblePos - the set of possible
                                                                   /* locations
    $destSC \leftarrow 0$     /* define lowestSC - the lowest relevant
                                                                     /* step count
    $relevantCells \leftarrow \xi^G$ with $\xi^G. s_1 =$ "settled" and $\xi^A = 0$
                                                                     /* subset of $\xi^G$ that is filled with beacons
                                                                     /* and with no mobile agents [5]
    If $|relevantCells| > 0$
      $destSC \leftarrow min\ (relevantCells.s_2 > s_2)$   /* lowest step count in the set relevantCells
                                                                     /* greater than current step count [5]
      $possiblePos \leftarrow relevantCells$ with $relevantCells.s_2 = destSC$
                                                                    /* the subset of relevantCells with beacons
                                                                    /* having a step count equal to lowestSC [6]
      If $|possiblePos| > 0$     /* if possiblePos is not an empty set, then
        $v \leftarrow random(possiblePos)$     /* arbitrarily select a cell from set
        $u \leftarrow v$;   $s_1 \leftarrow$ "mobile";   $s_2 \leftarrow destSC$   /* move to new location, update step count
      Else     /* if possiblePos is an empty set, then stay in
                                                                      /* place [8]
        $u \leftarrow u$;   $s_1 \leftarrow$ "mobile";   $s_2 \leftarrow s_2$
      End If
    Else     /* all surrounding cells occupied by 2 agents
      $u \leftarrow u$;   $s_1 \leftarrow$ "mobile";   $s_2 \leftarrow s_2$   /* stay in place [9]
    End If
  End If
Elseif $s_1 = $ "settled"     /* if settled
  Project ($s_2$)     /* project the step count
End If

Figure 18   Dual Layer, Unlimited Gradient pseudocode

### 4.3.1 Example

An example of the difference in the coverage process between the Dual Layer, Unlimited Gradient and Dual Layer, Unlimited Gradient algorithms is shown below in Figure 19. At $t = 26$ the agents are distributed in the region as shown in Figure 19(a). The difference between the two algorithms effects the behavior of the agent surrounded by the blue circle. In the example, at $t = 27$ that agent wakes up before the agent over $sc = 5$. Since the DLLG algorithm limits the gradient steepness to 1, the agent can only stay in place as shown in Figure 19(b). However, using DLUG, it will move to the cell with $sc = 7$ (Figure 19(c)). As seen, due to a specific wake-up order the ability to move up a steeper gradient enabled the agent to move instead of staying in place. Simulation results show this is not very common and in many cases the termination times of the two algorithms are close.

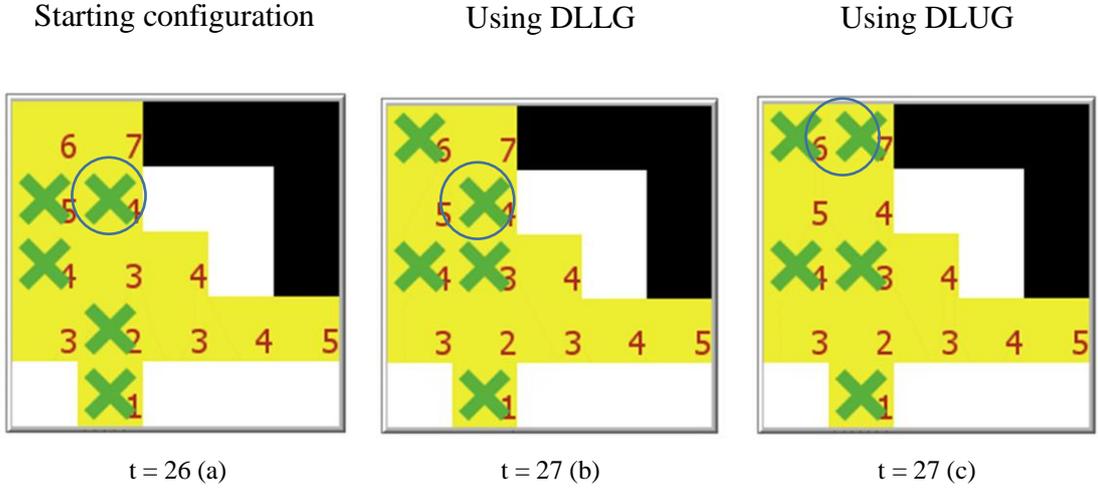

Starting configuration　　　Using DLLG　　　Using DLUG

t = 26 (a)　　　t = 27 (b)　　　t = 27 (c)

Figure 19　Comparison of the coverage process using DLLG and DLUG of a region with 18 empty cells. At the left is a snapshot of the simulation at t =26. At the center and right are snapshots of the coverage process using either the DLLG (sub-figure b) or DLUG (sub-figure c) at the 27[th] time step. The positions of the agent in the blue circle at the 27[th] time step illustrates the difference in the local action rules.

### 4.3.2 Algorithm Analysis & Performance

In linear regions the upper bounds are the same as those derived for the Dual Layer, Limited Gradient algorithm since in a linear region the gradient steepness is always one due to the single direction of advance. Experimental show that in general the termination time in DLUG is linear with the size of the region for $\Delta T \geq 2$ and regardless of the topology of the region. Moreover, for $\Delta T = 1$ the termination time is upper bounded by the termination time for $\Delta T = 2$. Additionally, the termination time is linear with $\Delta T$ for $\Delta T \geq 2$. Based on these results we propose the following conjecture.

**Conjecture 2**: The termination time of the Dual Layer, Unlimited Gradient algorithm over any connected region **R** is given by

$$T_C(\mathbb{G}) \leq (2n - 1) \times max(2, \Delta T) \quad (17)$$

The DLC family of algorithms solve the uniform coverage problem using simple agents and local action rules. Dual Layer, Random Walk requires minimal sensing capability by the agent but at the cost of a high coverage time. Dual Layer, Limited Gradient and Dual Layer, Unlimited Gradient exhibit significantly shorter upper bounds on the coverage time that are validated by experimental results. However, all the DLC algorithms have two major disadvantages. The first is the need for $2n$ agents in order to reach coverage along with the ensuing cost. The second is the significant energy expenditure of the mobile agents that fill the top layer especially since this layer is filled from the boundary of the region towards the entry point. The Single Layer Coverage family of algorithms discussed in the next section was developed in order to overcome these disadvantages. Hence, it might be asked "what do the agents in the top layer do after termination?". There are at least two possible answers – settle down next to the existing beacons and create a redundant coverage layer or exit the region according to some signal sent by the user. We leave that decision to the developer of the operational system.

## 5 Single Layer Coverage (SLC) Algorithms

Coverage in the Single Layer Coverage family of algorithms is achieved when every cell in the region contains one settled agent and this information is known to an external observer (i.e., the user). This definition extends the definition of the coverage process and the termination time of the coverage process because of the paramount importance of indicating that the process completed. All the SLC algorithms use the meta-concept of Dual Use Agents – as explorers and as beacons – with the algorithms differing one from another by the action rules. More importantly, all the SLC algorithms use the meta-concept of Backward Propagating Closure as the means to generate the indication of termination. In the following sub-sections, the algorithms are described in detail and their performance analyzed

**Definition 15**: The *termination time*, $T_C(\mathbf{R})$, of a coverage process done by a multi-agent system using a Single Layer Coverage algorithm with respect to a region $\mathbf{R}$ is the first time step at which (a) every cell in $\mathbf{R}$ is occupied by one settled agent and (b) external observer/s located at the entry point/s receive a deterministic indication of termination.

### 5.1 Top-level design

Achieving single layer coverage means changes to the functionality of the settled agents. In the DLC algorithms, the term settled agent is interchangeable with the term beacon. The term *settled agent* describes the physical state (i.e., on the ground) while the term *beacon* describes the functionality of the agent (i.e., a guiding light). This is possible since the (sole) function of a settled agent is to guide the mobile agents towards the relevant areas of expansion. In the SLC algorithms, we add another function to the settled agent – to signal to the mobile agents where NOT to move. A settled agent performing that function is called a *Closed Beacon* since it signals incoming mobile agents "this direction of advance is closed". Thus, a settled agent can be either a "Beacon" or a "Closed Beacon" depending on its locally sensed information. Transition of an agent to Closed Beacon is always from Beacon as depicted in Figure 20 and explained in the next section about the Backward Propagating Concept. In either case the transition is nonreversible. Note that in order to enable movement along the gradient a Closed Beacon continues to signal its step count.

In the SLC algorithms, the mobile agents move up or down the gradient of step counts projected by the settled agents. The rationale for advancing up the gradient is the same as in the DLLG and DLUG algorithms. In many instances during a single layer coverage process there are mobile agents in a sub-region in which all the settled agents are in the Closed Beacon state. We call these agents the "*superfluous agents*". Moving down the gradient both removes the superfluous agents away from the regions that are already uniformly covered and guides them towards the Entry Point/s (when all the region is in state Closed Beacon). Formally, the state of agent $a_i$, $S_i$, is according to Definition 11 where $s_1$ has a value from the set {mobile, Beacon, Closed Beacon} and $s_2$ is the step count (as defined in Definition 14) . The additional state requires no additional memory since at every time step the agent decides whether it is a Beacon or a Closed Beacon.

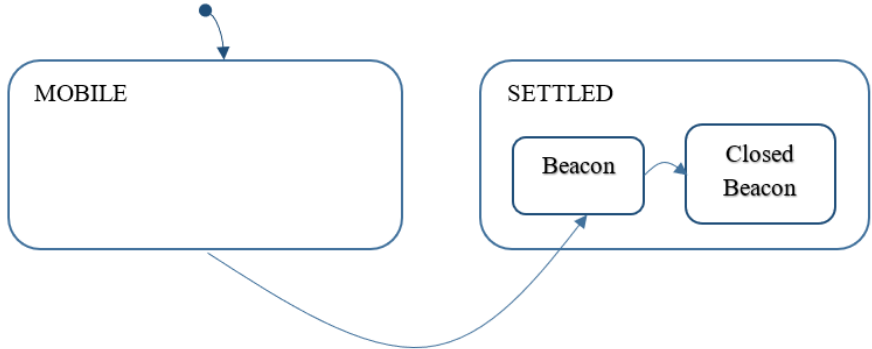

Figure 20    SLC algorithms – agent transition diagram. An agent always starts as mobile and may transition to Beacon or Closed Beacon. Beacons may transition to Closed Beacons.

## 5.2    Backward Propagating Closure and deterministic termination

In real life search and rescue scenarios, the sole source of information of the user into the coverage process is via the events at the entry point. While the user may conclude that the process has finished if the rate of entry slows down significantly it may be critical in some scenarios to definitely know that the process has finished. The purpose of the *Backward Propagating Closure* (BPC for short) meta-concept is to give the user explicit indication that the coverage process is finished. The combination of gradient-guided movement and the BPC concept results in "funneling" of mobile agents to still "un-Closed" parts of the region and the return of the superfluous agents back to the entry point/s after process termination. The funneling speeds up the coverage process while the return of the agents to the entry point decreases the number of agents engaged in the coverage process.

Essentially the BPC concept is a way of propagating information about on the state of settled agents in case the state is Closed Beacon. Intuitively, the indication an exploration direction is "closed" will start from the edges of the region. When all the relevant neighbors of settled agent $a_i$ signal "Closed" the agent will change its state to "Closed Beacon" as well. This process continues until all the settled agents are Closed Beacons[*]. The transition of a settled agent to the Closed Beacon state depends upon two conditions. The first is that there are no empty cells around. The second condition depends on the rules that define the neighboring cells to which a mobile agent may advance. Thus, the BPC meta-concept is defined differently for each SLC algorithm. One possible definition is that a settled agent will transition to Closed Beacon only when all the neighboring settled agents with a higher step count are also Closed Beacons. This is given in the definition below.

---

[*]This concept can be extended to propagate other types of information for example the energy state.

**Definition 16:** The state of the settled agent *a* located at $u$ will change to Closed Beacon if:

(1) $\check{\xi}^G = \hat{\xi}^G_u$ with
  a. $\check{\xi}^G = \{\xi^G_j . s_2 > \xi^G_0 . s_2 \mid j = 1:4\}$
  b. $\hat{\xi}^G_u = \{\xi^G_u . s_1 = \text{"Closed Beacon"} \mid j = 1:4\}$

(2) $\xi^G_j \neq 0 \mid j = 1:4$

Based on Definition 16 the flowchart (Figure 21) and the pseudocode (Figure 22) describing the local action-rules of a settled agent implementing the Backward Propagating Closure meta-concept are listed below.

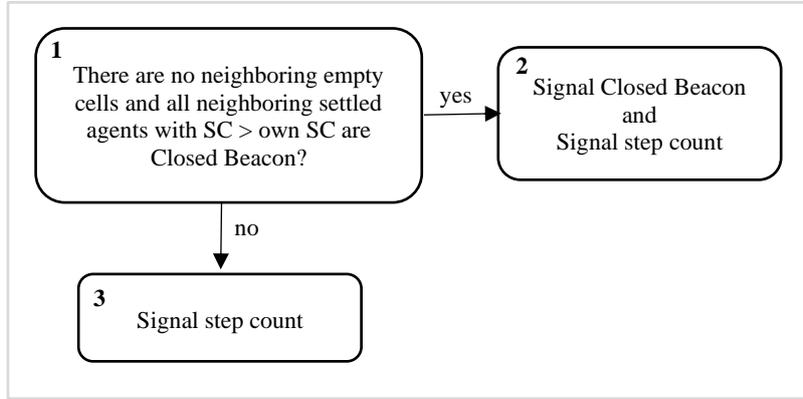

Figure 21  Backward Propagating Closure flowchart

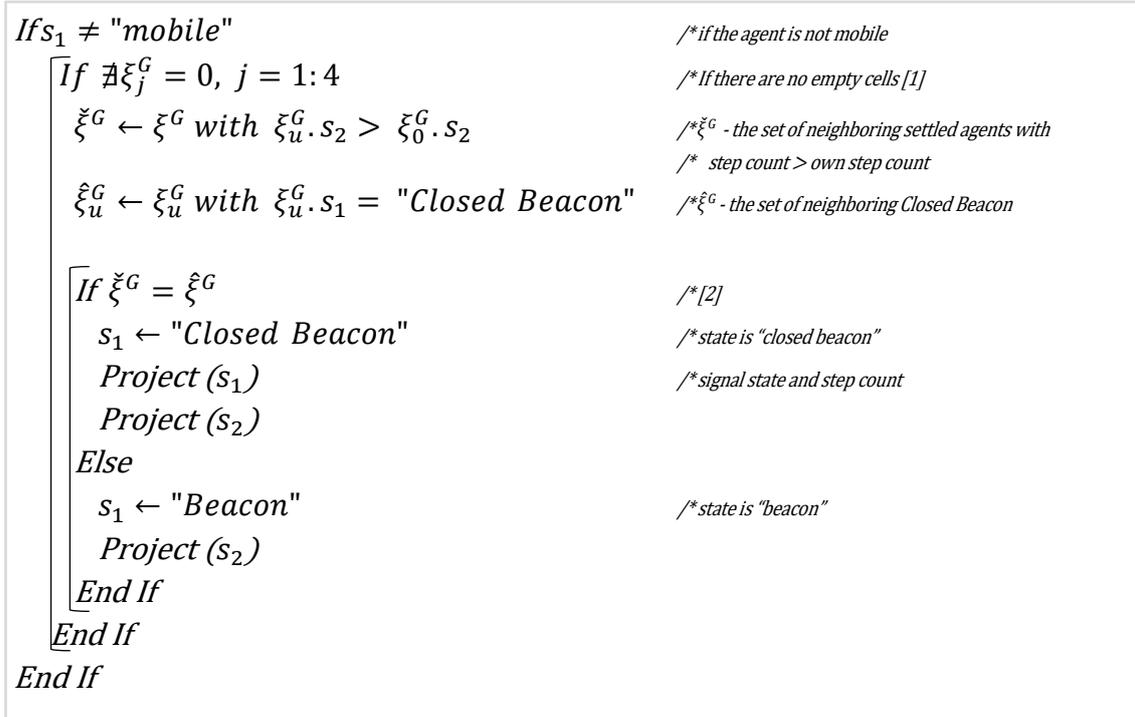

Figure 22  Backward Propagating Closure pseudocode

## 5.2.1 Example

An example of the Backward Propagating Closure process is given in Figure 23 for a region with $n = 7$ cells. Figure 23(a) shows the empty region while Figure 23(b)-(f) show the coverage process. Once an agent settles in a cell the cell is colored yellow and when a settled agent transitions to Closed Beacon the cell is colored brown. Focusing on the cell surrounded by the blue circle we see that it transitions to Closed Beacon once all the settled agents up-stream become Closed Beacons. The Close signal then propagates backwards, towards the entry point, until all the cells are Closed Beacons.

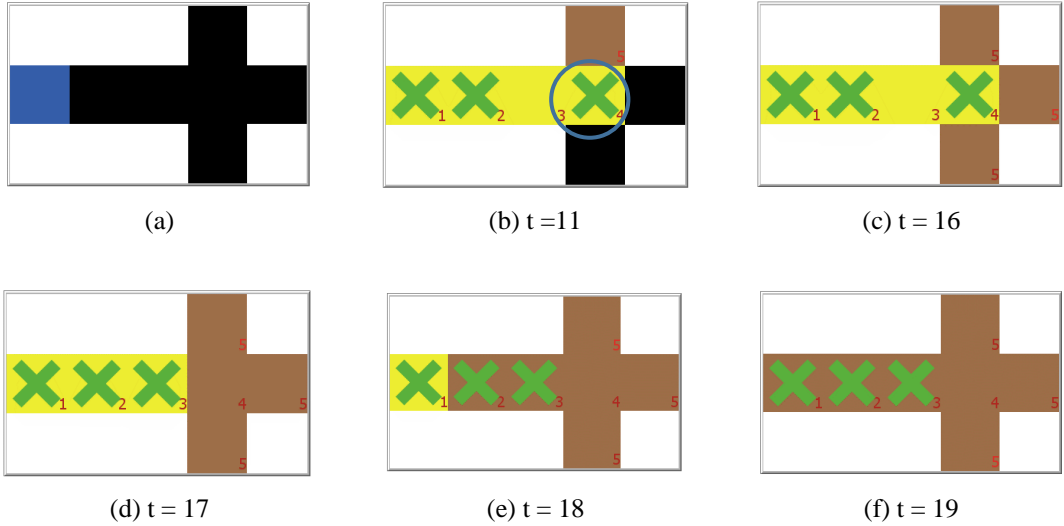

(a)  (b) t =11  (c) t = 16

(d) t = 17  (e) t = 18  (f) t = 19

Figure 23 Backward Propagating Closure – Example. Color code: empty cells are black, obstacles are white, cells filled by a Beacon are yellow and cells filled with a Closed Beacon are brown, the step count is in red and mobile agents are green "X".

## 5.2.2 Algorithm Analysis & Performance

Following from the definition of the termination time in the SLC algorithms is the need to bound the time of the BPC process. Intuitively, the upper bound on the run-time of the Backward Propagating Closure (BPC) process on $\mathbb{G}(V, E)$ is $n$ and in the following paragraphs we will prove this bound by deriving expressions for the maximum distance the signal must propagate and for the minimal propagation rate. The ratio of these expressions is the maximum propagation time.

**Definition 17**: $T_{BPC}(\mathbb{G})$ is the time span from the time step an agent settles in the last empty cell in $\mathbb{G}(V, E)$ until the settled agent at the Entry Point transitions to "Closed Beacon".

**Definition 18:** The maximal propagation distance is the maximal number of vertices the BPC signal must pass from the last leaf in $\mathbb{G}(V, E)$ that transitions to "Close Beacon" to the Entry Point. It is denoted by $D_{BPC}(\mathbb{G})$.

**Proposition 4:** For any given graph $\mathbb{G} = \mathbb{G}(V, E)$ and $|V| = n$

$$D_{BPC}(\mathbb{G}_{linear}) \geq D_{BPC}(\mathbb{G})$$

*Proof*: We denote by $\mathbb{G}_{linear}$ a linear graph with $n$ vertices and the Entry Point at one end. Trivially the last cell to be filled is the cell at the other end of the graph. Hence the BPC signal has to propagate along the entire graph and consequently $D_{BPC}(\mathbb{G}_{linear}) = n - 1$. Suppose that this is not the minimal distance and that $D_{BPC}(\mathbb{G}) = n$ (or greater). However, this means the signal propagated through $n + 1$ vertices. This contradicts that the cardinality of the graph is $n$. Hence

$$D_{BPC}(\mathbb{G}_{linear}) \geq D_{BPC}(\mathbb{G}) \quad \blacksquare \tag{18}$$

Given a settled agent $a_i$ at cell $u$ that at time step $t = k$ transitions to Closed Beacon. We ask what is the number of time steps until the BPC signal propagates to settled agent $a_j$ located in a neighboring cell $v$ and $u \in \hat{\xi}_v^G$. From the time scheme described above it is clear the signal may propagate within the same time step if $t_{i,k} < t_{j,k}$ and in the following time step if $t_{i,k} > t_{j,k}$. In other words, the minimum propagation rate is 1.

**Proposition 5:** For any given graph $\mathbb{G} = \mathbb{G}(V, E)$ and $|V| = n$ the upper bound on $T_{BPC}(\mathbb{G})$ is given by n.

$$T_{BPC}(\mathbb{G}) \leq n$$

*Proof*: An agent that settles always transitions to Beacon state hence transition to Closed Beacon can only be done in the following time step. The ratio of the maximal distance derived in Proposition 4 and the minimal propagation rate derived above we get that the propagation time of the BPC signal is $n - 1$. Summing up we get that $T_{BPC}(\mathbb{G}) = n$ when $\mathbb{G}$ is a linear graph and $T_{BPC}(\mathbb{G}) < n$ otherwise. $\quad \blacksquare$

Note that the above proof did not make use of Definition 16 and is thus correct for other definitions.

## 5.3 Single Layer, Limited Gradient (SLLG)

The SLLG algorithm combines the exploration strategy and traversal rules used in Dual Layer, Limited Gradient algorithm with the Backward Propagating Closure meta-concept in order to fill the region by a single layer of agents. Preference is given by the mobile agents to settling versus advancing along already covered cells. When settling is not possible, agents advance up a gradient projected by the settled agents. At every time step, agent $a_i$ with step count $sc_i$ senses its neighboring region. If there are Beacons signaling a step count that is equal to $sc_i + 1$ the agent will attempt to move to those cells that are also unoccupied by mobile agents. If all the Beacons with $sc_i + 1$ are occupied, then it will stay in place. The rationale being that the mobile agents may move in the next time step/s. However, if there are no Beacons with $sc_i + 1$ then the agent will try to retrace its movement by going down the gradient projected by Closed Beacons. By limiting the retracing to Closed Beacons there is no possibility that a retracing agent will block the advance of incoming agents. That is because as stated above the advancing mobile agents move only over Beacons. Accordingly, an agent that determines it cannot advance will check if there are any Closed Beacons with a step count that is smaller than its own and no mobile agents. If there are, it will arbitrarily select one and move to it. The transition of a settled agent to the Closed Beacon state in SLLG follows directly from the action rules governing the advance up the gradient and is defined below.

**Definition 19:** The state of the settled agent $a$ located at $u$ will change to Closed Beacon if:

(1) $\check{\xi}^G = \hat{\xi}_u^G$ with
   a. $\check{\xi}^G = \{\xi_j^G . s_2 = \xi_0^G . s_2 + 1 | \ j = 1:4\}$
   b. $\hat{\xi}_u^G = \{\xi_u^G . s_1 = \text{"Closed Beacon"} | \ j = 1:4\}$

(2) $\xi_j^G \neq 0 | \ j = 1:4$

The flowchart in Figure 24 shows the different behaviors of the mobile agents with the expansion behavior in blocks (1)-(4), advancing up the gradient in blocks (5)-(8) and retracing in blocks (9)-(13). The detailed local action-rules implemented by the mobile agents are described in the pseudocode below (Figure 26 and Figure 27). Based on Definition 19 the flowchart (Figure 25) and the pseudocode (Figure 28) describing the local action-rules of a settled agent implementing the Backward Propagating Closure meta-concept are listed below.

The resulting macro behavior is the uninterrupted flow of the mobile agents to the parts of the region that are still not filled by agents in Closed Beacon state. The explored region using the Single Layer, Limited Gradient local action rules is very similar to that of explored using (centralized) BFS with one major difference. Since SLLG has no (centralized) queue and the movement is partly arbitrary, at any time before completion the exploration may progress in certain parts more than others. Moreover, the step count of a settled agent may be greater than its minimal distance from the entry point.

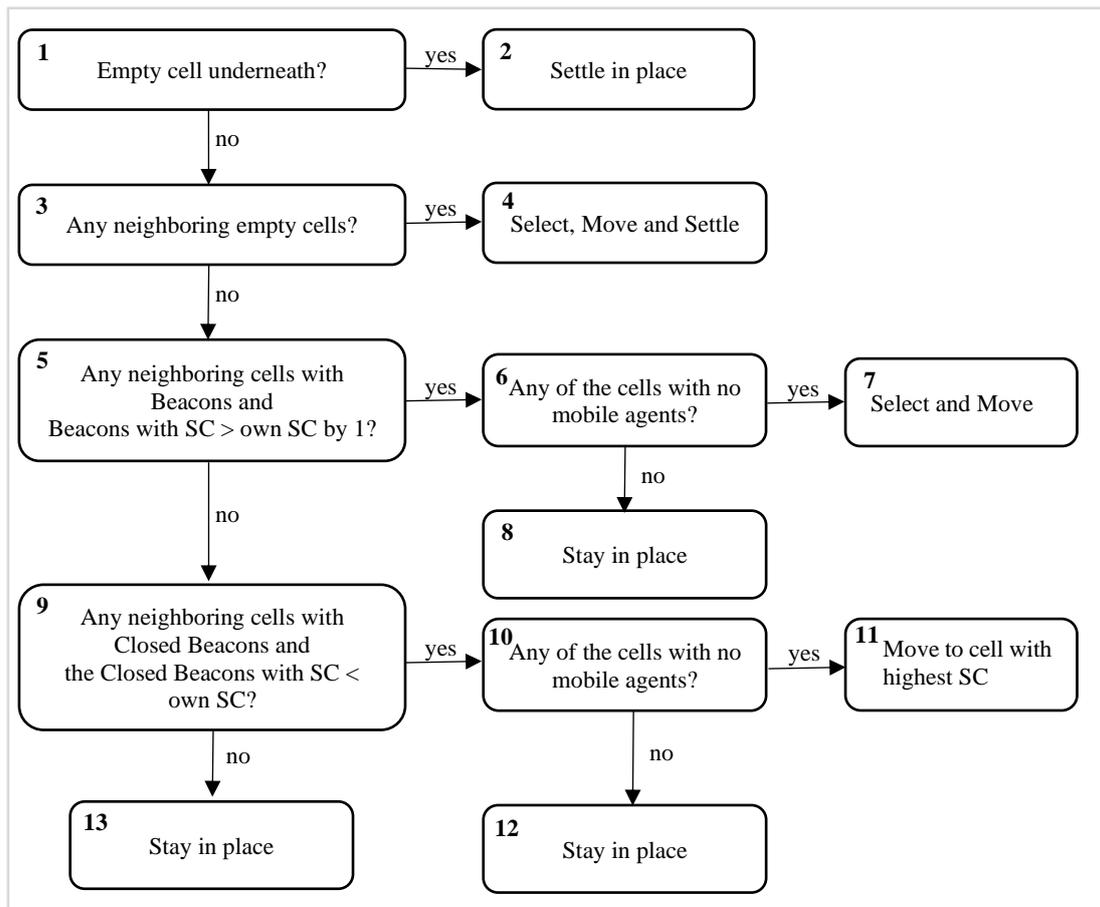

Figure 24 Single Layer, Limited Gradient algorithm – mobile agent flowchart
Each block is numbered and referred to by this number in the pseudo-code below

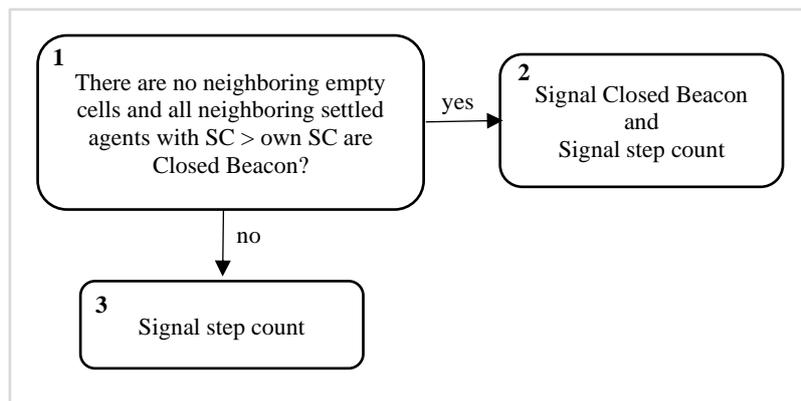

Figure 25  Single Layer, Limited Gradient algorithm – settled agent flowchart
Each block is numbered and referred to by this number in the pseudo-code below

**Initialization**: for all agents $a \in A$ located at $t=0$ at the agent source set $S_i = (s_1, s_2) = ("mobile", 0)$

```
If s_1 = "mobile"
    If ∃(ξ^G) = 0 then                                      /*if there are empty, neighboring cells
        If ξ_0^G = 0 then                                    /*if the cell directly underneath is empty [1]
            u ← u;    s_1 ← "settled";    s_2 ← 1            /*set step count as 1 [2]
        Else                                                 /*if not [3]
            ξ̃^G = {ξ_j^G = 0: j = 1:4}                       /* define set ξ̃^G as all empty neighboring
                                                             /*cells
            v ← random(|ξ̃^G|)                                /*select a cell from ξ̃^G
                                                             /* settle at selected empty cell, update
                                                             /*state, increase step count by 1 [4]
            u ← v;    s_1 ← "settled";    s_2 ← s_2 + 1
        End If
    Else
        possiblePos ← ∅                                      /*initialize possiblePos - the set of possible
                                                             /*locations
        destSC ← s_2 + 1                                     /*define the step count of the destination
                                                             /*cell
        relevantCells ← ξ_u^G with ξ_u^G.s_1 = "Beacon" and ξ_u^G.s_2 = destSC
                                                             /*the subset of ξ̃^G that is filled with agents in
                                                             /*Beacon state and have a SC eq. to destSC
        If |relevantCells| > 0                               /*if there are neighboring cells with Beacons
                                                             /* and step count > own step count by 1 [5]
            possiblePos ← relevantCells with ξ_u^A = 0       /*the subset of relevantCells with no mobile
                                                             /*agents
            If |possiblePos| > 0                             /*if there are cells to move-to [6]
                v = random(|possiblePos|)                    /*select a cell from possiblePos [7]
                u ← v;    s_1 ← "mobile";    s_2 ← destSC    /*move to selected cell and increase Step
                                                             /*Count by 1
            Else
                u ← u;    s_1 ← "mobile";    s_2 ← s_2       /*stay in place [8]
            End If
        Else
            destSC ← 0                                       /* initialize the step count of the destination
                                                             /*cell
            relevantCells ← ξ_u^G with ξ_u^G.s_1 = "Closed Beacons and ξ_u^G.s_2 < s_2
                                                             /* search for cells that are Closed Beacons
                                                             /* and have a step count lower than own
                                                             /* step count
```

Figure 26  Single Layer, Limited Gradient pseudocode – mobile agent - part 1

```
            If|relevantCells| > 0                                    /* [9]
              possiblePos ← relevantCells with ξ_u^A = 0             /*the subset of relevantCells with no
                                                                     /*mobile agents
              If|possiblePos| > 0                                    /*if there are cells to move-to [10]
                 destSC = max (possiblePos.s_2 < s_2)                /*find in possiblePos the cell/s with the
                                                                     /*highest step count [11]
                 possiblePos ← possiblePos with possiblePos.s_2
                            = destSC and ξ_u^A = 0
                                                                     /* find all the cells in possiblePos with Step
                                                                     /*Count eq. to destSC and no mobile agents
                 v = random(|possiblePos|)                           /*select a cell from possiblePos [11]
                 u ← v;   s_1 ← "mobile";   s_2 ← destSC
                                                                     /*move to selected cell and set step count
                                                                     /*to destSC
              Else                                                   /*[12]
                 u ← u;   s_1 ← "mobile";   s_2 ← s_2                /*stay in place
              End If

           Else                                                      /* [13]
              u ← u;   s_1 ← "mobile";   s_2 ← s_2                   /*stay in place
           End If
        End If
     End If
Else                                                                 /* if the agent is not mobile
     Do SLLG - settled agent action rules (Figure 28)
End
```

Figure 27     Single Layer, Limited Gradient pseudocode – mobile agent - part 2

```
If s_1 ≠ "mobile"                                          /*if the agent is not mobile
   If ∄ ξ_j^G = 0, j = 1:4                                  /*If there are no empty cells [1]
      ξ̌^G ← ξ^G with ξ^G.s_2 = ξ_0^G.s_2 + 1                /*ξ̌^G - the set of neighboring settled agents with
                                                             step count = own step count +1
      ξ̂^G ← ξ^G with ξ^G.s_1 = "Closed Beacon"              /*ξ̂^G - the set of neighboring Closed Beacon
      If ξ̌^G = ξ̂^G                                          /*[2]
        s_1 ← "Closed Beacon"                                /*state is "closed beacon"
        Project (s_1)                                        /*signal state and step count
        Project (s_2)
      Else
        s_1 ← "Beacon"                                       /*state is "beacon"
        Project (s_2)
      End If
   End If
End If
```

Figure 28     Single Layer, Limited Gradient pseudocode – settled agent

*5.3.1 Example*

A possible coverage scenario with 18 empty cells is described in Figure 29. The Figure 29 shows snapshots of the progression of the coverage process at different times steps. In order to distinguish between the different entities, colors are used as follows - empty cells are black, obstacles are white, and mobile agents are green "X"-s. Cells filled by a Beacon are yellow while cells filled by a Closed Beacon are brown. In both cases the step count is in red. In Figure 29(b), a settled agent has transitioned to Closed Beacon because further expansion is not possible. The indication however has not propagated since each of its neighbors has neighboring cells that are empty and/or filled by a Beacon. The back propagation, once all the cells are filled, is shown in Figure 29(e) -(f) as is the return of the eight superfluous agents to the entry point. When comparing both the locations of the mobile agents and the regions of Closed Beacons in Figure 29(e) and Figure 29(f) it is apparent that the Close indication propagates faster than the agents move.

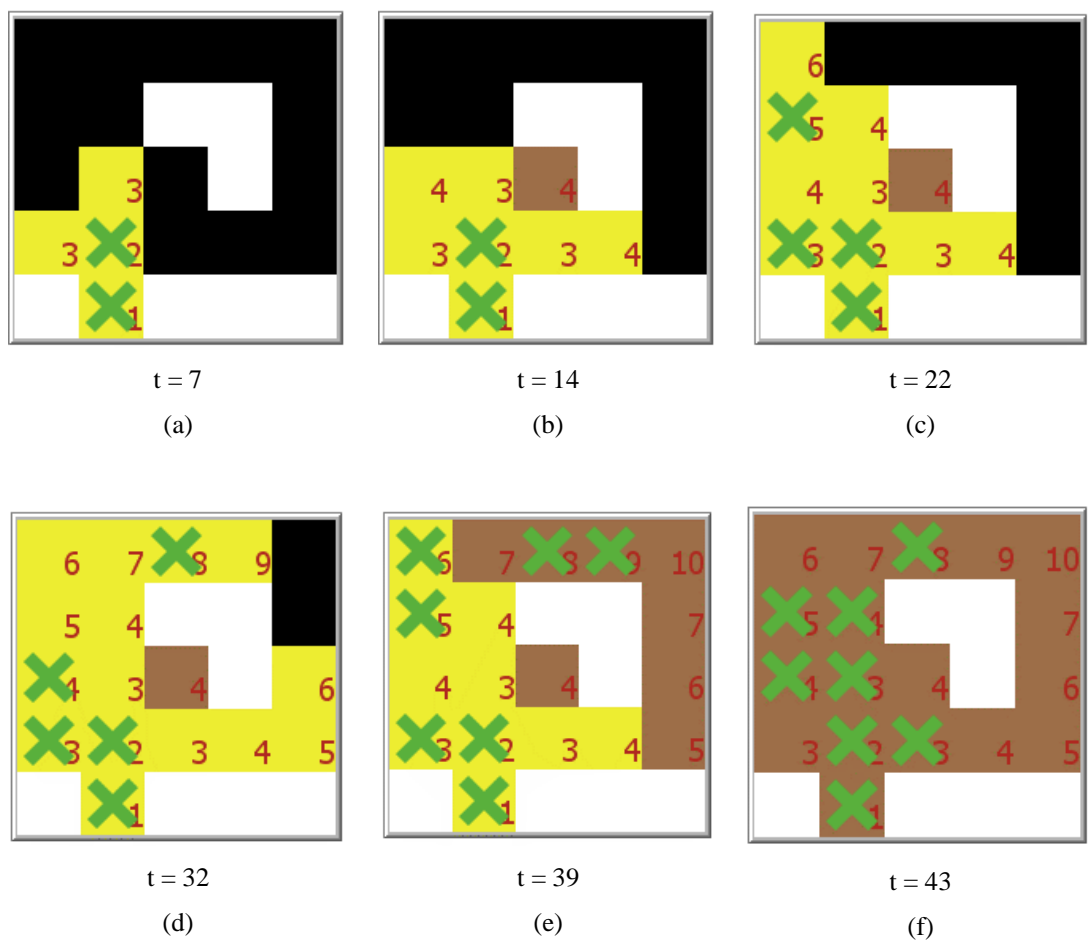

Figure 29    Single Layer, Limited Gradient – Example. The snapshots of the process are ordered left to right, top to bottom. Color code: empty cells are black, obstacles are white, cells filled by a Beacon agent are yellow and cells filled with a Closed Beacon are brown, the step count in red and mobile agents are green "X"-s.

*5.3.2 Algorithm Analysis & Performance*

**Lemma 1**: The termination time using Single Layer, Limited Gradient algorithm of a region **R** with $n$ cells is upper bounded by

$$T_C(\mathbb{G}) \leq (2n-1) \times max(2, \Delta T) + n$$

and from below by

$$(n-1)\Delta T$$

*Proof*: During the coverage process agents deploy into the region, uniformly fill it by settling while concurrently indications of the process completion propagate back to the entry point. Although the coverage is by a single layer, the number of agents that enter the region may be larger than $n$ since agents will continue to enter the region as long as the entry point is not a Closed Beacon. This number is bounded by $2n$ as no more than 2 agents may populate a cell. Hence, in order to derive the upper bound we can ask what is the time it will take to fill the region with $2n$ agents with termination being achieved when all the settled agents are in the Closed Beacons state and there two layers of agents (one settled and one mobile). The termination time in this case will necessarily be longer than that of the original problem since $2n$ agents must enter the region regardless of the Backward Propagating Closure process. This is equivalent to asking what is the termination time of the Dual Layer, Limited Gradient algorithm when the termination condition is that $2n$ enter the region and that the agent at the entry point is at state Closed Beacon. We call this new, notional algorithm "DLLG with BPC". We upper bound the termination time of this algorithm by (artificially) dividing the process into a deployment and filling phase followed by a "completion propagation" phase and using the upper bound on the time of each phase. That is possible for the following reasons: (1) The BPC process propagates regardless of the mobile agents (2) The coverage using DLLG is not affected by the BPC process since in the Closed Beacon state the settled agents continue to signal their step counts and guide the mobile agents. The mobile agents implementing the DLLG local action rules will disregard the Close indication and continue moving up the gradient as before. Consequently, the upper bound on the termination time of the "DLLG with BPC" algorithm is the sum of the termination times of the two constituent processes. The termination time of the DLLG algorithm according is $2n - 1)\Delta T + 1$ for $\Delta T \geq 2$ and from Proposition 5 we know that $T_{BPC}(\mathbb{G}) \leq n$ hence

$$T_C(\mathbb{G}) \leq (2n-1) \times max(2, \Delta T) + n \quad \blacksquare \qquad (19)$$

For the sake of the lower bound, we assume a best-case scenario in which the exploration process and the backward propagating closure process are concurrent process or in other words when the longer of the two finishes so does the other. The minimal time for the BPC process is one in the case the wake-up sequence of the agents is from the last order to settle to the entry point. The minimal time for the deployment and filling phase depends on the minimal number of agents that must enter the region, $n$, and the (minimal) time between successive entries, $\Delta T$. Thus, the lower bound of the deployment and filling phase is given by $(n-1)\Delta T$. Hence:

$$T_C(\mathbb{G}) = (n-1)\Delta T \quad \blacksquare \tag{20}$$

The main drawback of limiting the gradient steepness to 1 in SLLG is that may stop the advancement of the mobile agents prematurely. This is best explained using an example. We focus in Figure 30 on the two black ovels that surround the cells occupied by both a mobile agent and a settled agent with a step count of 10. In both cases the mobile in the cell can't move up the gradient although the region is not nearly filled since the gradient is 5 rather than 1. This slows the advance of other, incoming mobile agents. It is compounded by the fact that the local rules give precedence to staying in place over retracing when the movement up the gradient is prevented due to a mobile agent (see block 8 in the SLLG flowchart). The resulting "traffic jam" is released only when the settled agents in the surrounding cells become Closed Beacons at which time movement down the gradient is possible and the condition preferring staying in place becomes irrelevant. This problem is unique to SLLG and obviated in the following algorithms. This unwanted consequence increases $N$ and $T_C$, however, it is most noticeable in its effect on the $E_{Total}$. Naturally, its severity grows with $\Delta T$ since as $\Delta T$ increases so does the termination time and accordingly so does the time the mobile agents in the "traffic jam" are stuck with the accompanying energy consumption. This phenomena is also seen in DLLG however it is much less noticeable since inherently the required coverage is by two layers.

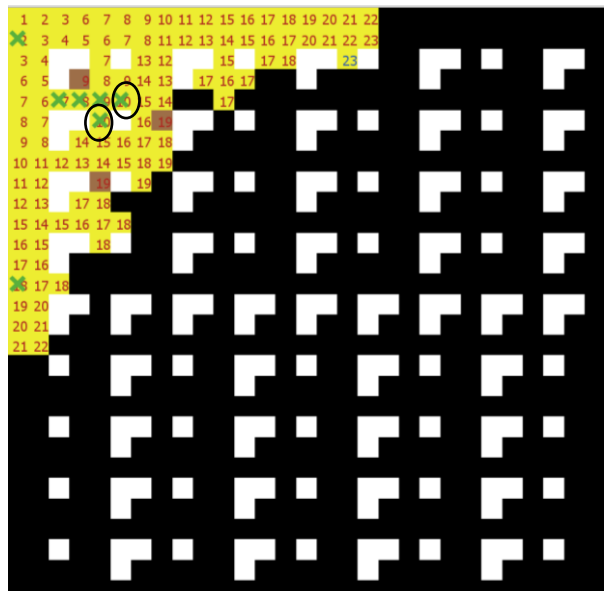

Figure 30   Single Layer, Limited Gradient – The drawback of limiting the movement to gradient steepness of one

## 5.4 Single Layer, Unlimited Gradient (SLUG)

The SLUG algorithm differs from the SLLG in two crucial aspects. The first is that the gradient steepness the agents can move on is not limited and the second is the ability to move down the gradient over Beacons as well as Closed Beacons. These two differences reflect the preference to move in any given time over staying-in-place. Moving up the gradient without limiting its value to one is used in the Dual Layer, Unlimited Gradient algorithm and results in a faster ascent up the gradient. Moving down the gradient over Beacons as well Closed Beacons means the superfluous agents can continue exploring. This is a significant change from retracing only over Closed Beacons which prevents the superfluous agents from moving over Beacons. However, movement over Beacons is mandatory if a (superfluous) agent is to explore the region. The main drawback is that such agents, when exiting a Closed Beacon sub-region, may temporarily block advancing mobile agent. Experimental results show this rarely occurs and the performance of this algorithm is significantly better than that of the Single Layer, Limited Gradient algorithm.

The transition of a settled agent to the Closed Beacon state in SLUG follows directly from the action rules governing the advance up the gradient. Thus, a settled agent will transition to Closed Beacon only if all neighboring settled agents with a higher step count are already Closed Beacon. This is described in Definition 16 and the relevant flowchart and pseudocode are given in Figure 21 and Figure 22 respectively.

The result of the SLUG algorithm, as in the Dual Layer, Unlimited Gradient algorithm, is a directed acyclic graph (DAG) with its root at the entry point and its vertices denoted by their step count. Since a mobile agent $a_i$ in cell $v$ may move to any neighboring cell with a higher step count the outgoing degree of cell $v$ will be equal to the number of cells with a higher step count.

The flowchart in Figure 31 shows the different behavior types of a mobile agents with the expansion behavior in blocks (1)-(4), advancing up the gradient in blocks (5)-(8) and retracing in blocks (9)-(13). Since the gradient is not limited, the presence of mobile agents is checked first (5). The lowest relevant step count is found only if there are relevant cells to move to and following that the corresponding possible cells to move to (6). The same concept is used when moving down the gradient.

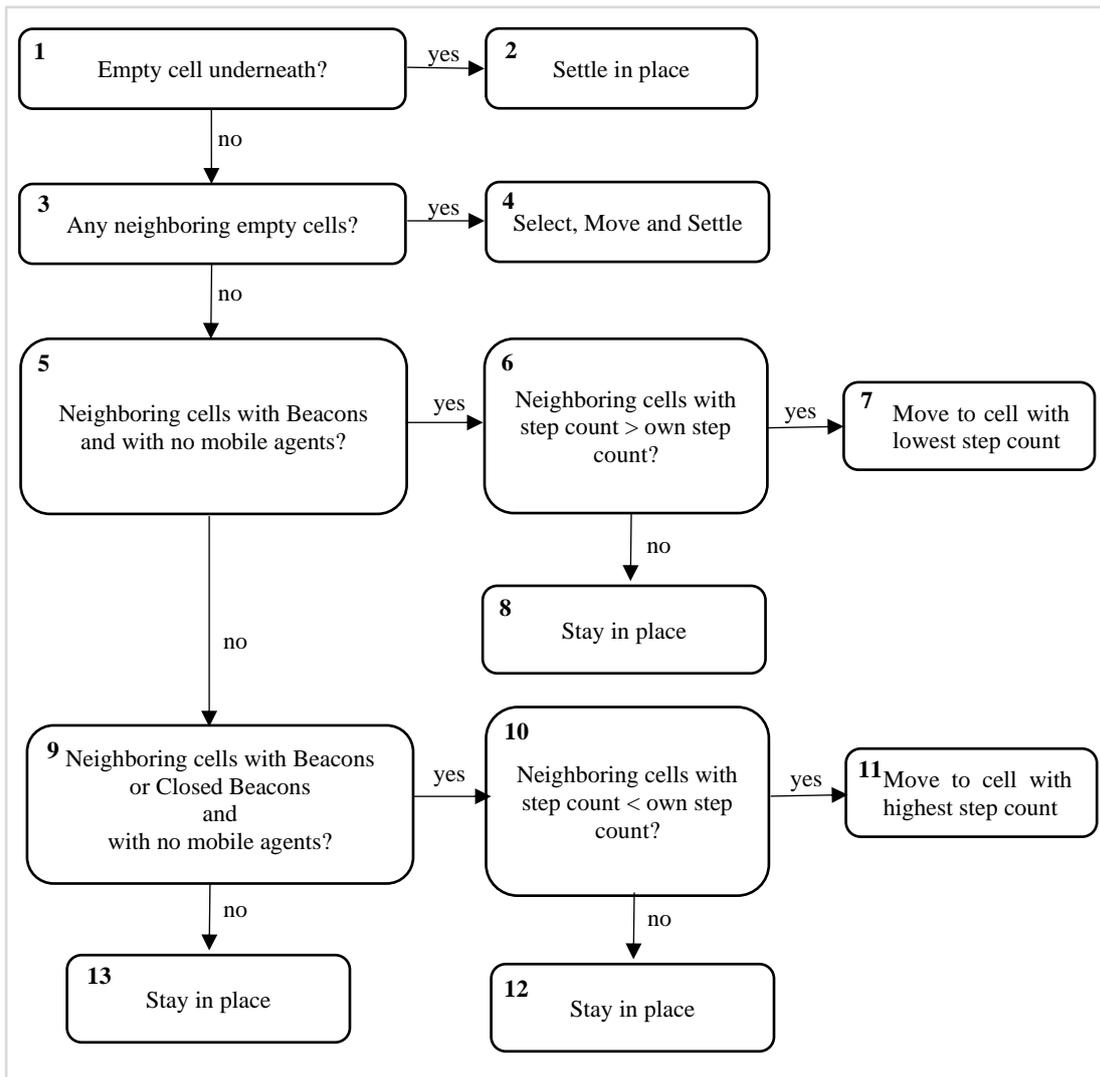

Figure 31  Single Layer, Unlimited Gradient – mobile agent flowchart
Each block is numbered and referred to by this number in the pseudocode below

**Initialization**: for all agents $a \in A$ located at *t=0* at the agent source set $S_i = (s_1, s_2) =$ ("mobile",0)

$If\ s_1 = \text{"mobile"}$
    $If\ \exists(\xi^G) = 0$                                                        /*if there are empty, neighboring cells

        $If\ \xi_1^G = 0$                                                     /*if the cell directly underneath is empty [1]

           $u \leftarrow u;\quad s_1 \leftarrow \text{"settled"};\quad s_2 \leftarrow 1$         /*set step count as 1 [2]

        $Else$                                                                       /*if not [3]

           $\tilde{\xi}^G = \{\xi_j^G = 0: j = 1:4\}$                    /* define set $\tilde{\xi}^G$ as all empty neighboring
                                                                                      /*cells

           $v \leftarrow random(|\tilde{\xi}^G|)$                            /*select a cell from $\tilde{\xi}^G$

           $u \leftarrow v;\quad s_1 \leftarrow \text{"settled"};\quad s_2 \leftarrow s_2 + 1$    /*settle at selected empty cell, update state,
                                                                                        /*increase step count by 1 [4]

        $End\ If$

    $Else$

        $possiblePos \leftarrow \emptyset$                                /*define possiblePos - the set of possible
                                                                                  /*locations

        $destSC \leftarrow s_2 + 1$                               /*define the step count of the destination
                                                                                 /*cell

        $relevantCells \leftarrow \xi_u^G\ with\ \xi_u^G.s_1 = \text{"settled"}\ and\ \xi_u^A = 0$
                                                                               /*subset of $\xi_u^G$ that is filled with Beacons and
                                                                               /*with no mobile agents [5]

        $If\ |relevantCells| > 0$                     /*if there are neighboring cells with Beacons
                                                                              /*and no mobile agents

           $destSC \leftarrow \min(relevantCells.s_2 > s_2)$       /* lowest step count in the set relevantCells
                                                                              /*greater than current step count

           $possiblePos \leftarrow relevantCells\ with\ relevantCells.s_2 = destSC$
                                                                              /*the subset of relevantCells with beacons
                                                                              /*having a step count equal to destSC

           $If\ |possiblePos| > 0$                         /*if there are cells to move-to [6]

              $v \leftarrow random(|possiblePos|)$                 /*select a cell from possiblePos [7]

              $u \leftarrow v;\quad s_1 \leftarrow \text{"mobile"};\quad s_2 \leftarrow destSC$      /*move to selected cell and increase Step
                                                                                   /* Count by 1

           $Else$                                                                   /*[8]

              $u \leftarrow u;\quad s_1 \leftarrow \text{"mobile"};\quad s_2 \leftarrow s_2$           /*stay in place [8]

           $End\ If$

        $Else$

           $destSC \leftarrow 0$                                   /*initialize the step count of the destination
                                                                               /*cell

           $relevantCells \leftarrow \xi_u^G\ with\ (\xi_u^G.s_1 = \text{"Beacons"}\ or\ \xi_u^G.s_1$
                                $= \text{"Closed Beacons"})\ and\ \xi_u^A.s_1 = 0$
                                                                               /*search for cells that are filled by settled
                                                                               /*agents and no mobile agents

Figure 32        Single Layer, Unlimited Gradient pseudocode – mobile agent - part 1

```
            If |relevantCells| > 0                                    /* [9]
              destSC ← max (relevantCells.s₂ < s₂)
              possiblePos ← relevantCells with relevantCells.s₂ = destSC
                                                                      /* the subset of relevantCells with Step
                                                                      /* count eq. to destSC no mobile agents
              If |possiblePos| > 0                                    /*if there are cells to move-to [10]
                  v ← random(|possiblePos|)                           /* select a cell from possiblePos [11]
                  u ← v;   s₁ ← "mobile";   s₂ ← destSC               /*move to selected cell and set step count
                                                                      /* to destSC
              Else                                                    /* [12]
                  u ← u;   s₁ ← "mobile";   s₂ ← s₂                   /*stay in place
              End If

            Else                                                      /* [13]
                u ← u;   s₁ ← "mobile";   s₂ ← s₂                     /*stay in place
            End If

          End If
      End If
  Else                                                                /* if the agent is not mobile
      Do SLUG – settled agent action rules (Figure 22 )
  End
```

Figure 33      Single Layer, Unlimited Gradient pseudocode – mobile agent - part 2

### 5.4.1 Example

The coverage process using the SLUG algorithm in a region with 18 cells is described in Figure 34. Figure 34 shows the agents' distribution at different times during. As in previous examples, in order to distinguish between the different entities, colors are used as follows - empty cells are black, obstacles are white, and mobile agents are green "X"-s. Cells filled by a Beacon are yellow while cells filled by a Closed Beacon are brown. In both cases the step count is in red. Of note is the funneling of the mobile agents to the empty parts of the region in Figure 34(d), and the return of the superfluous agents toward the entry point in Figure 34(f).

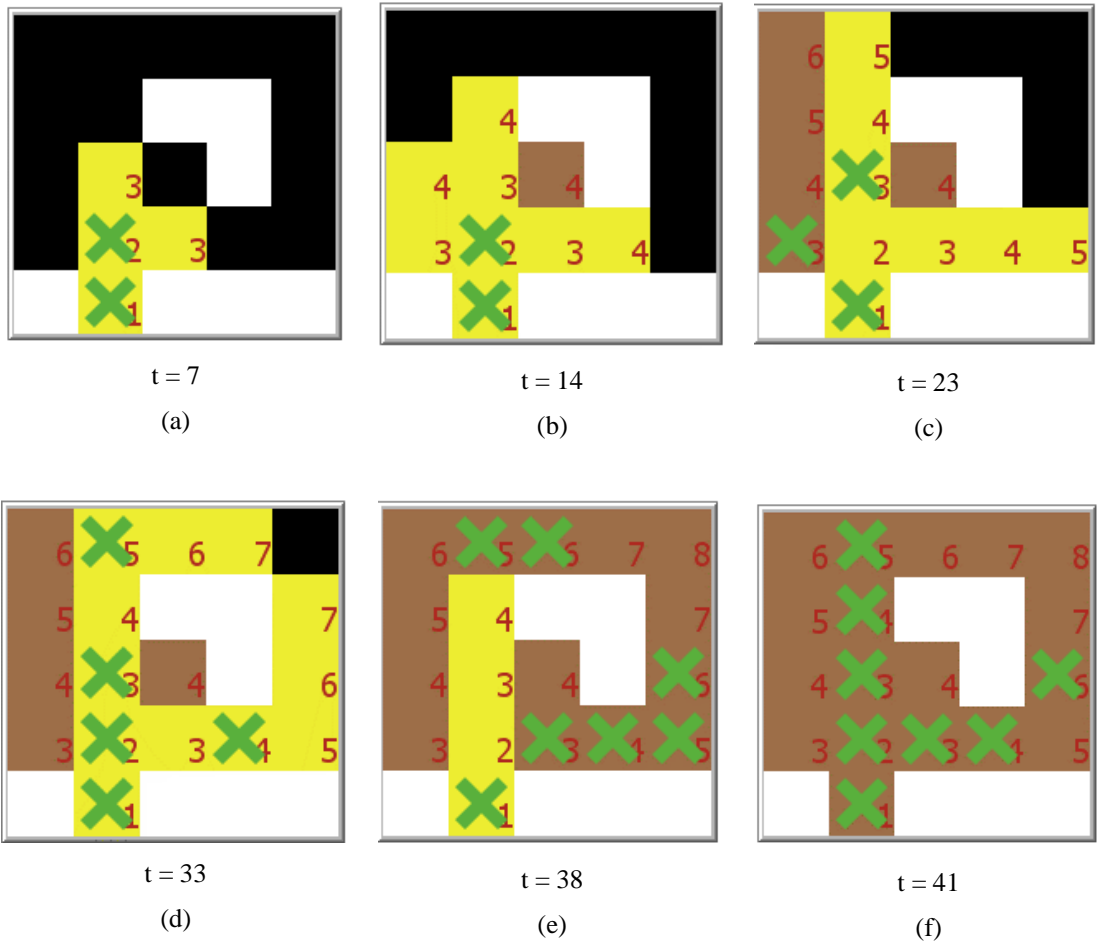

Figure 34    Single Layer, Unlimited Gradient – The snapshots of the process are ordered left to right, top to bottom. Color code: empty cells are black, obstacles are white, cells filled by a Beacon agent are yellow and cells filled with a Closed Beacon are brown, the step count in red and mobile agents are green "X"-s.

*5.4.2 Algorithm Analysis & Performance*

The extensive experimental show that the termination time in SLUG is linear with the size of the region for $\Delta T > 1$ and regardless of the topology of the region. Moreover, for $\Delta T = 1$ the termination time is upper bounded by the termination time for $\Delta T = 2$. Additionally, the termination time is linear with $\Delta T$ for $\Delta T > 1$ and indeed $T_{BPC} \leq n$. Based on these results we propose the following conjecture.

**Conjecture 3**: The termination time of the SLUG algorithm over any connected region **R** is upper bounded by

$$T_C(\mathbb{G}) \leq (2n - 1) \cdot max(2, \Delta T) + n$$

## 5.5 Single Layer, Depth First (SLDF)

Depth first search was used in previous research to explore regions but with different assumptions about an agent's capabilities or region. For example in [39] a multi robot DFS is used to explore only tree-type regions. This approach is expanded in [40] to more general regions but with robots having unlimited memory. In [41] the authors present an upper bounds for the number of moves traversed by $k$ agents implanting a distributed DFS protocol that solves a related problem - the Labelled Map Construction problem. The agents in that research are anonymous, asynchronous but again have unlimited memory and assumed to be initially dispersed in the graph. Similarly, in [42] strategies for the exploration of a graphs by $Dn^{1+\varepsilon}$ agents (D denoting the diameter of the graph) are given. Again, the agents are assumed unlimited memory, as do the vertices. A similar problem, deploying agents in a region such that all points of the region are visible to at least one agent, is discussed in [43]. The region is assumed simply connected and the agents to have a unique ID thus simplifying the traversal of the resulting tree. In this research, however, the region's topology is not limited, the agents are initially located outside the region and the agents have limited memory and sensing ranges.

The depth-first exploration strategy used in this algorithm differentiates it from all the algorithms described previously. In depth-first exploration the objective is to search "deeper" first and consequently the agents move as far as possible along the gradient before settling and expanding the covered region. In this algorithm the two meta-concepts described above are combined with the depth-first exploration strategy. Agents fill (and thus explore) the region along a single trajectory until further exploration is impossible. At which point the exploration continues from the closest branching point (i.e., a cell with a Beacon inside). This process repeats itself until the entire region is covered. Thus, at any time there is exactly one directed path along which the agents move up the gradient. Consequently, the step counts of settled agents are equal to the distance from the empty point. Explicitly, mobile agents move up the gradient of step counts projected by the settled agents as much as possible before settling and becoming Beacons. When a mobile agent reaches the end of the gradient, (i.e. the last settled agent projecting a step count) and senses an empty cell, it will settle in it. Thus, a main difference in the action rules implementing SLDF and those of other Single Layer Coverage algorithms is which action is done as a first priority – advancing up the gradient in SLDF versus settling in the SLUG and SLLG. When an agent can neither advance nor settle, the agent will move to a cell with a lower step count thereby retracing its path. When retracing, the agent will only move above settled agents whose state is Closed Beacon in order to prevent the mutual interference between the agents moving up and those moving down the gradient. The transition of a settled agent to the Closed Beacon state in SLDF follows directly from the action rules governing the advance up the gradient and is thus the same as in SLLG (a mobile agent may move only to neighboring cells with a step count greater by exactly one than its current step count) and thus described in Definition 19. Intuitively the result of Single Layer, Depth First algorithm is a spanning tree of the region with its root at the entry point (or a spanning forest if there is more than one entry point). This is because at any time step the mobile agents advance over exactly one path.

In the flowchart of the Single Layer, Depth First algorithm (see Figure 35) the different order of actions compared to the SLUG and SLLG is apparent. Blocks (1) to (4) describe movement up the gradient while blocks (5) to (8) describe the settling logic. The behavior of mobile agents at each time step $t$ at location $u$ is according to the pseudocode in Figure 36 while the behavior of the settled agents is according to Figure 40 and Figure 43.

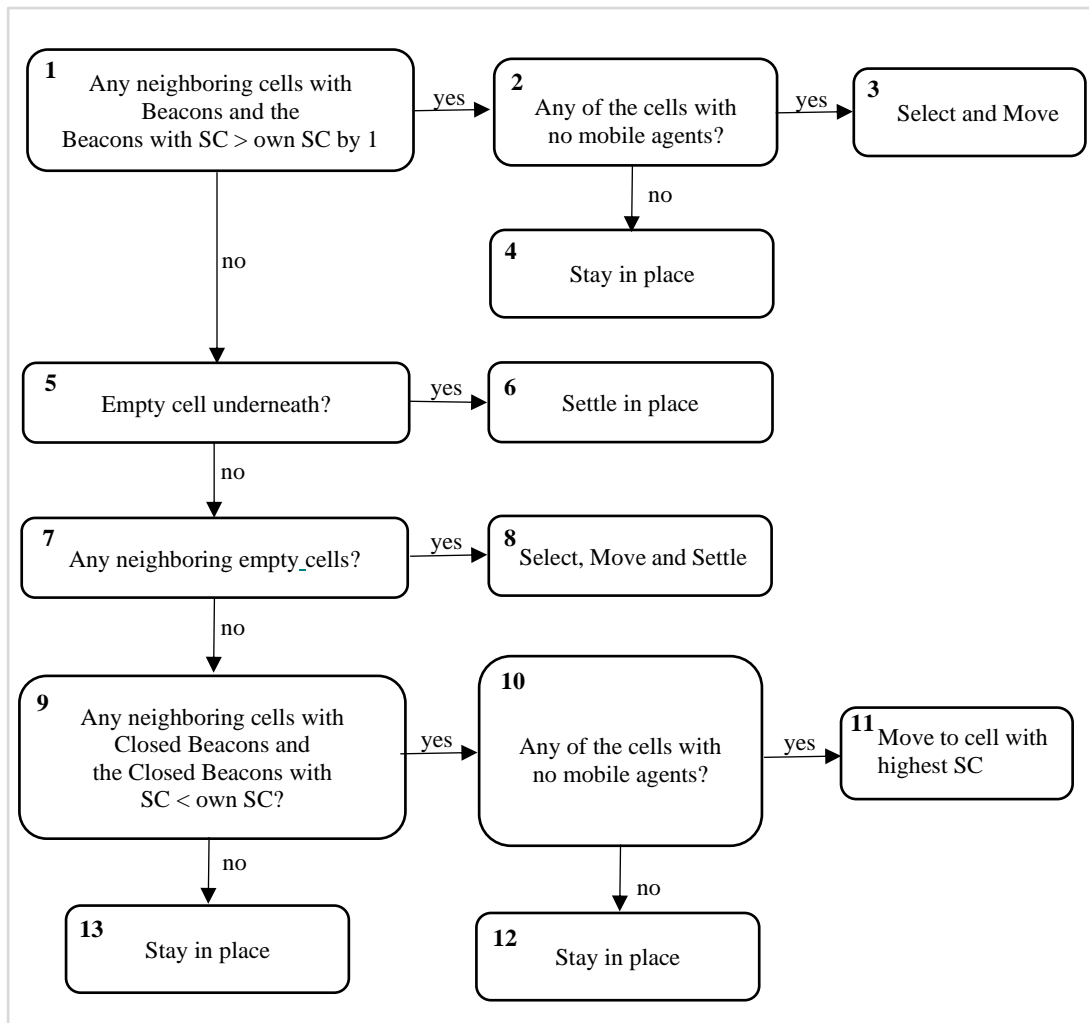

Figure 35     Single Layer, Depth First – mobile agent flowchart
Each block is numbered and referred to by this number in the pseudocode below

**Initialization**: for all agents $a \in A$ located at $t=0$ at the agent source set $S_i = (s_1, s_2) = ("mobile", 0)$

```
If s₁ = "mobile
    possiblePos ← ∅                                          /*initialize possiblePos - the set of possible
                                                             /*locations
    destSC ← s₂ + 1                                          /*define the step count of the destination
                                                             /*cell
    ξᴳ = {ξⱼ: j = 0: 4}
    relevantCells ← ξᴳ with ξᴳ.s₁ = "Beacon" and ξᴳ.s₂ = destSC
                                                             /*the subset of ξᴳ that is filled with Beacons
                                                             /* and have a step count eq. to destSC
    If |relevantCells| > 0                                   /*if there are neighboring cells with Beacons
                                                             /*and step count>own step count by 1[5]
        possiblePos ← relevantCells with ξᴬ = 0              /*the subset of relevantCells with no
                                                             /* mobile agents
        If |possiblePos| > 0                                 /*if there is a cell to move-to [6]
            u ← v;   s₁ ← "mobile";   s₂ ← destSC            /*move to cell and increase step count by 1
        Else
            u ← u;   s₁ ← "mobile";   s₂ ← s₂               /* stay in place [8]
        End If
    Else
        ξ̃ᴳ = {ξⱼᴳ = 0: j = 1: 4}                             /* define set ξ̃ᴳ as all empty neighboring
                                                             /*cells
        If ξ₁ᴳ = 0                                           /*if the cell directly underneath is empty [1]
            u ← u;   s₁ ← "settled";   s₂ ← 1                /* set step count as 1 [2]
        Elseif ∃ ξ̃ᴳ = 0                                      /*if not [3]
            v ← random(|ξ̃ᴳ|)                                 /*select a cell from ξ̃ᴳ
            u ← v;   s₁ ← "settled";   s₂ ← s₂ + 1           /* settle at selected empty cell, update state
                                                             /* increase step count by 1 [4]
        Else
            relevantCells ← ξᴳ with ξᴳ.s₁ = "Closed Beacon" and ξᴬ = 0
            destSC ← max (relevantCells with relevantCells.s₂ < s₂ )
            possiblePos ← relevantCells with relevantCells.s₂ = destSC
            If |possiblePos| > 0
                v= ← random(|possiblePos|)                   /*select a cell from possibplePos
                u ← v;   s₁ ← "settled";   s₂ ← destSC       /* settle at selected empty cell, update
                                                             /* state, increase step count by 1 [4]
            Else
                u ← u;   s₁ ← "mobile";   s₂ ← s₂            /*stay in place [8]
            End If
        End If
    End If
Else                                                         /* if the agent is not mobile
    Do SLLG – settled agent action rules (Figure 28)         /* see explanation above
End
```

Figure 36    Single Layer, Depth First algorithm – mobile agent pseudocode

### 5.5.1 Example

Using the same region as in the previous examples, an example of a coverage process using the Single Layer, Depth First is given Figure 37. The start of the backward propagating closure is clearly seen in Figure 37(c). The larger number of superfluous agents, compared to the SLLG example above, is result of the longer coverage time. The larger spatial distribution of those agents is common in runs in which the Backward Propagating Closure proceeds faster than the retracing by the superfluous agents.

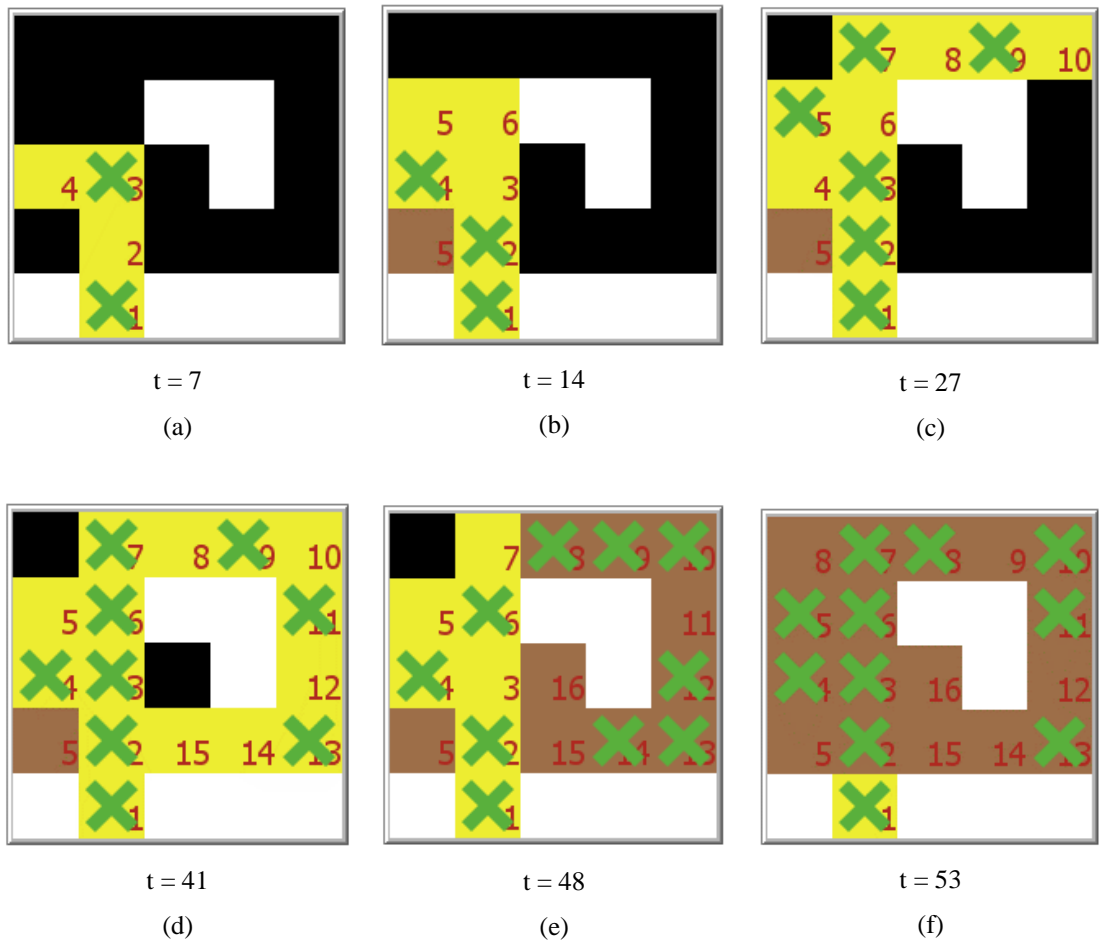

Figure 37    Single Layer, Depth First – Example. The snapshots of the process are ordered left to right, top to bottom. Color code: empty cells are black, obstacles are white, cells filled by a Beacon agent are yellow and cells filled with a Closed Beacon are brown, the step count in red and mobile agents are green "X"-s

*5.5.2 Algorithm Analysis & Performance*

**Lemma 2**: The upper bound on the termination time of a coverage process using the SLDF algorithm on the region **R** with a single entry point is given by:

$$T_C(\mathbb{G}) \leq \begin{cases} 5n - 5 & \Delta T = 1 \\ (2n - 1)\Delta T + n & \Delta T \geq 2 \end{cases}$$

We derive the expressions for the upper bound by first proving that SLDF generates a spanning tree over R, next show that the termination time of a spanning tree is upper bounded by the termination time of a linear graph with respect to a dual layer coverage process. Finally, define a new "virtual algorithm" composed of Dual Layer, Limited Gradient and Backward Propagating Closure (see proof of Lemma 1) over a linear graph and show its termination time upper bounds the termination time when using SLDF.

**Proposition 6:** The SLDF algorithm generates a spanning tree, **T**, of a region **R** with single entry point at time $T_C$.

*Proof*: Agents enter the region R from the entry point. At time $t = 1$, the tree **T**(1) includes only the entry point. At every subsequent time step mobile agents advance up the gradient. Suppose at time $t = k$ there is an empty cell, $v$. that is adjacent to cell $u \in \mathbf{T}(k)$ occupied by a settled agent, in state Beacon, with the largest step count at time $t = k$. According to the SLDF action rules the mobile agents advance from the entry point to $u$ hence by $t + \Delta T$ a mobile agents will move from $u$ to $v$ and we'll get $v \in \mathbf{T}(k + \Delta T)$. Note that $v$ can't be adjacent to a Closed Beacon since this contradicts the definition of Closed Beacon (i.e., no empty neighboring cells). Suppose the empty cell, $v$ also neighbors the cell $w \in \mathbf{T}(k)$ that is occupied by a settled agent, in state Beacon, but not with the largest step count at time $t = k$. Advance of agents from $w$ to $v$ is prohibited by the SLDG action rules. Hence the cell $v$ will have an in-degree of exactly 1 which is the definition of tree and $V(\mathbb{G}) = V(\mathbf{T}(T_C))$. ∎

**Proposition 7:** The termination time of the coverage process over a spanning tree $\mathbf{T}(V, E_T)$ using the DLLG is upper bounded by:

$$T_C(\mathbf{T}) \leq \begin{cases} 4n - 5 & \Delta T = 1 \\ (2n - 1)\Delta T & \Delta T \geq 2 \end{cases}$$

*Proof*: The DLLG action rules allow agents to move only up the gradient. In a spanning tree the movement of agent $a_i$ at time step $t = k$ up the gradient depends solely on the relative wake up times of the agents ahead of it and the spaces (i.e., settled agents without a mobile agent) between them. The filling of a branch in a tree does not delay the filling of any other branch since the agents continue moving up the gradient and superfluous agents cannot enter the vertex at which the branching occurs. Using the adversarial scheduler defined in Definition 12: The distance of agent $a_i$ from the entry point is given by $\delta(\text{EP}, j)$.

**Definition 13**, the termination time of the DLLG algorithm over a directed tree is maximal when the tree is a linear graph (tree). The reason, in a linear graph all the agents ahead of agent $a_i$ affect its movement whereas in all other trees the number is smaller by definition. From Proposition 3 we have the expression for the termination time of the DLLG algorithm over a linear graph. ∎

In the proof of Lemma 1 we defined a notional algorithm called "DLLG with BPC" that is an upper bound on the termination time of a region. We apply this notional algorithm on a linear region since according to Proposition 7 its termination time upper bounds that of a spanning tree of the same cardinality. The termination time of our notional algorithm is the sum of the termination time of the DLLG algorithm and run time of the BPC algorithm over said region. We thus get:

$$T_C(\mathbb{G}) \leq \begin{cases} 5n - 5 & \Delta T = 1 \\ (2n - 1)\Delta T + n & \Delta T \geq 2 \end{cases} \quad \blacksquare \tag{21}$$

## 5.6 Single Layer, Tree Traversal (SLTT)

The SLTT algorithm differs from the other algorithms described above in the way mobile agents are guided. Instead of moving on a gradient of step counts, in the SLTT algorithm the mobile agents follow "arrows" as described in [44]. The "arrows" are projected by the settled agents and signal the last direction an agent moved prior to settling in its own local reference frame. The SLTT extends the algorithm in [44] and generates a single layer coverage instead a dual layer coverage using the Backward Propagating Closure meta-concept described above. As in the SLLG and SLUG algorithms, the top priority of an agent is to settle and become a beacon. At every time step a mobile agent that cannot settle, tries to move forward according to the arrows. A mobile agent will only move to a cell with its arrow indicating that the settled agent inside it also moved in that direction. When a settled agents has several neighboring empty cells, following mobile agents will branch out in the different directions. In order to describe the SLTT algorithm explicitly we first define the state of an agent using arrows (instead of step counts).

In accordance with Definition 2, the state, $S_i$, of agent $a_i$ is determined by the combination of its physical state (i.e., mobile or settled) and Arrow.

**Definition 20**: The Arrow of agent $a_i$ at time step $t = k$, denoted by $\mathcal{A}_i(t)$, is the direction the agent selected at step $t = k$. The Arrow can take any of the 4 values 1 to 4 depicted in Figure 4. Note that the Arrow signaled by agent $a_i$ is the last direction selected by $a_i$ prior to settling.

**Definition 21**: The state of an agent $a_i$ using the SLTT algorithm is defined by the 2-tuple $S_i = (s_1, s_2)$ where $s_1$ has a value from the set {mobile, Beacon, Closed Beacon} and $s_2$ is the Arrow.

Based on the above definitions we can now write that an agent will advance when in at least one of its neighboring cells the Arrow projected by neighboring settled agent is pointing in the same (relative) direction as the enumerated cell with respect to agent $a_i$. This removes the need for a common reference frame (or even chirality) between the agents. In Figure 38 this is the case with the neighboring cells 1 and 2. An agent will retrace when in at least one of its neighboring cells the Arrow signaled by a neighboring cell is in the opposite direction to the enumerated position of that cell. This geometric relationship is translated to whether the difference between the Arrow direction and the number of the cell is exactly two (in absolute terms). In Figure 38 this is the case with neighboring cell 3. The flowchart in Figure 39 resembles that of SLLG with 2 major changes. First, the use of Arrows instead of step counts (mainly) manifests itself in blocks (5) and (9) that describe the action rules for advancing either from or towards the entry point. The second is a change in the mechanism of the Backward Propagating Closure meta-concept. In the BPC meta-concept described above a settled agents changes its state depending on the step count of its neighboring settled agents. While this definition is irrelevant (since there are no step counts in SLTT) the basic idea is kept. A settled agent using SLTT transitions to Closed Beacon when all its children are Closed Beacon and there are no empty neighboring cells as defined below. The dependency only on the children and not all neighbors means that transition to Closed Beacon using SLTT can be faster compared to the previous implementations of the Backward Propagating Closure. This behavior is corroborated by the experimental results.

**Definition 22:** The state of the settled agent $a$ located at $u$ will change to Closed Beacon if:

(1) $\check{\xi}^G = \hat{\xi}^G_u$ with
  a. $\check{\xi}^G = \{\check{\xi}^G \leftarrow \xi^G \text{ with } \xi^G_j.s_2 = j \mid j = 1:4\}$
  b. $\hat{\xi}^G \leftarrow \xi^G \text{ with } \xi^G.s_1 = \text{"Closed Beacon"} \mid j = 1:4\}$

(2) $\xi^G_j \neq 0 \mid j = 1:4$

The change is in the first condition and the updated flowchart and algorithm are given in Figure 40 and Figure 43 respectively. The algorithm analysis and specifically the upper bound on the time are relevant since the underlying concept is not changed. Nevertheless, this modified definition greatly effects the performance of SLLT as is shown in the following section.

The result of the SLTT algorithm is a spanning tree since, trivially, the directions of movement from two cells can't match the direction of movement of a single destination cell.

Direct implementation of the SLTT action-rules requires agents that can store as well as project their Arrow. Since the signaling is by the beacons to the flying agents, multiple lights facing upwards controlled by color, intensity, etc., seems the most logical solution [7, p. 252]. In the case of a quadcopter, the lights at the ends of the rotor booms can be used for this. Moreover, a finite memory of 2 bits is required to store the Arrow.

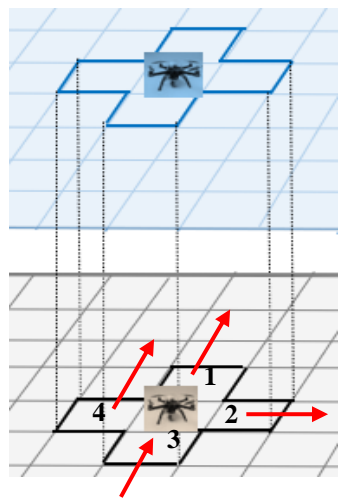

Figure 38      Single Layer, Tree Traversal algorithm – relative direction of movement. The black numbers enumerate the neighboring cells. The red arrows denote the last direction the agent settled in each of the neighboring cells moved.

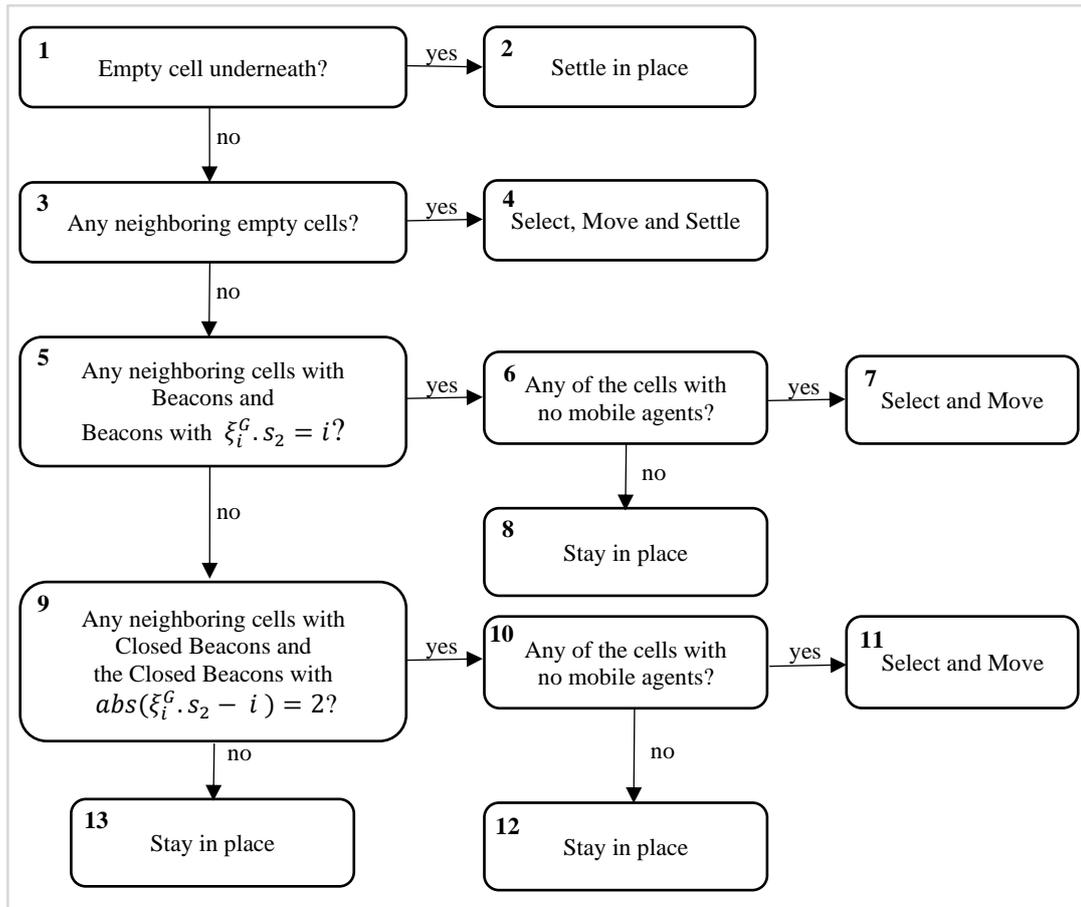

Figure 39    Single Layer, Tree Traversal algorithm– mobile agent flowchart
Each block is numbered and referred to by this number in the pseudocode below

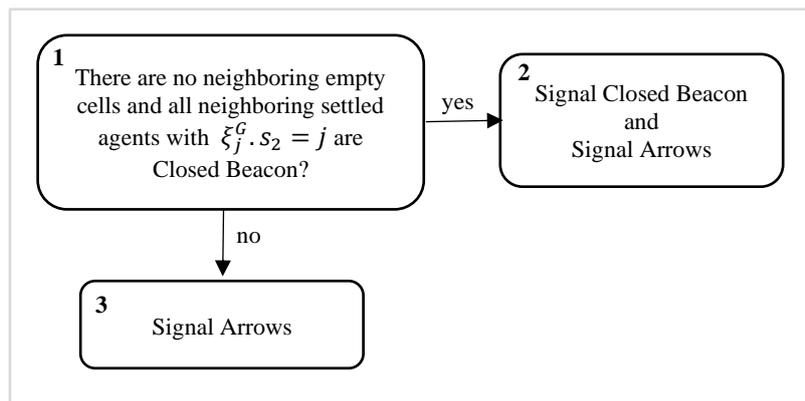

Figure 40    Single Layer, Tree Traversal algorithm– settled agent flowchart

**Initialization**: for all agents $a \in A$ located at $t=0$ at the agent source set $S_i = (s_1, s_2) = ("mobile", 0)$

```
If s_1 = "mobile"
    If ∃(ξ^G) = 0 then                                    /*if there are empty, neighboring cells
        If ξ_1^G = 0 then                                 /*if the cell directly underneath is empty [1]
            u ← u;   s_1 ← "settled";   s_2 ← 0           /*[2]
        Else                                              /*if not [3]
            ξ̃^G = {ξ_j^G = 0: j = 1:4}                    /* define set ξ̃^G as all empty neighboring
                                                          /*cells
            v ← random(|ξ̃^G|)                             /*select a cell from ξ̃^G
                                                          /*settle at selected empty cell, update state,
                                                          /*set A as the direction to move to v
            u ← v;   s_1 ← "settled";   s_2 ← A(v)
        End If
    Else
        possiblePos ← ∅                                   /*initialize possiblePos - the set of possible
                                                          /* locations
        relevantCells ← ξ^G with ξ^G.s_1 = Beacon and ξ_i^G.s_2 = i (for i = 1:4)
                                                          /*the subset of ξ^G that is filled with Beacons
                                                          /*and the direction they are signaling is eq.
                                                          /* to the relevant position in the sensing
                                                          /* neighborhood

        If |relevantCells| > 0                            /*[5]
            possiblePos ← relevantCells with ξ^A = 0      /*the subset of relevantCells with no mobile
                                                          /*agents
            If |possiblePos| > 0                          /*if there are cells to move-to [6]
                v = random(|possiblePos|)                 /*select a cell from possiblePos [7]
                u ← v;   s_1 ← "mobile";   s_2 ← A(v)     /*move to selected cell and set A as the
                                                          /*direction to move to v
            Else
                u ← u;   s_1 ← "mobile";   s_2 ← s_2      /*stay in place [8]
            End If

        Else
            relevantCells ← ξ^G with ξ^G.s_1
                       = "Closed Beacons and abs(ξ_i^G.s_2 − i) = 2
                         (for i = 1:4)
                                                          /*search for cells that are Closed Beacons
                                                          /*and the direction they are signaling is
                                                          /*opposite to the relevant position in the
                                                          /*sensing neighborhood
```

Figure 41      Single Layer, Tree Traversal algorithm – mobile agent pseudocode – part 1

```
        If |relevantCells| > 0                                       /* [9]
          possiblePos ← relevantCells with $\xi^A = 0$    /*the subset of relevantCells with no mobile
                                                                              /*agents
            If |possiblePos| > 0                                     /*if there are cells to move-to [10]
              possiblePos ← possiblePos with $\xi^A = 0$
                                                                              /* find all the cells in possiblePos with no
                                                                              /* mobile agents
                v = random(|possiblePos|)                       /*select a cell from possiblePos [11]
              u ← v;    $s_1$ ← "mobile";    $s_2$ ← $\mathcal{A}(v)$
                                                                              /*move to selected cell and set $\mathcal{A}$ as the
                                                                              /*direction to move to v
            Else                                                             /* [12]
                u ← u;    $s_1$ ← "mobile";    $s_2$ ← $s_2$     /*stay in place
            End If

          Else                                                               /* [13]
                u ← u;    $s_1$ ← "mobile";    $s_2$ ← $s_2$     /*stay in place
          End If

        End If
    End If
Else                                                                         /* if the agent is not mobile
    Do Backward Propagating Closure algorithm (see Figure 43)
End
```

Figure 42  Single Layer, Tree Traversal algorithm – mobile agent pseudocode – part 2

```
If s₁ ≠ "mobile"                                          /* if the agent is not mobile
    If ∄ξⱼᴳ = 0, j = 1:4                                  /* If there are no empty cells [1]
        ξ̌ᴳ ← ξᴳ with ξⱼᴳ.s₂ = j                           /* ξ̌ᴳ - the set of neighboring settled agents with
                                                           /* he direction they are signaling is  eq.
                                                           /* to the relevant position in the sensing
                                                           /* neighborhood

        ξ̂ᴳ ← ξᴳ with ξᴳ.s₁ = "Closed Beacon"              /* ξ̂ᵤᴳ - the set of neighboring Closed Beacon
        If ξ̌ᴳ = ξ̂ᴳ                                        /* [2]
            s₁ ← "Closed Beacon"                           /* state is "closed beacon"
            Project (s₁)                                   /* signal state and Arrow
            Project (s₂)
        Else
            s₁ ← "Beacon"                                  /* state is "beacon"
            Project (s₂)                                   /* project Arrow
        End If
    End If
End If
```

Figure 43    Single Layer, Tree Traversal algorithm– settled agent pseudocode

*5.6.1 Example*

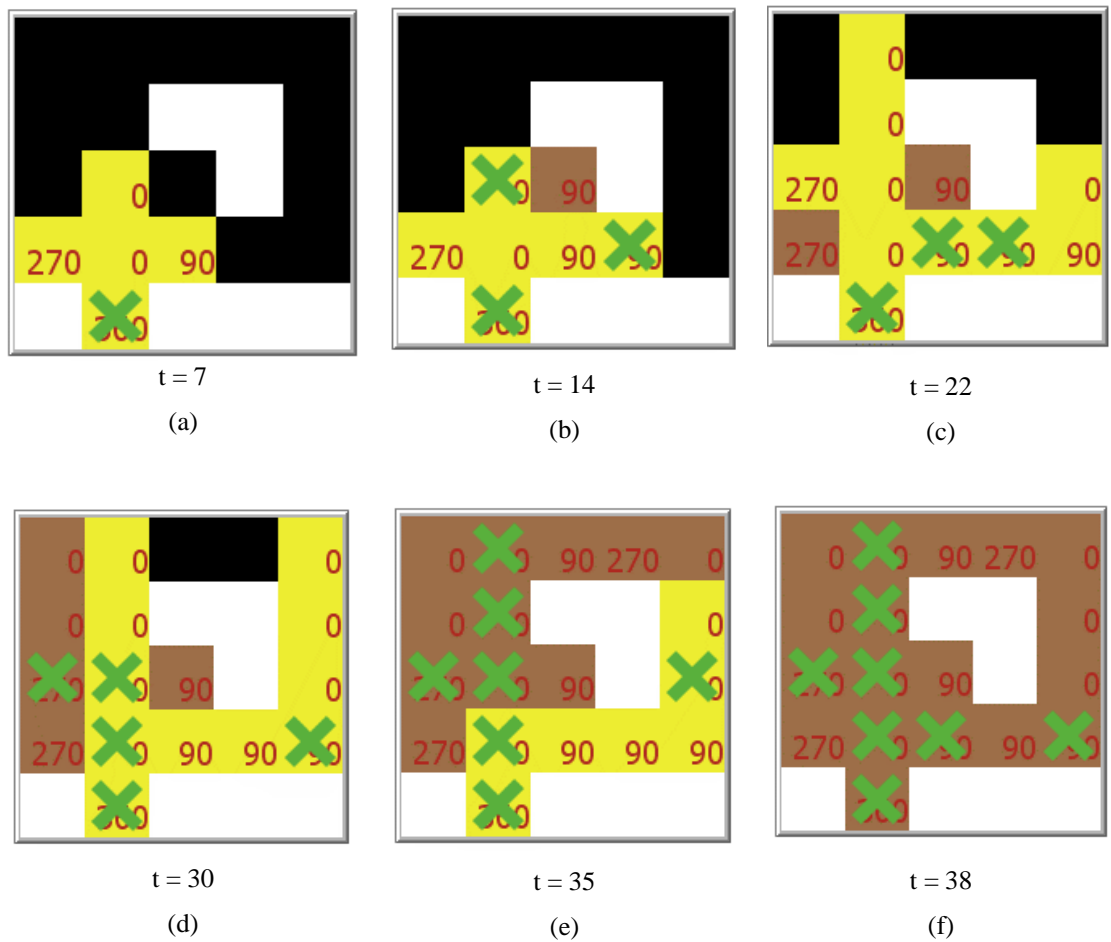

Figure 44　　Single Layer, Tree Traversal– Example. The snapshots of the process are ordered left to right, top to bottom. Color code: empty cells are black, obstacles are white, cells filled by a Beacon agent are yellow and cells filled with a Closed Beacon are brown, the directions are denoted in degrees for ease of understanding and mobile agents are green "X"-s

*5.6.2 Algorithm Analysis*

　　The Single Layer, Tree Traversal algorithm generates a spanning tree as described above and the action rules dictate that mobile agents retrace their movement only along settled agents in Closed Beacon state. Hence the analysis performed of the Single Layer, Depth First algorithm applies to this algorithm. Consequently, the upper bound is described in Lemma 2.

## 5.7 The traversal rules and the underlying graph

Different local action rules are used in the various algorithms to define the movement of the mobile actions. They differ from one another by the how loose they are in the sense of how much they limit the movement of an advancing mobile agent to adjacent neighboring cells. At one extreme is the random walk in which an agent does not have a notion of what "advance" is and can therefore move to all neighboring cells. At the other extreme are the Single Layer, Tree Traversal and Depth First algorithms in which the agent is limited to a gradient steepness of one and can move only to children of the current settled agent. Another way to look at the traversal rules is based on the outdegree of each settled agent. The outdegree of each settled agent, in the Dual Layer, Random Walk algorithm, is the same as that of the cell in **R** and it is bounded by 4. In all other algorithms, however, the outgoing degree of a settled agent is bounded by three and depends on the aforementioned action rules. The exploration and coverage process, in all algorithms, implicitly generates a directed acyclic graph (DAG) over the region **R** (in the case of the Single Layer, Tree Traversal and Depth First the underlying graph is a spanning tree which is a subset of the DAG). The number of edges in a DAG is related to the outdegree of each of its vertices which in this problem are the settled agents. Denoting by **D** (or **T**) the underlying DAG (or spanning tree) we can thus write the following general relationship:

$$|E(\mathbf{D}_{SLUG})| \geq |E(\mathbf{D}_{SLLG})| \geq |E(\mathbf{T}_{SLTT})| = |E(\mathbf{T}_{SLDF})| = n - 1 \qquad (22)$$

In the following section we will review this relationship using simulation results.

## 5.8 On the fault tolerance of the Single Layer Coverage algorithms

Based on the discussion above of the Single Layer Coverage algorithms we can now discuss the fault tolerance of these algorithms. When discussing search and rescue scenarios the tolerance to such failures is critical. Not only are the drones themselves relatively fragile, the environment, in the form of ceiling collapse etc. may greatly increase the probability of such failures. We define a fault as a mechanical or electrical failure of the drone that results in its ceasing to operate. If at the time of the failure the drone is mobile it will fall and may cause the failure of a settled agent[†].

The objective of the fault tolerance capability of the algorithms is for the coverage process to terminate correctly (as defined above) in finite time. In the following paragraphs we review possible failure modes, the adaptations to the local action-rules and resultant swarm behavior.

1. Failure of mobile drones. The failed drones will fall however the coverage process will only be delayed. That is because mobile drones sense the step count at every action cycle rather than store it. Moreover, mobile agents only sense the environment and do not signal any other agent. Therefore, the failure of a mobile agent will not affect the underlaying graph structure being generated nor any agent.

---

[†] This is not always the case since as described above the drones are significantly smaller than the cells they occupy

2. A mobile agent is blown aside to another cell. As explained above the mobile agent will not affect the settled agents however this failure mode is further divided into three separate cases. In the first, the new position is an empty cell in which case the agent will settle down and project a step count of "1" (the same behavior of the first agent to enter the region). The process will continue correctly since agents always move up the gradient and the BPC depends only on neighboring agents with a higher step count. In the second case, the agent is blown over a cell occupied only by a settled agent. The mobile agent will behave correctly in its next action-cycle since decisions are made using only locally sensed information. In the third case the agent is blown into a cell occupied by a mobile agent. In this case they will both fail however as explained above the proves will continue correctly. Summing, in all three cases, when a mobile agent is blown off the path the process terminates correctly however the termination time may be increase.
3. Failure of a settled agent. This failure mode is divided into two cases – depending on the current state of the process. Specifically, whether the settled agent in an area of Beacons or Closed Beacons. In the first case there are still mobile agents advancing in the region hence a mobile explorer will (at some time step) sense the (now) empty cell. This raises the question what will this mobile project after settling? The specific answer depends on the type of algorithm – is it gradient based or arrow based. However, in either case its crucial to maintain the connectivity of the region. If it's a gradient based algorithm the step count will be equal to the lowest step count of the surrounding settled agents incremented by 1. This assures movement is possible from the failure site to the entry point. If it is SLTT then the mobile agent will select the arrow direction leading to the failure site.

   In the second case the area surrounding the failed settled agent is filled by Closed Beacons. When a settled agent senses that a previously occupied neighboring cell is now empty it will change its state to Beacon and the failure will propagate to the entry point due the Backward Propagating Closure that will cause a cascading of the change through the region. As a result, mobile agents will start to enter the region (since now the entry point is in state Beacon) and advance up the gradient until reaching the failure site. Note that the above discussion of failures of settled is indifferent as to the cause (partial ceiling collapse or being hit by a falling mobile agent).

The above discussion shows the Single Layer Coverage algorithms are fault tolerant to failures of drones – mobile and settled – alike. The mutual interaction between different failures is only partially discussed and merits further research.

## 6    Results and Discussion

Seven different exploration algorithms were presented in sections 4 and 5 above. In this section the performance of the algorithms according to the metrics defined above is experimentally evaluated. In the first two sub-sections, the effects of the region (size and topology) and time between successive entries (i.e., *ΔT*) on the performance are evaluated for each family of algorithms separately. Putting aside the Dual Layer, Random Walk algorithm, the algorithms differ from one another by the traversal rules, whether or not the BPC meta-concept is used and the rules for retracing. Dual Layer, Limited Gradient (DLLG) and Dual Layer, Unlimited Gradient (DLUG) use different traversal rules (limited versus unlimited movement on the gradient) however neither uses Backward Propagating Closure nor allows moving down the gradient. Single Layer, Unlimited Gradient (SLUG), Single Layer, Limited Gradient (SLLG) and Single Layer, Tree Traversal (SLTT) all use greedy type expansion but with different traversal rules. SLUG enables unlimited movement on the gradient, in SLLG the movement is limited to a gradient steepness of 1 and in SLTT the movement is further limited to only those originating from the current cell. In addition, all use BPC but with different retracing rules – in SLUG movement down the gradient is possible on both Beacons and Closed Beacons while in SLLG and SLTT it is possible only on Closed Beacons. The fourth SLC algorithm, Single Layer, Depth First (SLDF), uses a depth first exploration strategy unlike the other three single layer algorithms with retracing limited to Closed Beacons. The effect of these factors along with the effect of the entry point location on the performance is investigated in the third part of this section.

The results are derived from a simulator developed and implemented by the authors using the NetLogo programming language. Three different regions– a line, a square and a repeating multi-connected region - were investigated for both the effect of size and topology. An additional three regions, representing more complex and real-life regions were also investigated. The topologies are presented in Figure 45 and the list of the regions (size and topology) tested is detailed in Table 1. Each data point in the all the following graphs is the average of fifty runs unless specifically noted.

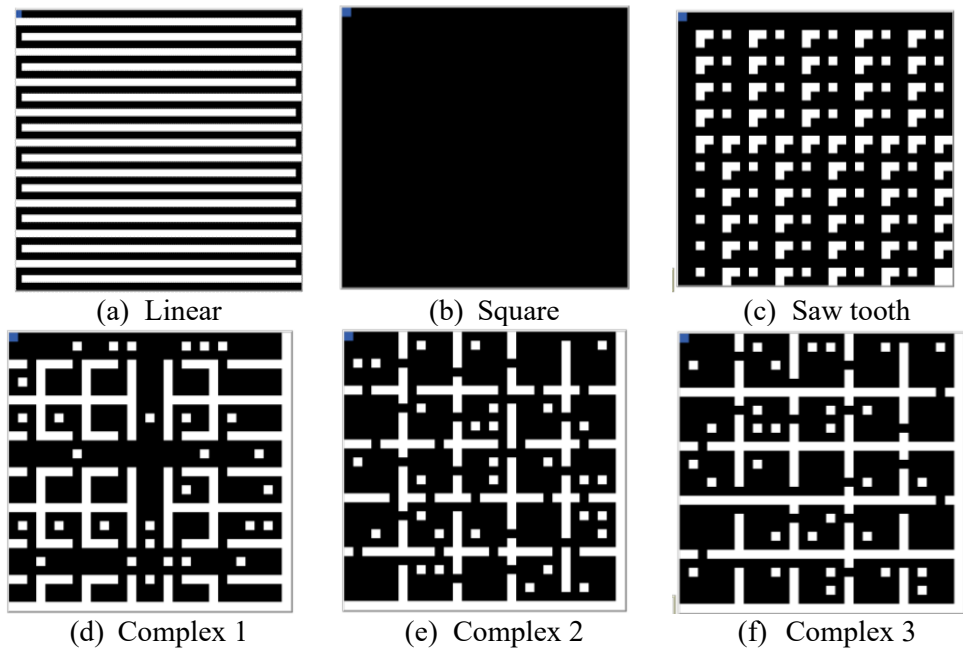

Figure 45 Regions used in the simulations. The regions in sub-figures (a)-(c) are used to investigate the effect of region shape. The bottom three regions represent more realistic regions and first appeared in [45]. The walls are marked in white, the empty cells in black and the entry point is at the upper-left corner and marked in blue.

| Region Type | Size |
|---|---|
| Linear | 120 |
| Square | 225 |
| Saw Tooth | 247 |
| Linear | 250 |
| Saw Tooth | 422 |
| Square | 441 |
| Linear | 571 |
| Complex 1 | 636 |
| Saw Tooth | 670 |
| Complex 3 | 670 |
| Complex 2 | 686 |
| Square | 961 |
| Linear | 987 |
| Square | 1681 |

Table 1 List of the regions (size and topology) evaluated

## 6.1 Dual Layer Coverage algorithms

The number of agents as a function of time, $N(t)$, from sample runs of the three Dual Layer Coverage algorithms – DLRW, DLLG, DLUG - is presented in Figure 46. A linear region with 120 cells, the entry point at the left terminal cell and $\Delta T = 1$ is used. The horizontal axis is the time while the vertical axis is the number of agents. The major difference between the algorithms is in the effective rate of agents' entry. The slope is linear-like in DLUG and DLLG indicating that the number of times the entry of an agent to the region is blocked (because the entry point is occupied) is nearly constant. On the other hand, in DLRW the slope (i.e., the entry rate) monotonously decreases indicating that as the number of agents inside the region increases so does the probability that there will be a mobile agent in the entry point blocking the entry of additional agents. Table 2 summarizes the termination time and total energy of the three DLC algorithms in the coverage process of linear regions. Clearly, the termination times and total energy consumption of the DLRW algorithm are significantly larger than those of the DLLG and DLUG algorithms. Consequently, in the rest of this section the discussion on the DLC algorithms will focus on the DLLG and DLUG algorithms.

| Region Size | DLRW | | DLLG | | DLUG | |
|---|---|---|---|---|---|---|
| | Termination Time | Total Energy | Termination Time | Total Energy | Termination Time | Total Energy |
| 120 | 16221.3 | 1.47E+06 | 472.7 | 21714.5 | 472.4 | 21696.5 |
| 250 | 81163.1 | 1.60E+07 | 992.1 | 94321.1 | 992 | 94370.2 |
| 415 | 223615 | 7.10E+07 | 1652.1 | 259464.6 | 1651.8 | 259423.2 |

Table 2  Termination Time and Total Energy for the three DLC algorithms over three linear regions ($\Delta T = 1$)

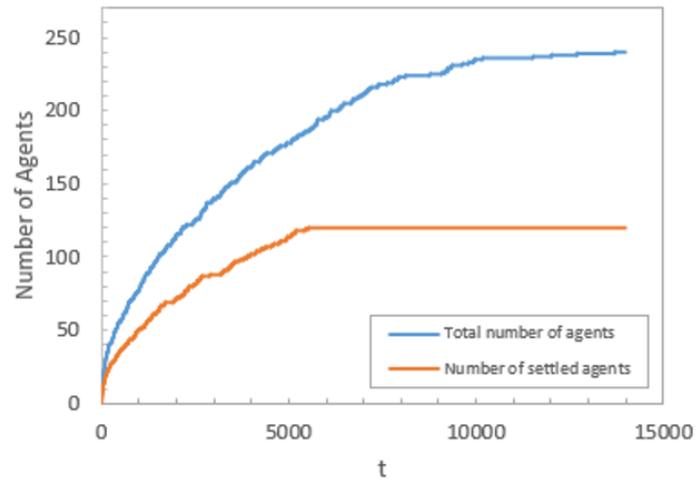

(a)

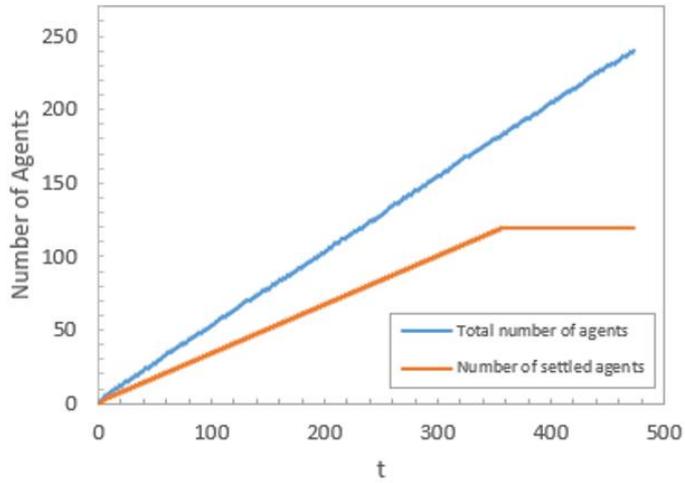

(b)

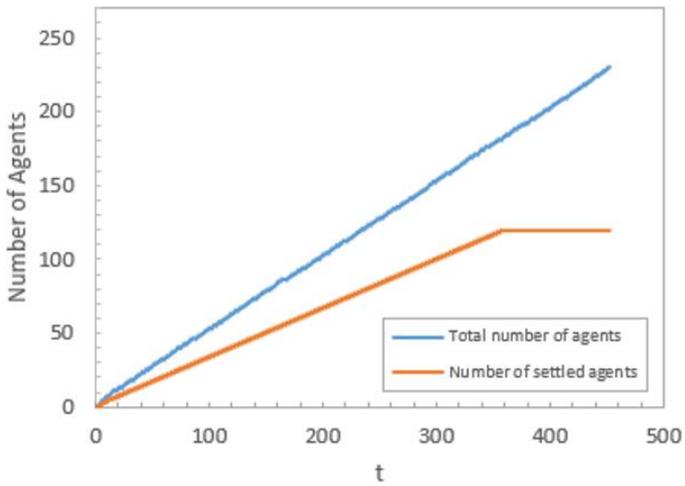

(c)

Figure 46     Total number of agents and number of settled agents as a function of time using the three Dual Layer Coverage algorithms on a linear region with 120 cells and $\Delta T = 1$. (a) DLRW; (b) DLLG; (c) DLUG

The theoretical analysis in section 4 predicts the termination time is linear with region size and $\Delta T$ for $\Delta T \geq 2$. Thus, we start by looking experimentally into the differences in the performance of the Dual Layer Coverage algorithms for $\Delta T = 1$ and $\Delta T = 2$. The effect of the region size on the termination time and the total energy with $\Delta T = 1$ is depicted in Figure 47. The lower bound and the upper bound (based on Conjecture *1* and *2*) are shown in Figure 47(a) in black. With respect to the termination time, all the results are within the bounds supporting both Conjectures. It is noteworthy that the termination times of the linear regions are the longest for a given size and very close to the upper bound. Moreover, the termination times of DLLG and DLUG are very close and in most regions the difference is imperceptible. In Figure 47 (b) the total energy consumption is shown and while in most regions $E_{Total}$ is similar there is a definite difference in the square regions. Given the random nature of the process, in a large, unobstructed region such as a square there is a higher probability that a mobile agent will move over a gradient greater than one compared to more obstructed regions such as the saw-tooth. Such movement will shorten the trajectories of the agent and reduce their energy consumption. Consequently, the main insight from Figure 47 is that the looser traversal rules in DLUG mostly effect the $E_{Total}$.

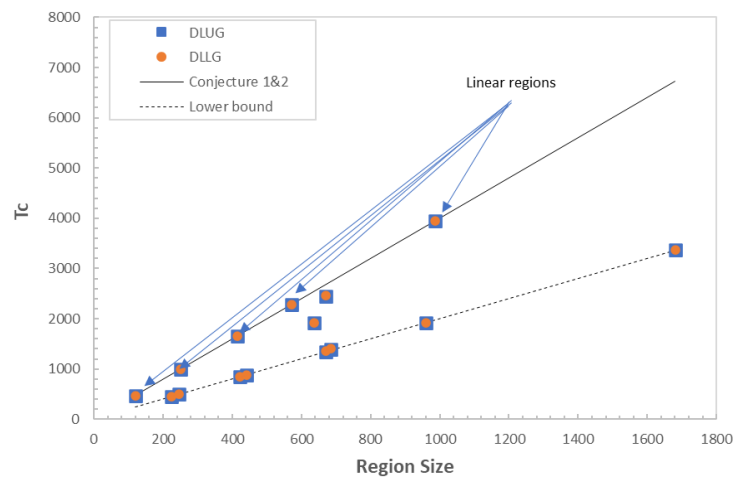

(a)

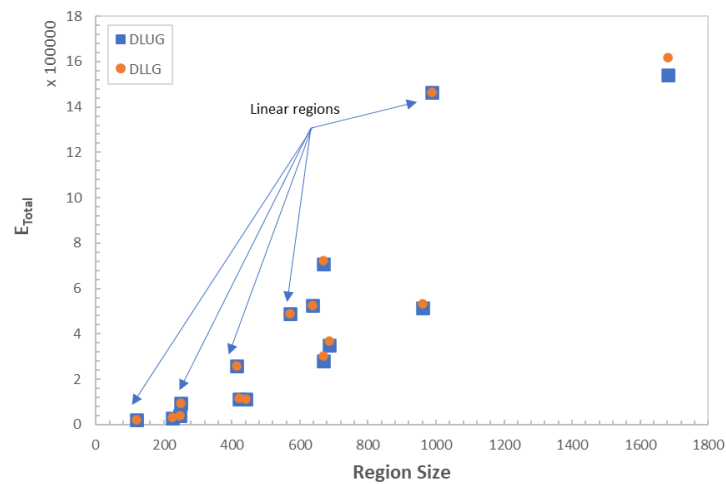

(b)

Figure 47    Dual Layer Coverage algorithms - Termination time and total energy for $\Delta T = 1$

When $\Delta T \geq 2$ the termination times, for a given region size, should all depend exclusively on the regions size. This understanding is supported by the results in sub-Figure 48(a) for all the regions. Additionally, from Figure 48(b) the difference in the total energy between DLLG and DLUG is largest in the square regions and this fact supports the insight gained from the plots for $\Delta T = 1$. Polynomial trendlines of the 2nd order with an intercept of 0 are shown for the total energy. The fit is good indicating the total energy may be a quadratic function of the region size.

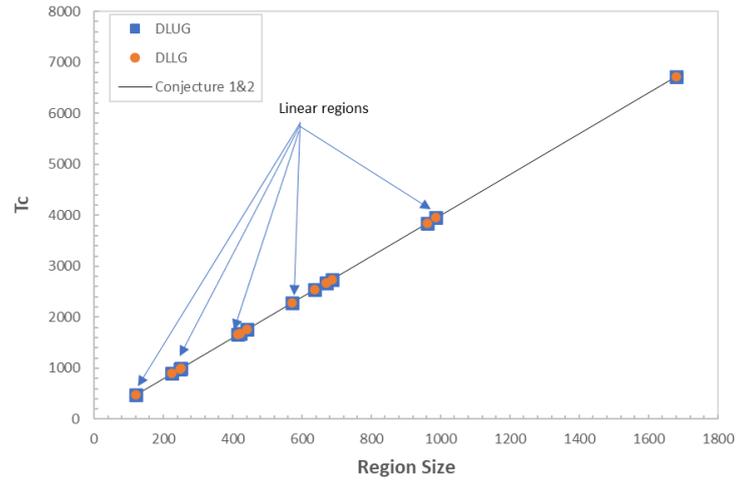

(a)

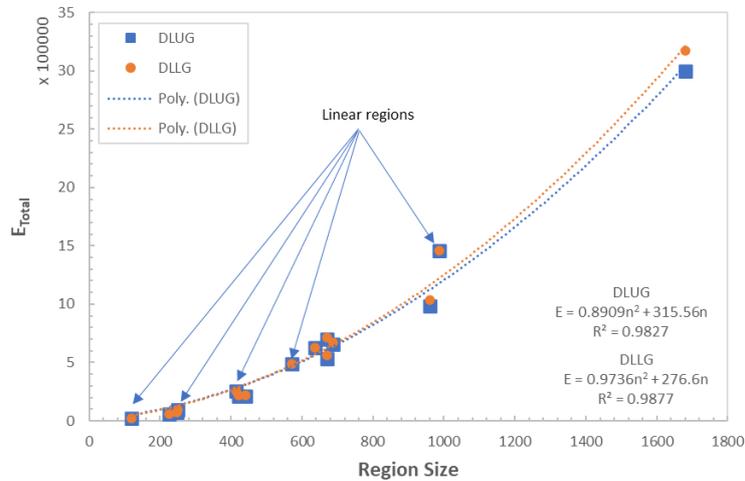

(b)

Figure 48　　Dual Layer Coverage algorithms - Termination time and Total Energy for $\Delta T = 2$

The effect of $\Delta T$ on the termination time in the DLC algorithms is shown in Figure 49. Figure 49(a) shows the termination time as a function of $\Delta T$ using Dual Layer, Limited Gradient algorithm and similarly Figure 49(b) for Dual Layer, Unlimited Gradient algorithm. Both the linear correlation between the termination time and $\Delta T$ (for $\Delta T \geq 2$ ) and the dependency of the slope on the region size are clear and fit the theoretical model of the process. Moreover, the termination time for $\Delta T = 1$ is upper bounded in all the regions by the relevant termination time for $\Delta T = 2$. An alternative presentation of the data is to show the correlation between the ratio of the termination time to region size and $\Delta T$ as is shown in shown Figure 50. The convergence of all the plots into a single line in both DLLG and DLUG indicates the good correlation between the theoretical and experimental results.

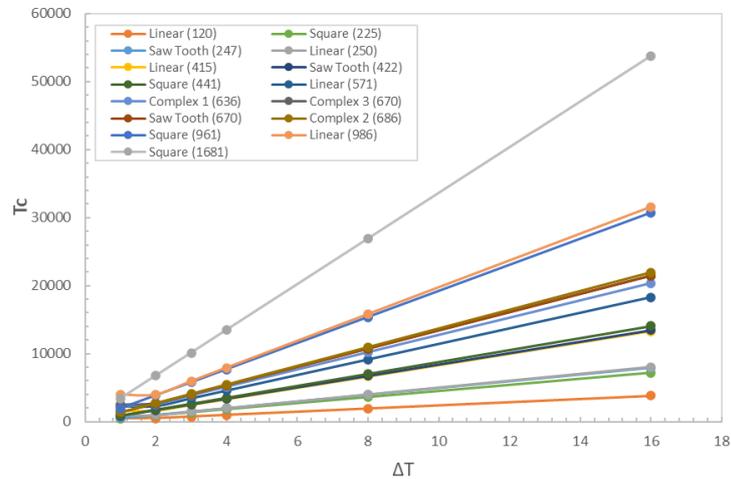

(a)

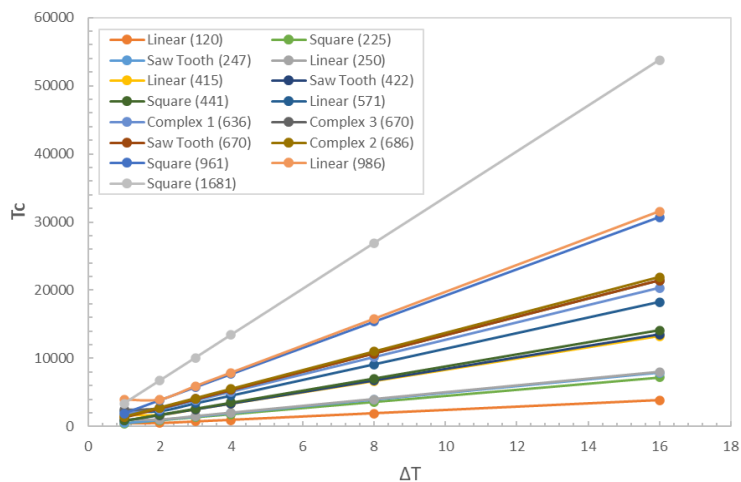

(b)

Figure 49  Termination time as a function of $\Delta T$ using Dual Layer Coverage algorithms for the various regions used. (a) DLLG; (b) DLUG

Yet another perspective on the results is to view the ratio of the termination time to $\Delta T$. This ratio is approximately $2n$ hence should result in parallel, horizontal lines when viewed for various values of $\Delta T$ and $\Delta T \geq 2$. This is exactly the result as seen in Figure 51(a) for DLLG and Figure 51(b) for DLUG.

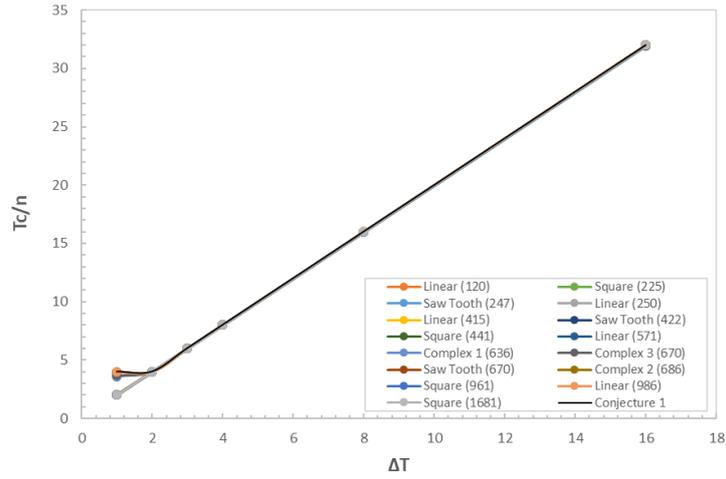

(a)

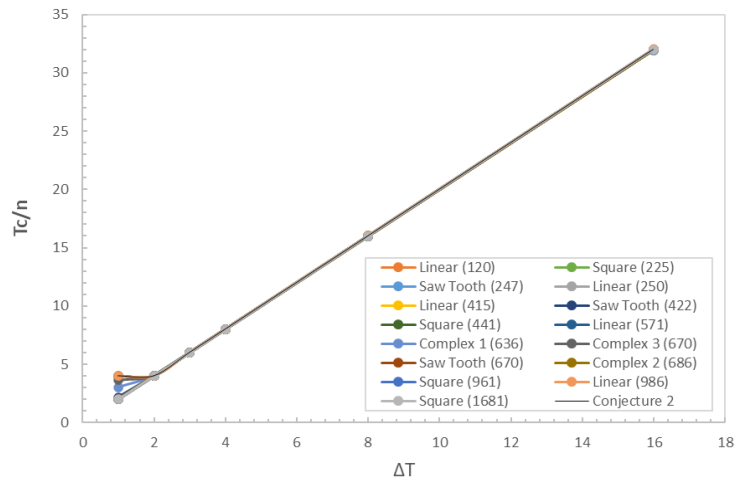

(b)

Figure 50　　Ratio of termination time to region size as a function of $\Delta T$ using Dual Layer Coverage algorithms (a) DLLG; (b) DLUG

Figure 52 shows the total energy for the different regions using $\Delta T = 4, 8$ and $16$. The energy consumption of the different algorithms is denoted by the markers and trendlines for both algorithms are denoted by the dotted lines. A 2nd order polynomial is used with the intercept point set to 0 hence of the form: $E_{Total} = C_2 n^2 + C_1 n$. This model was selected based on the notion that at least $n$ agents participate in the process and the time an agent is mobile is approximately the distance it travels which is bounded by n and thus the total energy is proportional to $n^2$. The good approximation is clear from the high values of $R^2$ in Figure 52 and Figure 48(b) . As evident, doubling of $\Delta T$ results in close to a doubling of the $n^2$ coefficient hinting that $E_{Total}$ is a linear function of $\Delta T$ as well as a quadratic function of $n^2$ when $\Delta T \geq 2$. These results merit further research.

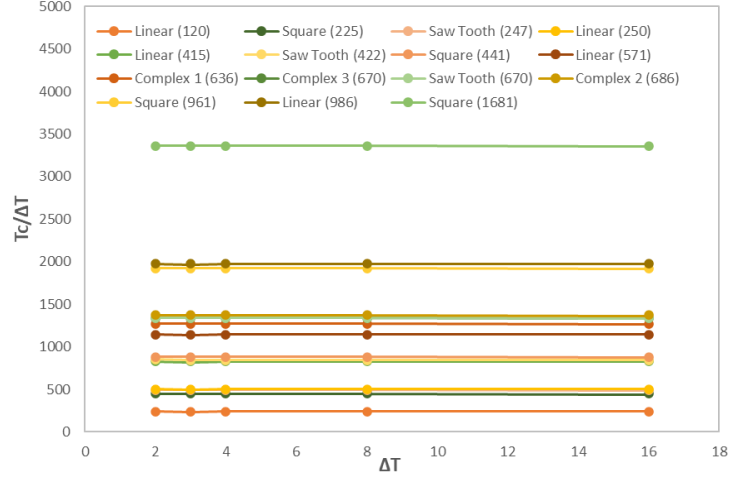

(a)

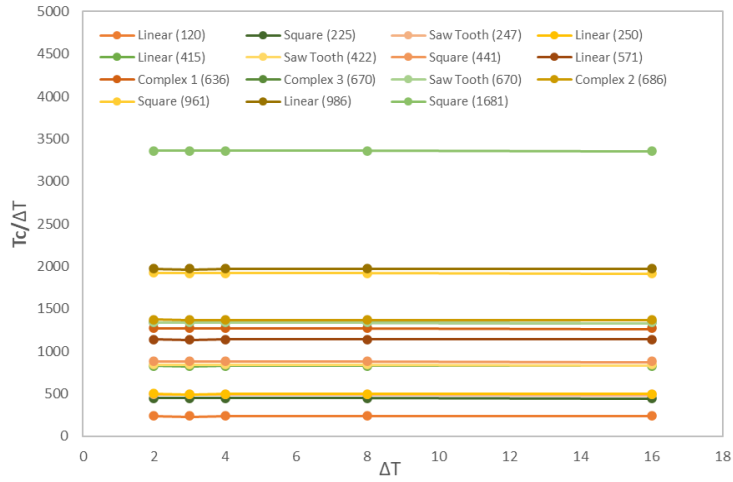

(b)

Figure 51    Ratio of termination time to $\Delta T$ as a function of $\Delta T$ using Dual Layer Coverage algorithms  (a) DLLG; (b) DLUG

Figure 52 shows the total energy for the different regions using $\Delta T = 4, 8$ and $16$. The energy consumption of the different algorithms is denoted by the markers and the trendlines for both algorithms are denoted by the dotted lines. A 2nd order polynomial is used with the intercept point set to 0 hence of the form: $E_{Total} = C_2 n^2 + C_1 n$ . This model was selected based on the notion that at least $n$ agents participate in the process and the time an agent is mobile is approximately the distance it travels which is bounded by n and thus the total energy is proportional to $n^2$. The good approximation is clear from the high values of $R^2$ in Figure 52 and Figure 48(b). As evident, doubling of $\Delta T$ results in close to a doubling of the $n^2$ coefficient hinting that $E_{Total}$  is a linear function of $\Delta T$  as well as a quadratic function of $n^2$ when $\Delta T \geq 2$. These results merit further research.

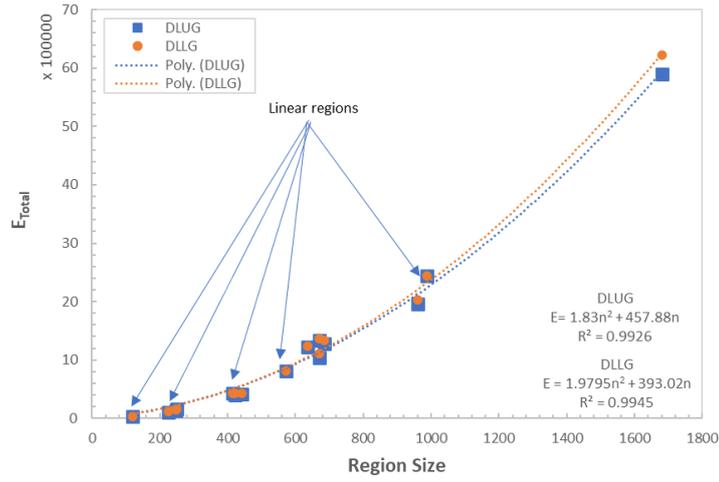

(a)

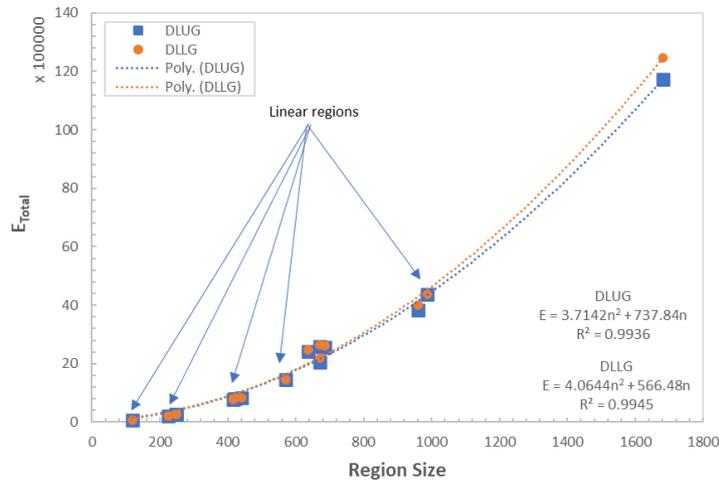

(b)

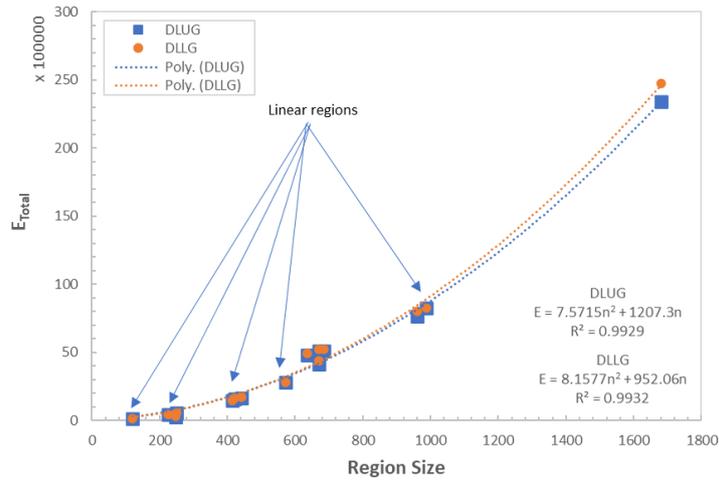

(c)

Figure 52  Total energy as a function of region size using Dual Layer Coverage algorithms
(a) $\Delta T = 4$; (b) $\Delta T = 8$; (c) $\Delta T = 16$
The equations and $R^2$ values of the trendlines are in the bottom-left corner

Summing, the results indicate that when $\Delta T \geq 2$ the termination time of the Dual Layer Coverage algorithms is linear with $n$ and $\Delta T$ while $E_{Total}$ is linear with $\Delta T$ and quadratic with respect to $n$. Furthermore, when $\varDelta T = 1$ both the termination time and $E_{Total}$ are upper bounded by the respective values for $\varDelta T = 2$.

## 6.2 The Single Layer Coverage algorithms

The bounds on the termination time of the Single Layer Coverage algorithms differentiate between the case of $\varDelta T = 1$ and $\varDelta T \geq 2$. We therefore start by examining the behavior of the algorithms for $\varDelta T = 1$ and for $\varDelta T = 2$ as shown in Figure 53 and Figure 54 respectively. The termination times in all the regions, as shown in sub-Figure 53(a), are lower than the upper bounds predicted by Lemma 1 and Conjecture 3 they are however further from the upper bound then in the Dual Layer Coverage algorithms (Figure 47(a) and Figure 48(a)). Additionally, for both $\varDelta T = 1$ and $2$ the $T_C, E_{Total}$ and $N$ in the linear regions are the same regardless of the algorithm used (also observed in the DLC algorithm). This is because in linear regions the gradient is always one and coincides with the directions of the arrows in the tree traversal algorithm. In the other regions the differences in performance between the four algorithms is clear, especially for $\varDelta T = 2$. SLUG shows the best performance i.e., lowest values) followed by SLTT, SLLG and lastly SLDF. The relative order in interesting since SLUG has the richest DAG while SLTT is a tree. The reason behind this phenomenon is discussed later in this section. In Figure 53(d) and Figure 54(d) the maximum specific energy consumption (max $(E_i)$) is presented. It is clear that SLLG out-performs SLTT in this metric in both $\varDelta T = 1$ and $2$. This means that in the examined scenarios the required minimal energy of an agent using SLLG is lower than that of SLLT however overall SLTT is more energy efficient compared to SLLG.

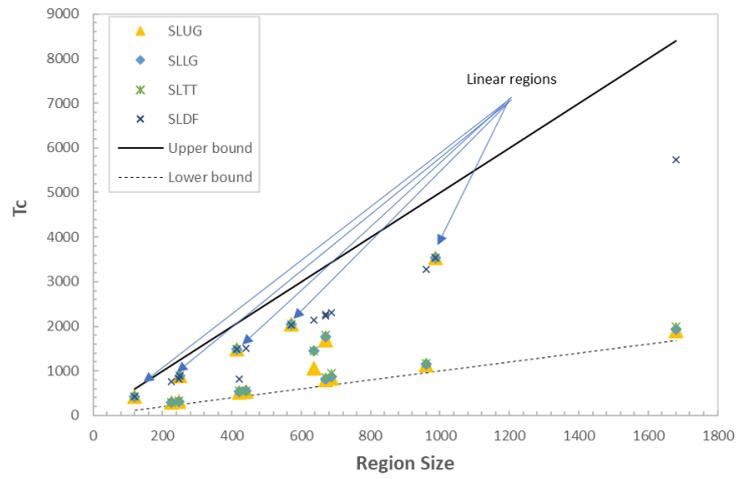

(a)

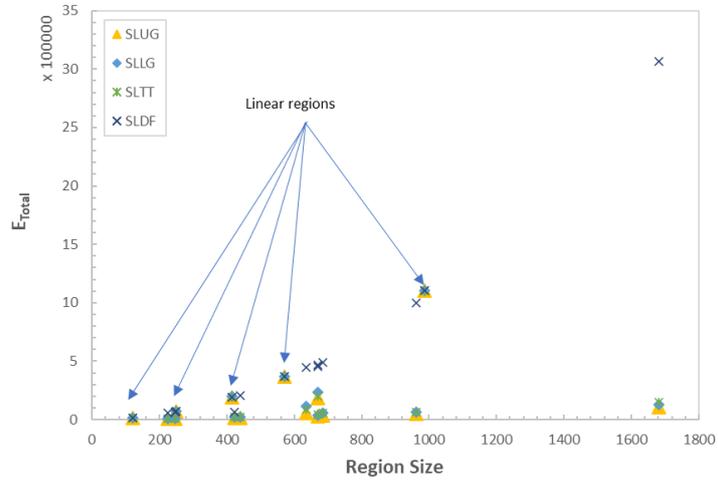

(b)

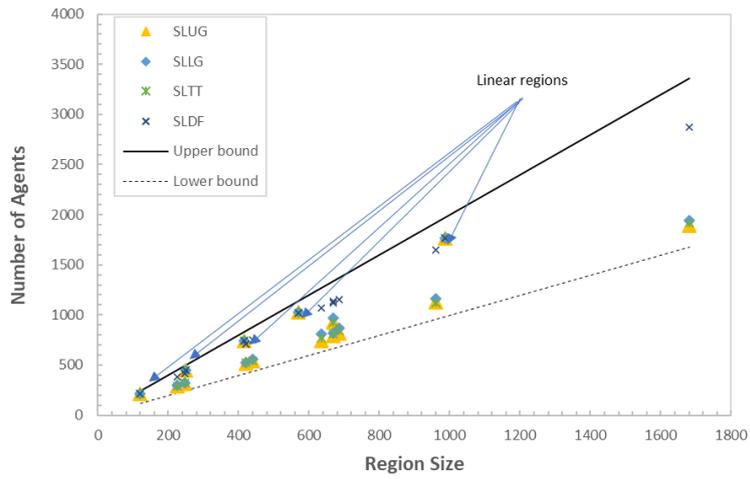

(c)

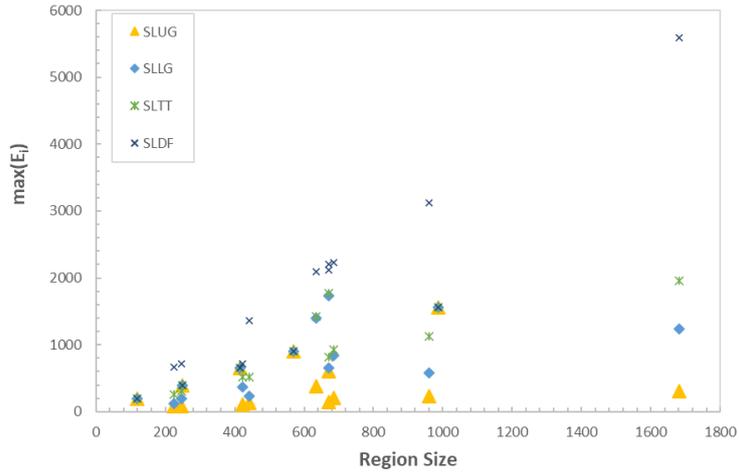

(d)

Figure 53  Single Layer Coverage algorithms performance for $\Delta T = 1$
(a) Termination time; (b) Total energy; (c) Number of agents; (d) $max(E_i)$

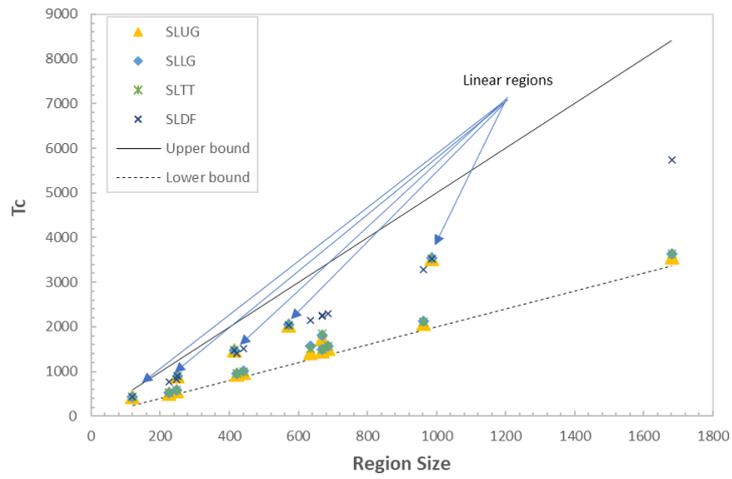

(a)

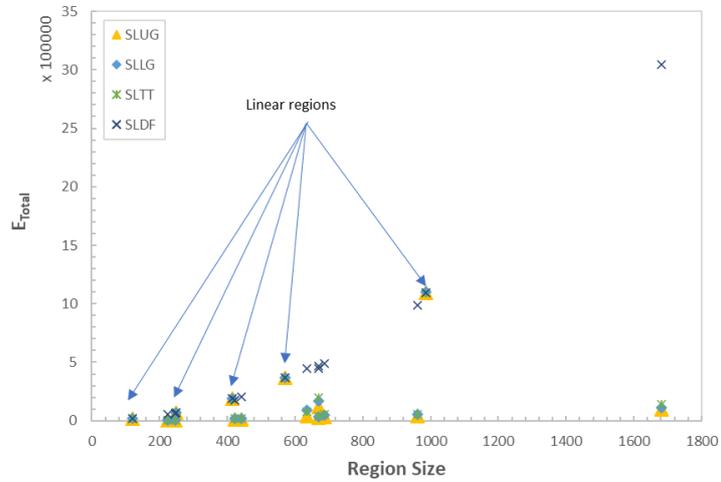

(b)

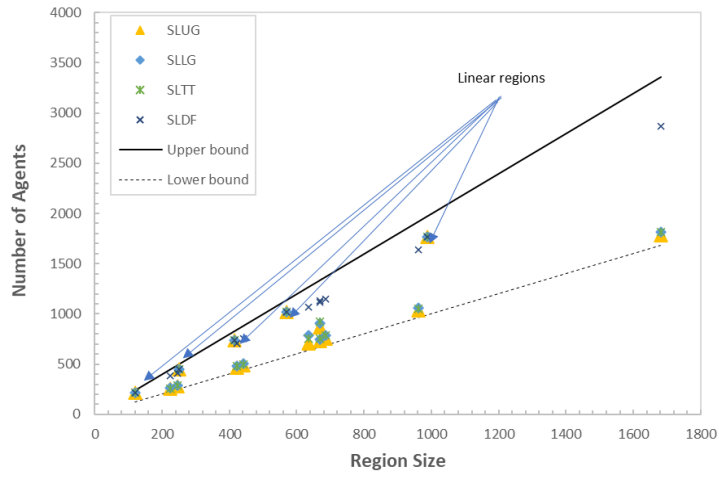

(c)

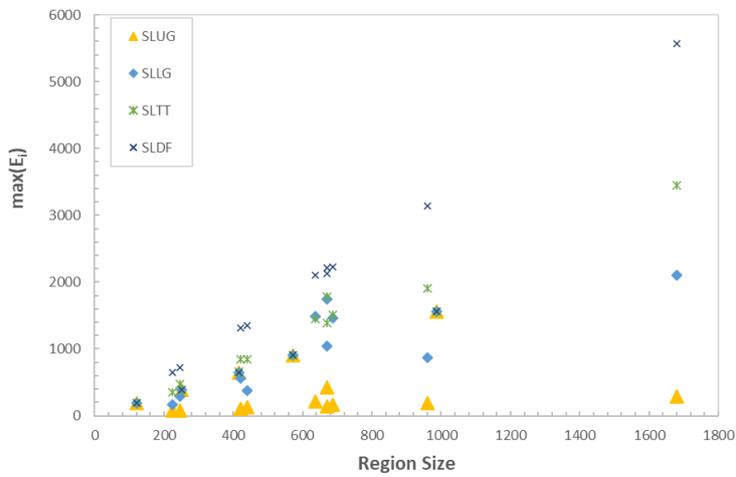

(d)

Figure 54    SLC algorithms performance for $\Delta T = 2$
(a) Termination time; (b) Total energy; (c) Number of agents; (d) $max(E_i)$

In the following figures the effect of $\Delta T$ on the three macro performance metrics is investigated. In Figure 55 the termination time is shown as a function of $\Delta T$ for the various Single Layer Coverage algorithms. While the linear dependency of the termination time on $\Delta T$ is clearly seen the termination times of the different algorithms are not the same. Figure 56 presents the same data only now given as the ratio of the termination time to the region size along with the theoretical upper bound. It is readily apparent that the ratio is linear with $\Delta T$. Furthermore, in all the algorithms that use greedy type exploration – SLUG, SLTT and SLDF - the lines corresponding to the linear regions overlap and are consistently higher than the lines for the other regions. This indicates that in both algorithms the linear case is the worst case. In SLDF, however, all the lines practically merge indicating weak sensitivity to the topology as is expected in depth-first exploration in which the underlying graph is a depth-first tree.

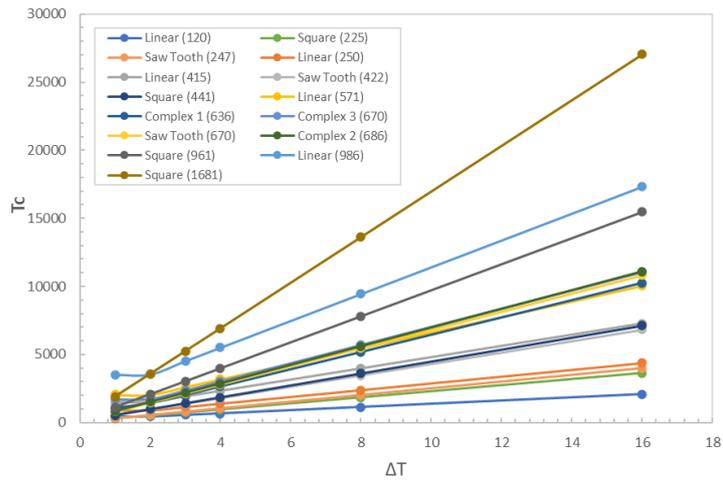

(a)

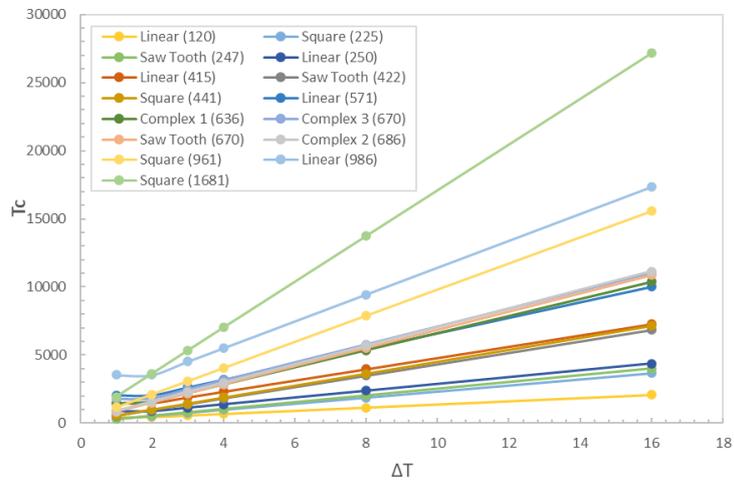

(b)

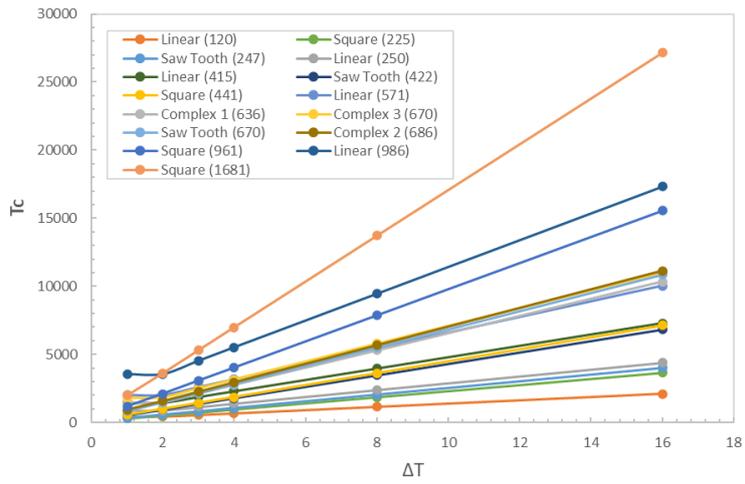

(c)

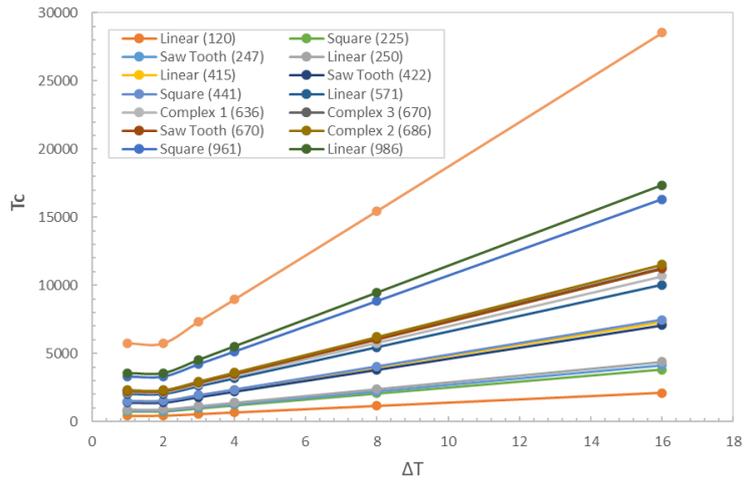

(d)

Figure 55      SLC algorithms - Termination time as a function of $\Delta T$
(a) SLUG; (b) SLLG; (c) SLTT; (d) SLDF

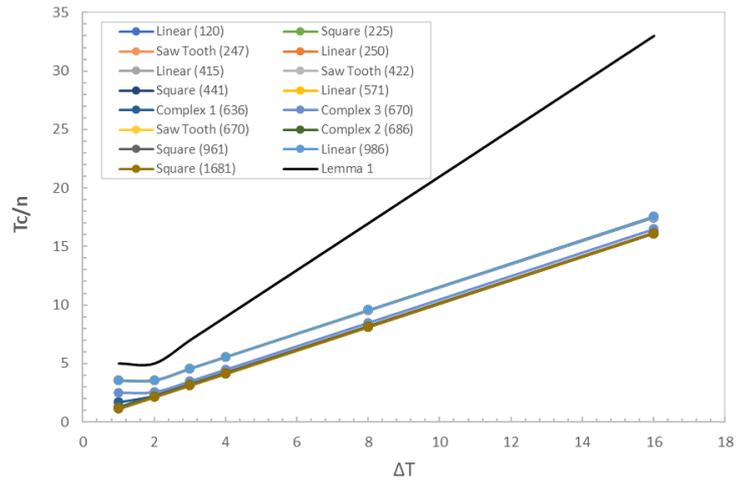

(a)

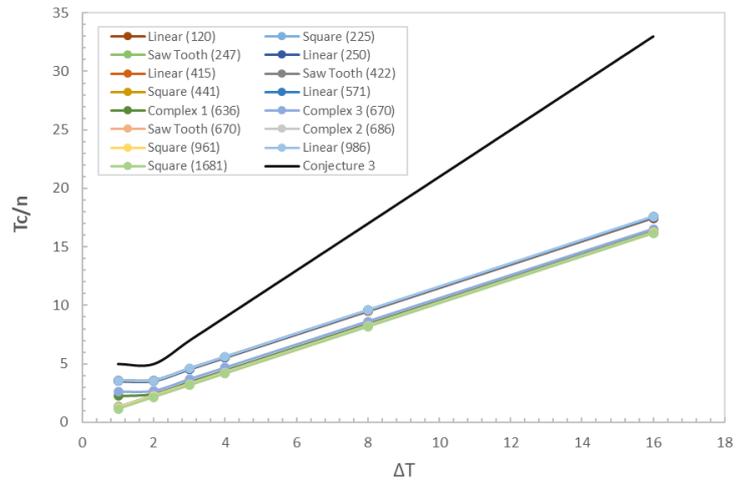

(b)

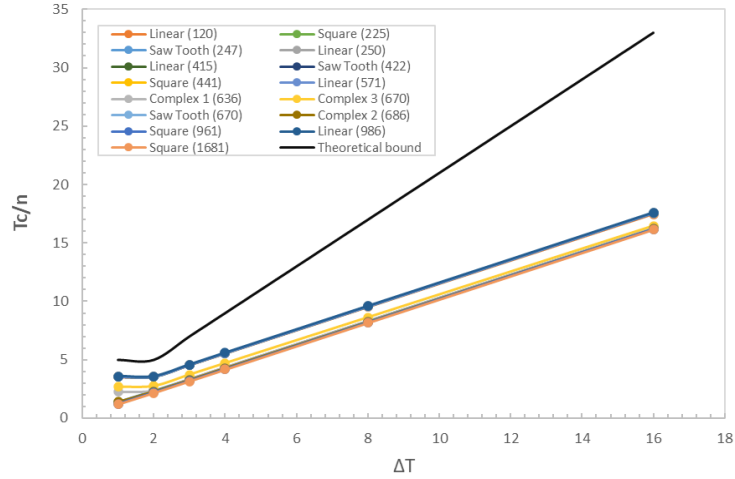

(c)

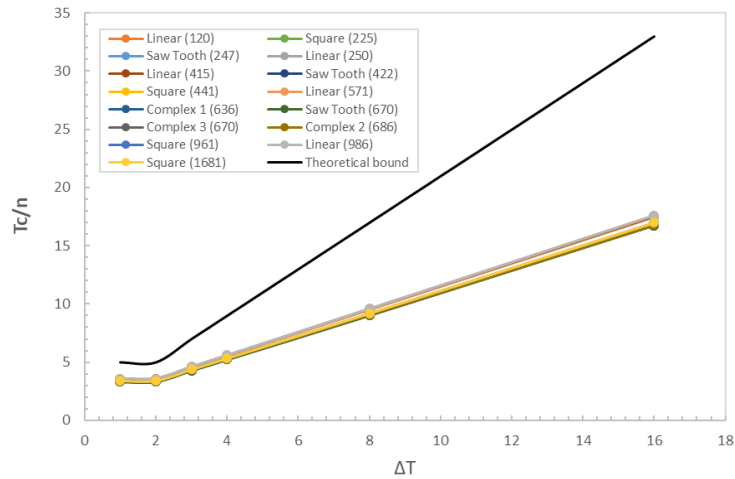

(d)

Figure 56  SLC algorithms – Ratio of Termination time to region size as a function of $\Delta T$
(a) SLUG; (b) SLLG; (c) SLTT; (d) SLDF

The effect of $\Delta T$ on the number of agents used in the process is shown in Figure 57 as the ratio between the number of agents used and the region size. This normalized representation of N enables easy comparison between the different regions. As all the linear regions used exhibit the same dependency on $\Delta T$ they are plotted as one curve. We see in all the sub-figures a similar trend of a decrease in the number of agents as $\Delta T$ increases. This is because the Close indication has more time to propagate as $\Delta T$ increases and thus better funneling of the entering agents to the empty parts of the region is achieved. Furthermore, the ratio converges to 1 (i.e., $N \approx n$) at $\Delta T = 16$. In other words, further increase in $\Delta T$ will yield negligible reductions in the number of agents used. It is interesting to see that the steepness of the gradient (i.e., the rate the number of agents decreases for a unit increase in $\Delta T$) differs from one algorithm to another algorithm and in some cases between the different types of regions in a given algorithm. The simpler the region is for exploration (with the linear region being an extreme case) the swifter the reduction in the number of agents and vis-versa. In addition, the effect of the region on the ratio is much smaller in SLDF and the behavior closely resembles that of the linear regions. Both are due to the depth-first exploration strategy.

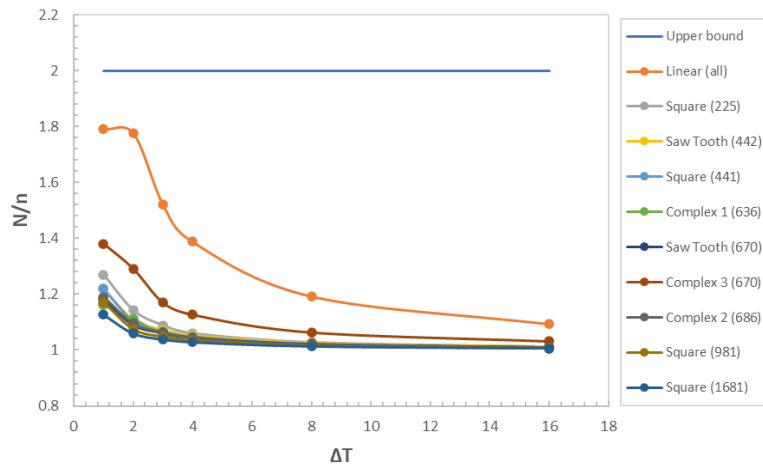

(a)

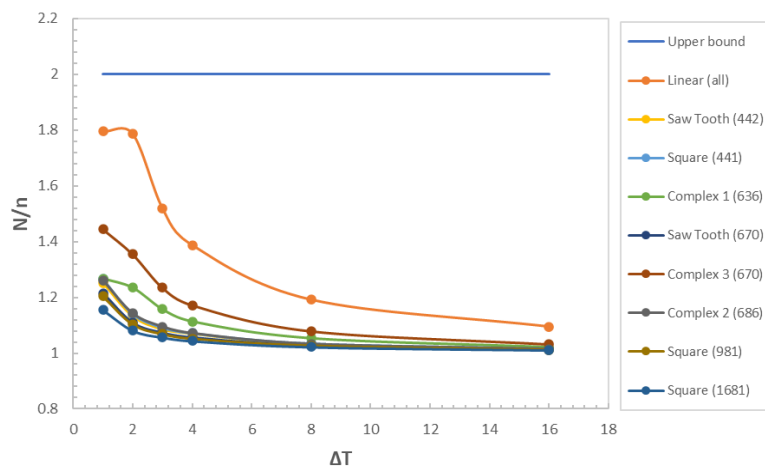

(b)

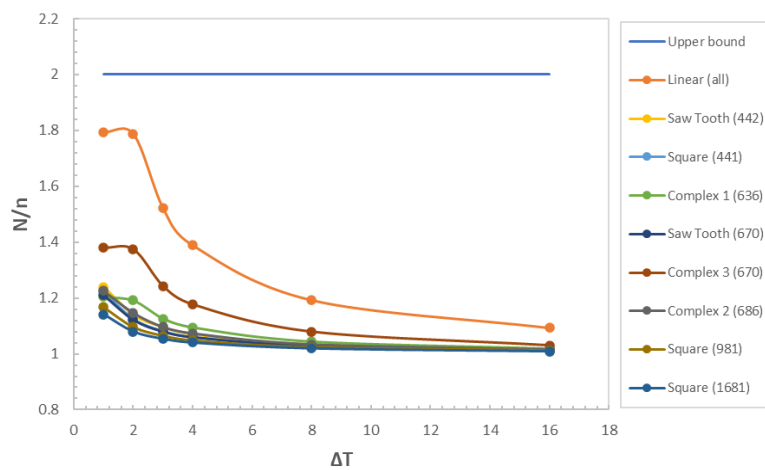

(c)

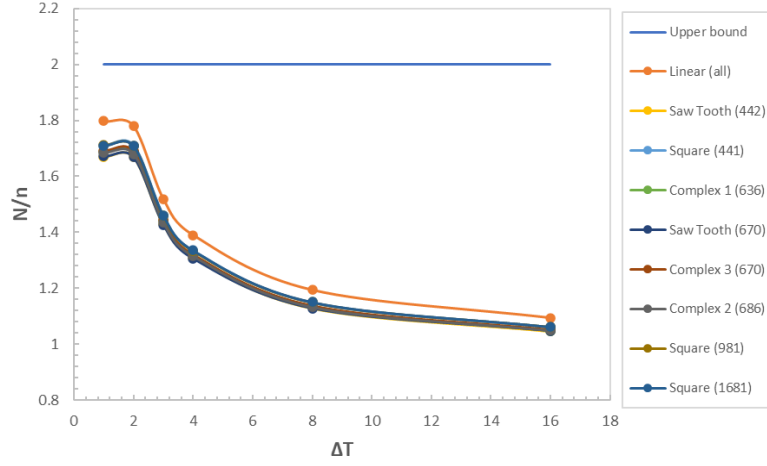

(d)

Figure 57 The ratio of number of agents used to region size as a function of $\Delta T$ in the various SLC algorithms. (a) SLUG; (b) SLLG; (c) SLDF; (d) SLTT. The legend details the types and sizes of the regions

$E_{Total}$ as a function of the region size is shown in Figure 58 for $\Delta T = 4, 8$ and $16$.. SLUG uses the smallest amount of energy followed closely by SLTT in all regions and regardless of $\Delta T$. This finding is surprising since SLUG has the loosest traversal rules and richest DAG compared to SLTT which is limited to traversal on a tree. Additionally, and regardless of $\Delta T$, all the SLC algorithms give the same results when the region is linear. In all sub-figures, a trendline of the total energy as a function of region size in the linear regions is presented. It is clear from the trendlines that the total energy in linear regions upper bounds the total energy in all other regions. Moreover, it is a quadratic function of $n$ (as was noted above for $\Delta T = 2$). Comparison of the results with those of the Dual Layer Coverage algorithms (Figure 52) shows that in the DLC algorithms $E_{Total}$ increases with $\Delta T$ while in the SLC algorithms it decreases. This is due to two reasons. First, in the DLC algorithms increasing $\Delta T$ increases the time that stationary mobile agents (i.e., those that advanced up the gradient and then stopped) consume energy. Second, in the SLC algorithms increasing $\Delta T$ results in the more effective funneling of the mobile agents to the relevant parts of the region. This is due to improved propagation of the Close indication.

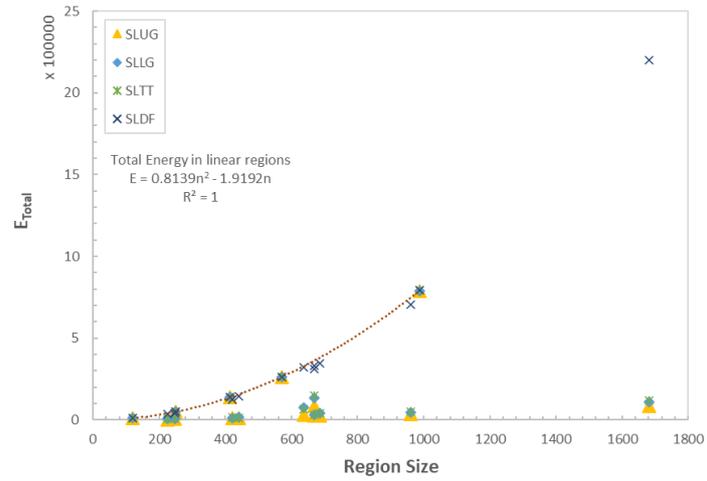

(a)

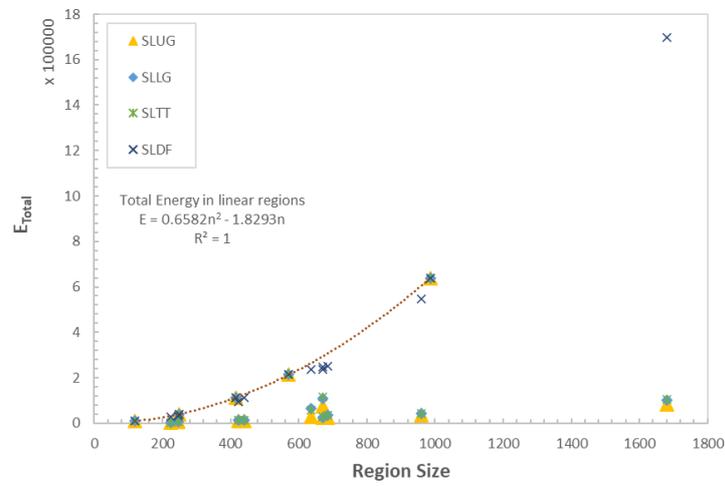

(b)

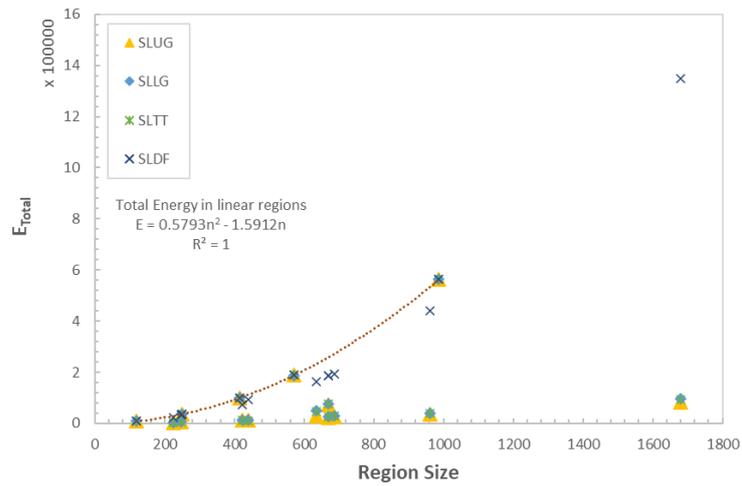

(c)

Figure 58　　Total energy as a function of region size using Single Layer Coverage algorithms
(a) $\Delta T = 4$; (b) $\Delta T = 8$; (c) $\Delta T = 16$

## 6.3 Effect of the entry point location

The effect of the entry point location on the performance was investigated using three regions – two simple, connected regions of approximately 420 cells and complex region #1 with 636 cells. To the best of the authors knowledge there are no previous results on the matter. The different locations in each region are marked and enumerated in the figures below with the original entry point denoted by "o". At each entry point location, the termination time and total energy were measured using the two Dual Layer Coverage algorithms and four Single Layer Coverage algorithms described above. Below each figure the termination time and total energy are given for each of the locations as both the absolute values and more importantly as the ratio of the measured quantity to the same quantity at the original location. Points (a)-(d) in region Complex #1 are characterized by an outdegree[‡] of 2, points (e)-(h) with an out degree of 3 and points (i)-(l) with an out degree of 4. The main takeaways are:

1. In a given region, SLDF is indifferent to the location of the entry point with regard to both the termination time and total energy.
2. The termination time and total energy are more effected by the proximity of the entry point location to the geometrical center of the region than the outdegree of the cell. This is clearly seen when comparing the termination times and total energies associated with entry points of **d** and **l** in the complex region. It's clear that while the out-degree of **d** is 2 compared to 4 in **l** the termination time and total energy are noticeably better in **d** compared to **l**. It should be pointed that in some regions (e.g., the square) the change in both metrics due to moving the entry point from the corner to the center of the region is quite small. Practically this means that, if possible, rescue workers should strive to place the entry point near the geometric center of the region.
3. Predominantly the DLC algorithms behave alike in all combinations of regions/entry points evaluated as do the four SLC algorithms.
4. In all cases the theoretical upper bound on the termination time is correct.

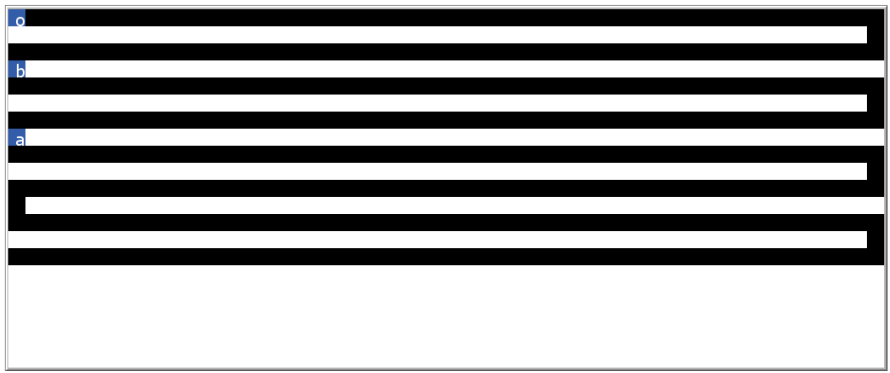

Figure 59        Linear region with 415 cells and three locations of the entry point

---

[‡] The outdegree of a vertex $v$ in a directed graph $\mathbb{G}$ is defined as the number of edges going out from vertex $v$.

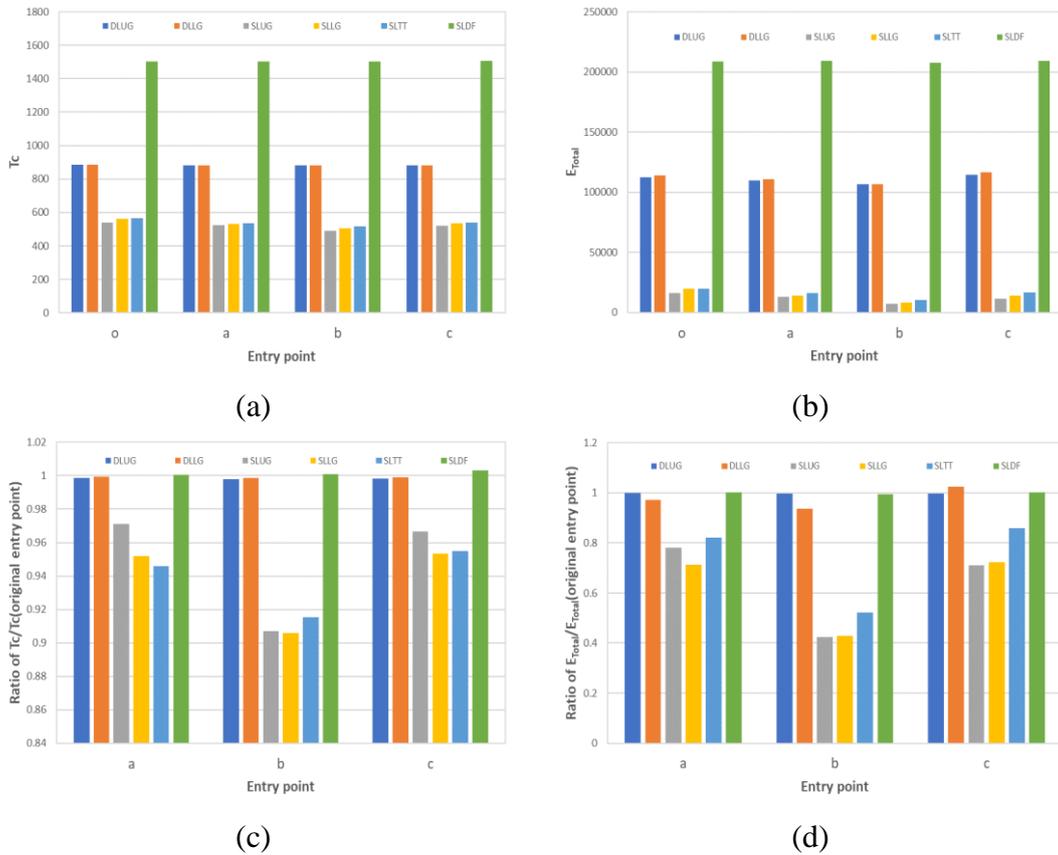

(a)            (b)

(c)            (d)

Figure 60     Effect of entry point location in a linear region with 415 cells points using the different DLC and SLC algorithms. (a) Termination time; (b) Total energy; (c) Ratio of termination time for an entry point to the termination time using the original entry point, (d) Ratio of total energy for an entry point to the total energy using the original entry point

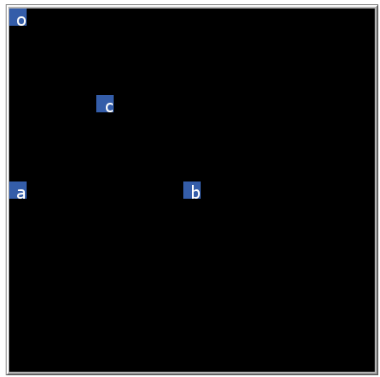

Figure 61     Square region with 441 cells and four locations of the entry point

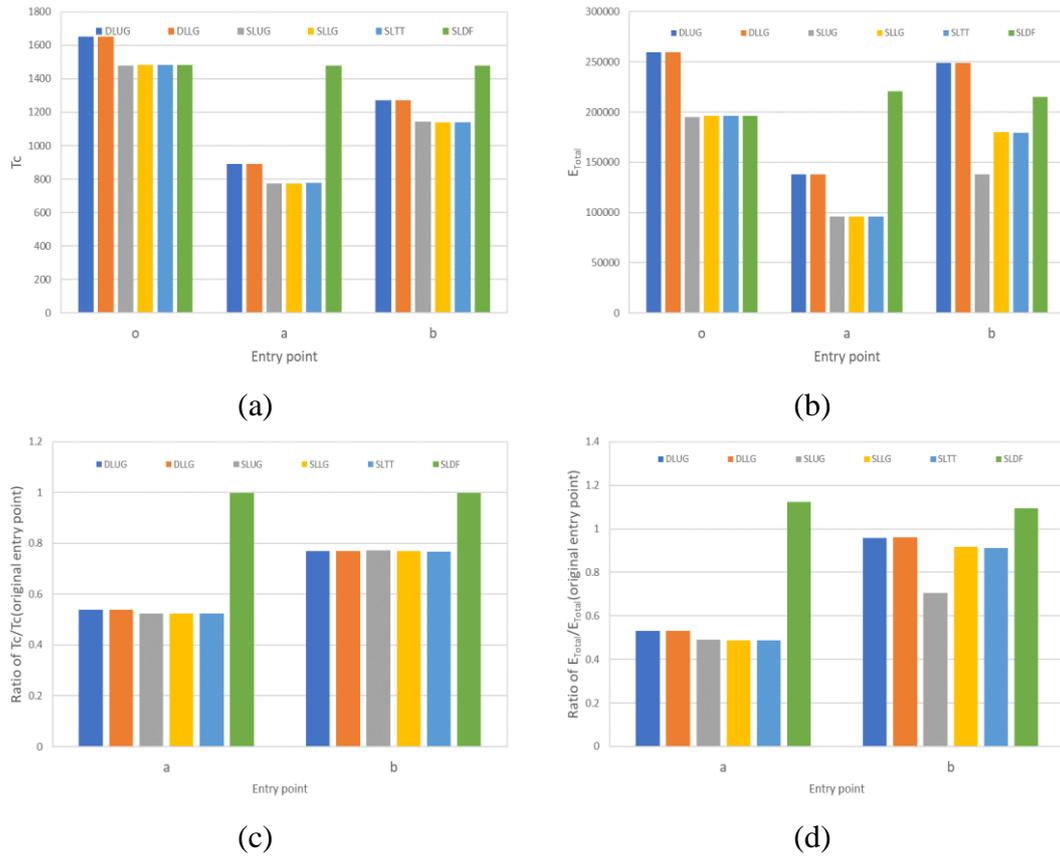

(a)                                   (b)

(c)                                   (d)

Figure 62        Effect of entry point location in a square region with 441 cells points using the different DLC and SLC algorithms. (a) Termination time; (b) Total energy; (c) Ratio of termination time for an entry point to the termination time using the original entry point, (d) Ratio of total energy for an entry point to the total energy using the original entry point

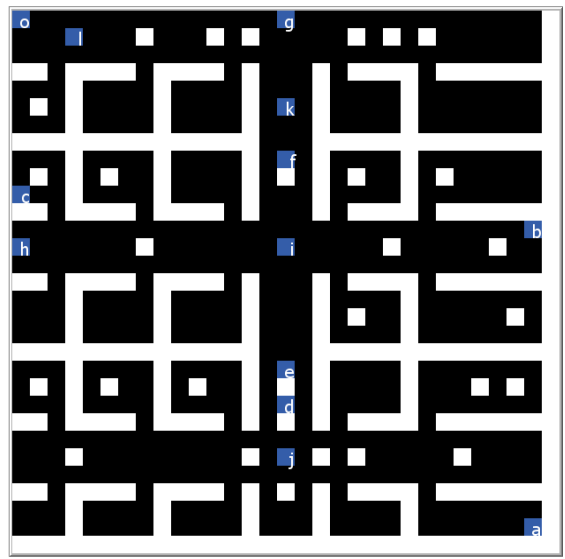

Figure 63        Complex region #1 with 636 cells and 13 locations of the entry point

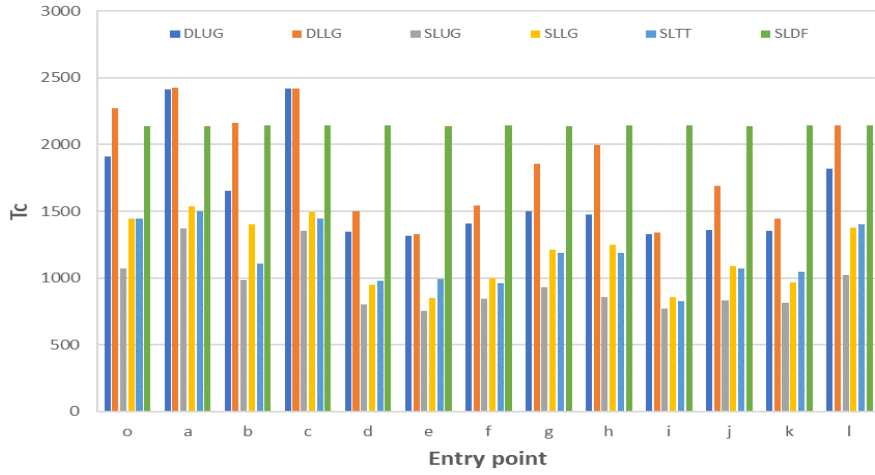

(a)

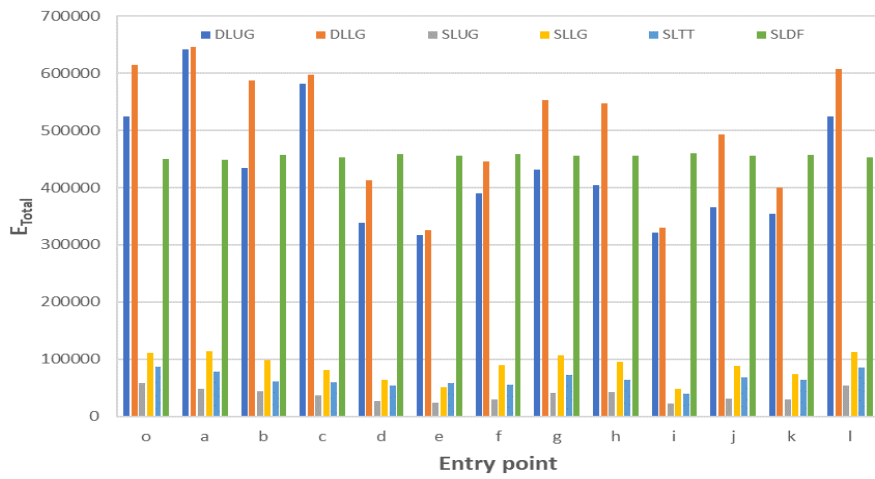

(b)

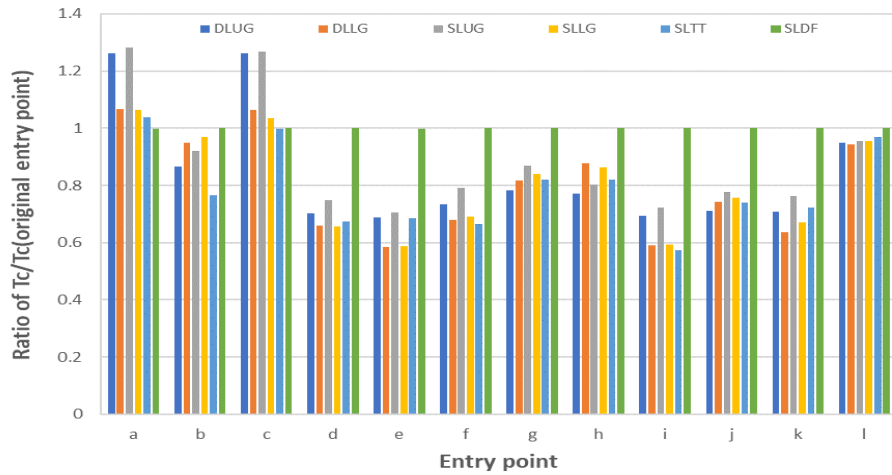

(c)

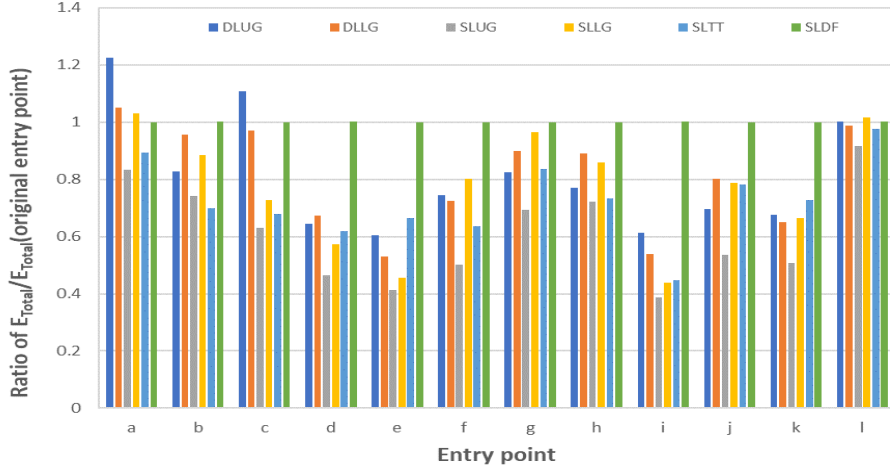

(d)

Figure 64    Effect of entry point location in a complex region with 636 cells points using the different DLC and SLC algorithms. (a) Termination time; (b) Total energy; (c) Ratio of termination time for an entry point to the termination time using the original entry point, (d) Ratio of total energy for an entry point to the total energy using the original entry point

## 6.4  Effect of the traversal rules

The effect of the looseness of the traversal rules on both $T_C$ and $E_{Total}$ is shown in Figure 65 as the ratio (of both metrics) between the different Dual Layer Coverage algorithms. Using the DLC algorithms gives the clearest perspective on the effect of the traversal rules since this is the only difference between the algorithms. In this comparison we also look at the two-layer coverage algorithm which inspired Single Later, Tree Traversal and is described in [44]. We naturally name it – Dual Layer, Tree Traversal (DLTT). The agents in that algorithm generate a directed tree over the region hence the traversal rules in this algorithm are the most limited of the three. In all the plots, the metric of the Dual Layer, Unlimited Gradient algorithm is in the denominator since DLUG has the loosest traversal rules hence when the ratio is greater the 1 it indicates DLUG has better performance and vis-versa when the ratio is less than 1. Figure 65(a) shows the ratio of the termination times when $\Delta T = 1$ ratio (when $\Delta T \geq 2$ the $T_C$ in all three DLC algorithms is the same as shown above). It is clear from the ratio of $T_C$ between DLLG and DLUG that the effect of the traversal rules is minor regardless of the region type. However, when looking at the ratio of $T_C$ between DLTT and DLUG the effect of the traversal rules is both noticeable and dependent upon the region type. In other words, traversal on a DAG decreases the $T_C$ compared to traversal on a tree while the effect of the number of edges in the DAG on $T_C$ is negligible. Similarly, Figure 65(b)-(d) show the effect of the looseness of the traversal rules on $E_{Total}$ for several values of $\Delta T$. The effect of the traversal rules on $E_{Total}$ is more pronounced between DLTT on the one hand and DLLG and DLUG on the other. Moreover, while the ratio is almost not dependent on $\Delta T$ it is upon the topology of the different regions. To summarize, the traversal rules effect both the termination time and the total energy and this is due to the associated underlying graph with the most significant change caused when the underlying graph structure is a tree versus a DAG.

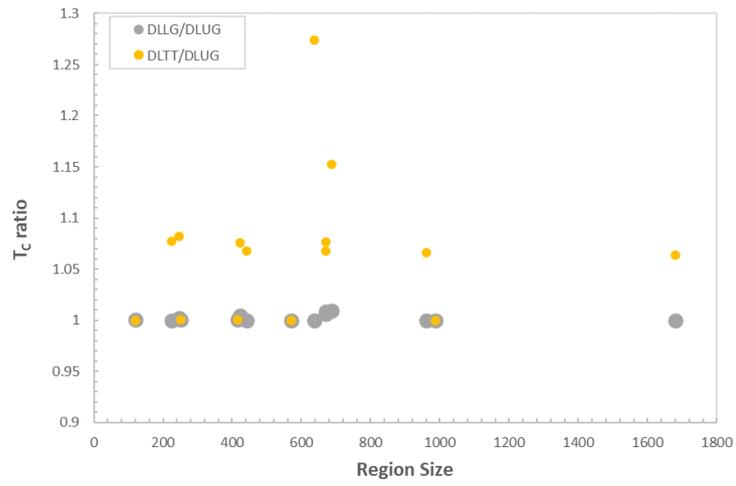

(a)

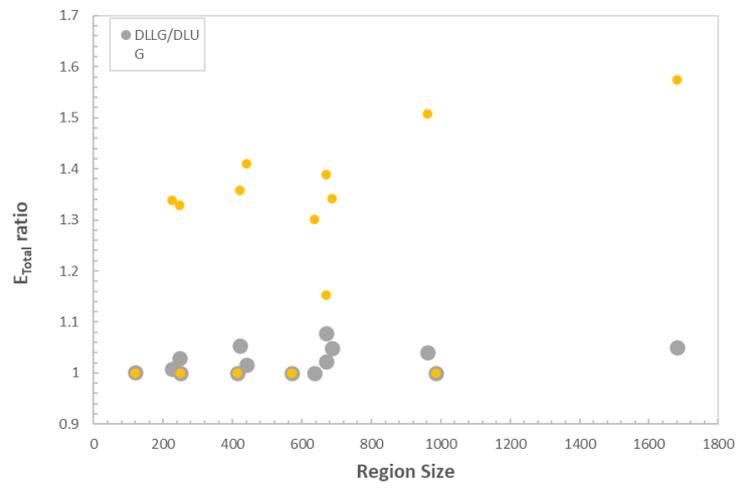

(b)

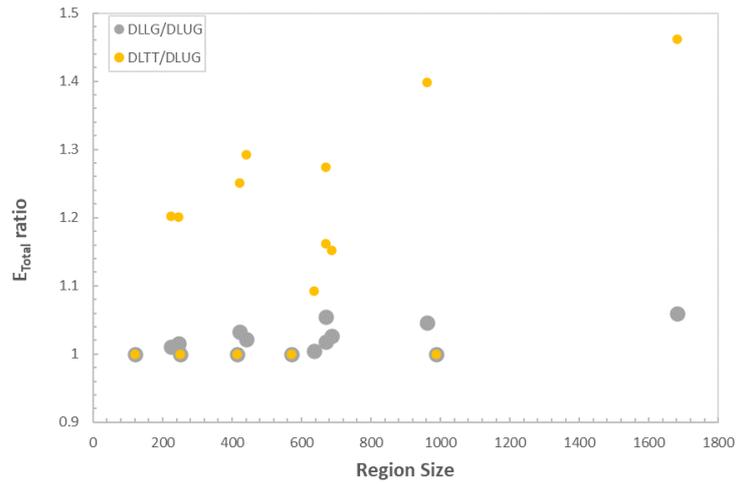

(c)

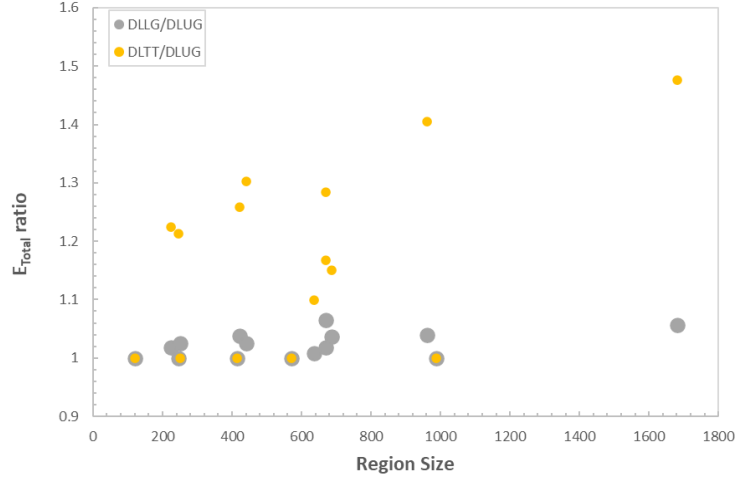

(d)

Figure 65          Dual Layer Coverage algorithms - Effect of traversal rules on $T_C$ and $E_{Total}$
(a) $T_C$ ratio for $\Delta T = 1$ ; (b) $E_{Total}$ ratio for $\Delta T = 1$; (c) $E_{Total}$ ratio for $\Delta T = 2$ ; (d) $E_{Total}$ ratio for $\Delta T = 4$

## 6.5 Effect of the Backward Propagating Closure

In this section the performance improvements achieved by using the Backward Propagating Closure meta-concept are reviewed. The main difference between the DLC and SLC algorithms is due to this concept that simultaneously generates a deterministic indication of termination to an external observer, funnels the advancing agents to the relevant parts of the region and enables the return of the superfluous agents to the entry point/s. The improvements due to the BPC concept are investigated by using the ratio of the three performance metrics when using a DLC algorithm and when using the complementary SLC algorithm (e.g., $T_C^{DLUG}/T_C^{SLUG}$) for $\Delta T = 1, 2$ and 4. By using the ratios rather than the absolute values the improvement is immediately obtained. Specifically, the higher the ratio the bigger the improvement in $T_C, E_{Total}$ and N due to the BPC meta-concept. Several facts are readily apparent from Figure 66 that shows the ratios of the number of agents used.

1. There is a significant reduction in N which increases as $\Delta T$ increases.
2. The reduction in the number of agents used in the linear regions is independent of region size or traversal rules but depends on $\Delta T$. This means that for a given $\Delta T$, $T_{BPC}$ is linear with size since the number of agents in the DLC algorithm is always $2n$.
3. For a given $\Delta T$, the ratio $N^{DLUG}/N^{SLUG}$ is always the largest of the three ratios. This means that the reduction in $N$ is maximal in SLUG (since the number of agents in the DLC algorithm is always $2n$) or in other words the contribution of the BPC is maximal in SLUG.
4. In most cases $N^{DLTT}/N^{SLTT}$ is greater than $N^{DLLG}/N^{SLLG}$ which means that the BPC is more effective (in reducing N) in SLTT then in SLLG. At first glance this is surprising since traversal in SLTT is more constrained compared to SLLG and it was shown above that DLLG out-performs DLTT (Figure 65). The reason for above result is the different definition of Closed

Beacon used in SLTT compared to SLTT. Transition of a settled agent in SLLG to Closed Beacon in most cases depends on at least as many neighboring cells as in SLTT if not more. This is a direct outcome of Eq. (22).

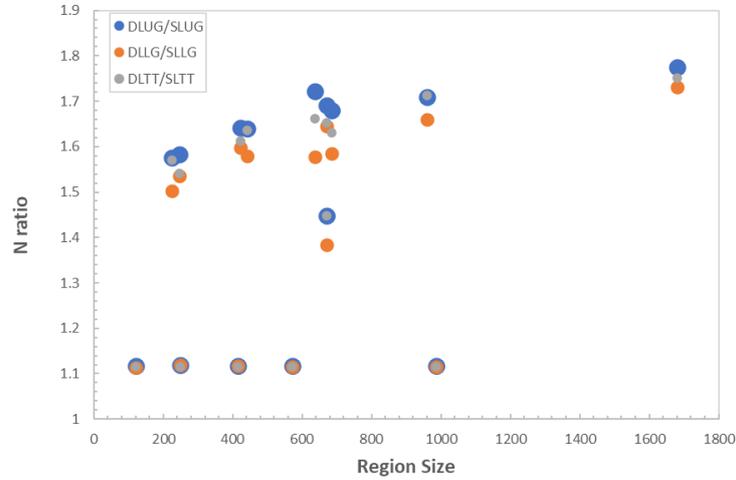

(a)

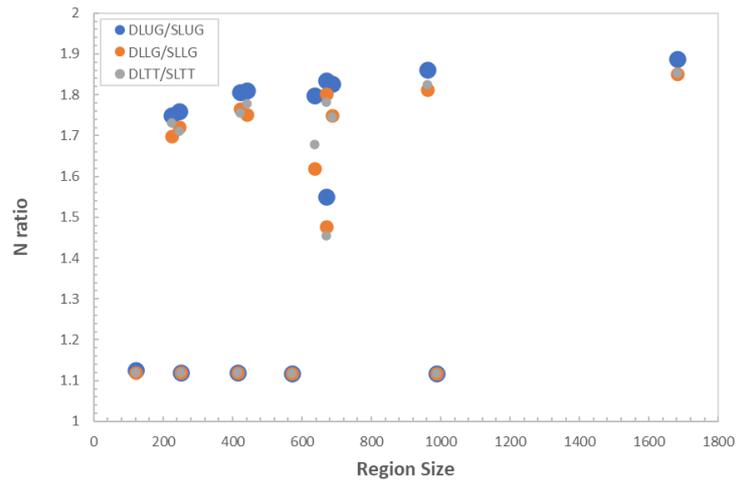

(b)

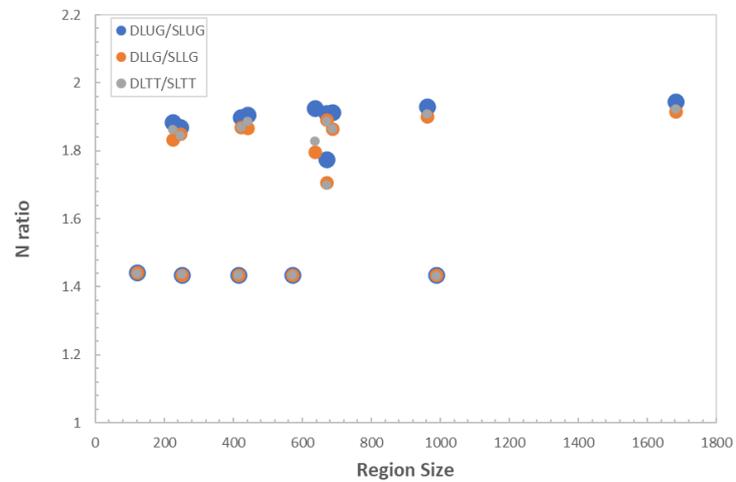

(c)

Figure 66      Backward Propagating Closure effect on $N$
(a) $\Delta T = 1$ ; (b) $\Delta T = 2$; (c) $\Delta T = 4$

Figure 67 shows the ratios of the termination times of comparable DLC and SLC algorithms. In the linear regions the ratio is constant and greater than one for a given $\Delta T$ which means that the SLC algorithms (on average) terminate faster than the DLC algorithms. However according to the ratio of upper bounds derived in the previous sections the ratio should be smaller than one. The reason for this discrepancy is the use of an adversarial scheduler in the derivation of the expressions. While this assumption gives the upper bounds its probability of occurring is extremely small. Hence, when the wake-up order is uniformly distributed the BPC propagates significantly faster and consequently both the process terminates faster and the number of agents participating the process is smaller. Additionally, for a given $\Delta T$ and region size, the ratio of $T_C^{DLUG}/T_C^{SLUG}$ is the largest of the three indicating (again) that the biggest improvement is achieved when using SLUG.

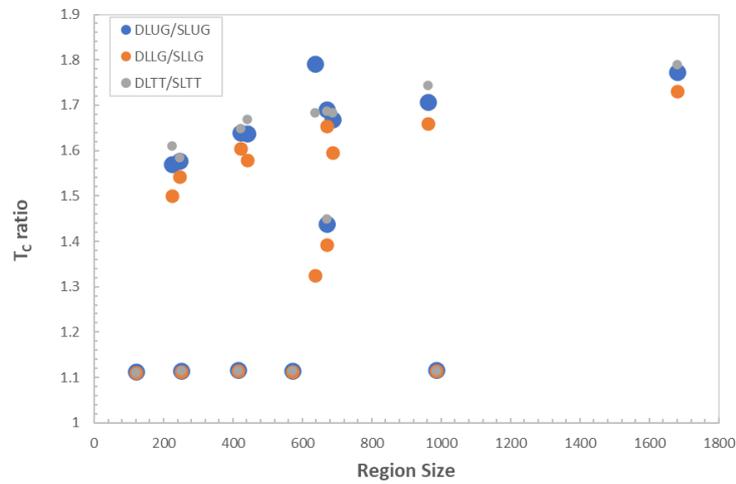

(a)

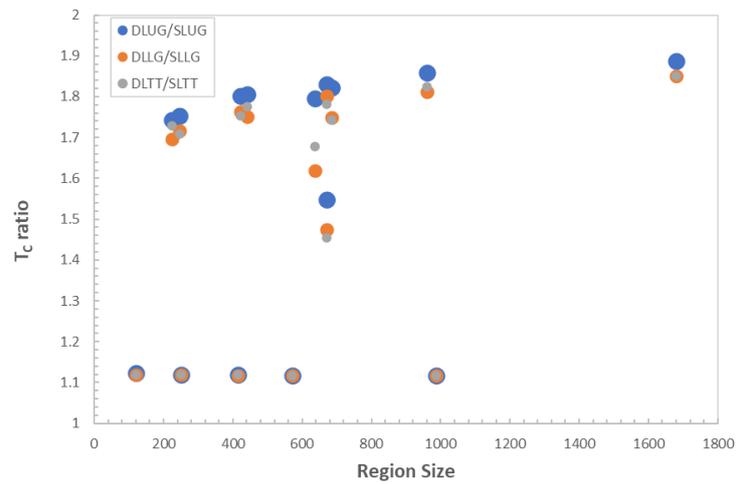

(b)

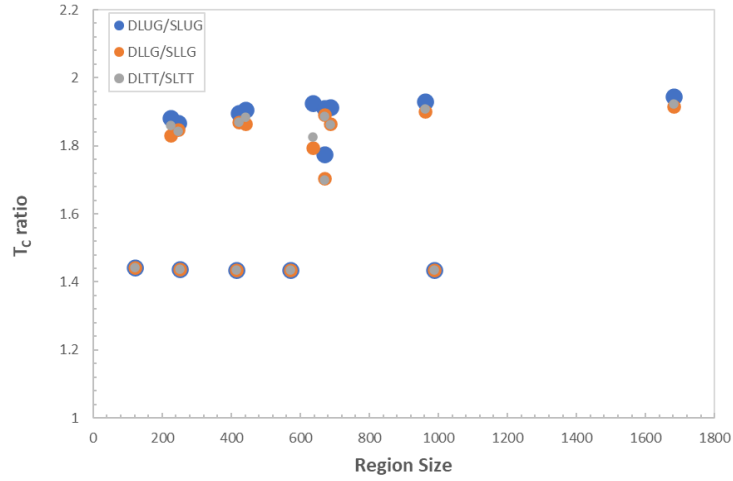

(c)

Figure 67    Backward Propagating Closure effect on $T_C$
(a) $\Delta T = 1$ ; (b) $\Delta T = 2$; (c) $\Delta T = 4$

The reduction in the total energy due to the BPC concept is clearly seen Figure 68. For a given region size the ratio of $E_{E_{Total}}$ of comparable DLC and SLC algorithms is greater than 1 when $\Delta T = 1$ and increases with $\Delta T$. This increase is the result of the increase in both the energy consumption by the stationary agents in DLC and the effectiveness of the funneling in the SLC algorithm. And again, the greatest improvement is when comparing the SLUG to DLLG. Summing we see that use of the BPC significantly improves performance of the SLC algorithms in addition to giving the user the deterministic indication of termination.

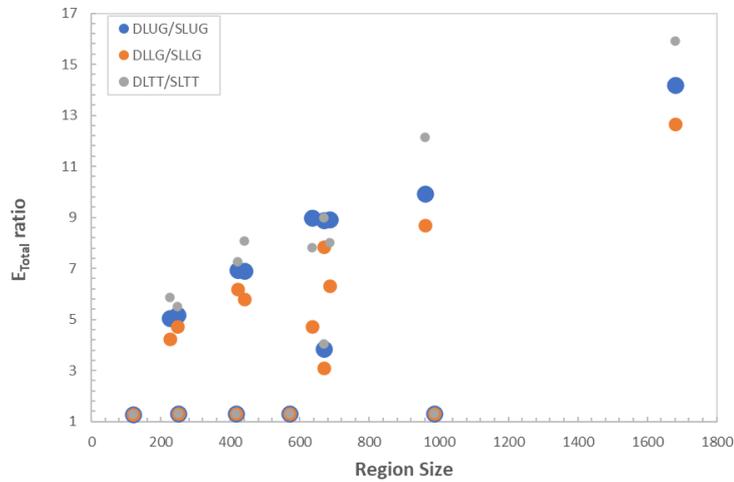

(a)

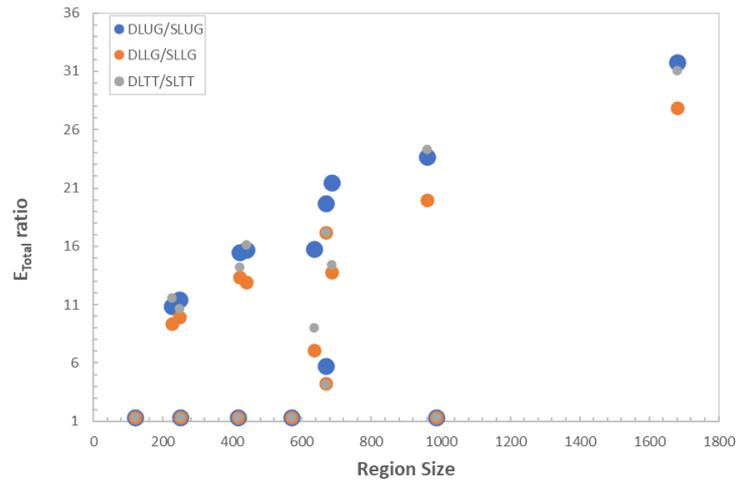

(b)

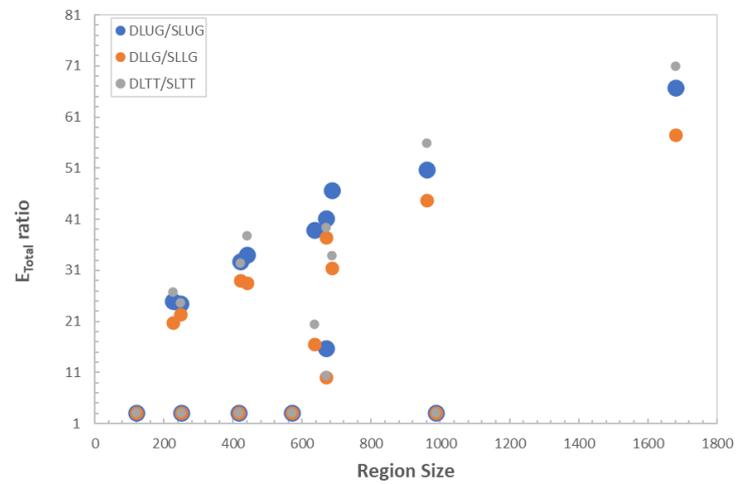

(c)

Figure 68  Backward Propagating Closure effect on $E_{Total}$
(a) $\varDelta T = 1$ ; (b) $\varDelta T = 2$; (c) $\varDelta T = 4$

In this section the performance of the each of the algorithms was reviewed and compared to the other algorithms. In the DLC algorithms the termination time is linear with respect to size and $\varDelta T$ for $\varDelta T \geq 2$ while the total energy is quadratic with the region size with the coefficient of the quadratic term being linear with $\varDelta T$. In the SLC algorithms the correctness of the upper bound on the termination time is corroborated by the experimental results. A correlation between the looseness of the traversal rules and the termination time was found. In addition, the effect of the entry point location depends mostly on its proximity to the geometrical center of the region and to a lesser degree on outdegree. Most importantly, the benefit of using the BPC meta concept was investigated quantitively and found to be significant.

# 7 Conclusions

Seven algorithms based on two meta-concepts are described in this paper. All the algorithms are based on the meta-concept of dual use agents - as mobile explorers and as settling agents that function as beacons. The Single Layer Coverage algorithms also utilize a second meta-concept – the Backward Propagating Closure.

Good correlation is found between the theoretical bounds of the termination time and the experimental results. Moreover, the experimental results led to several conjectures about the termination time in algorithms in which the upper bounds are yet to be analytically developed. The primary effect on the termination time is the region size, however in the SLC algorithms the effects of region topology (esp. the number of holes) and the entry point location are not negligible.

The termination time shortens as the sensed data and complexity of the algorithms increases. With the DLRW being the simplest algorithm and only the presence of the settled agents used to guide the mobile agents the termination times are correspondingly the worst. A large improvement is seen with the introduction of the step count as a means to guide the exploration with a corresponding increase in the sensing requirements and the algorithm's complexity. Use of the Backward Propagating Closure meta concept in the SLC algorithms further reduces the number of agents used, total energy and the termination time.

Further work remains in deriving explicit, theoretical bounds on the termination time. This is especially true regarding the case of $\Delta T = 1$ and/or the unlimited steepness version of the algorithms. Another issue that merits further research is the effect of multiple entry points on the various metrics. While the algorithms presented were designed with multiple entry points in mind, both the mathematical analysis and numerical experiments in this research dealt with the case of a single door.

The Single Layer Coverage algorithms are fault tolerant with a couple of minor modification however the topic as a whole and specifically the interaction between the different failure modes merits further research.